\documentclass[12pt, a4paper, oneside]{book}
\usepackage{graphicx}
\usepackage{amsmath}
\usepackage{fancyhdr}
\usepackage{psfrag}
\usepackage{cite}
\usepackage[latin2]{inputenc}
\usepackage{doublespace}
\newcommand{\td}{t_{\mathrm{d}}}
\newcommand{\tdiss}{t_{\mathrm{diss}}}
\newcommand{\be}{\begin{equation}}
\newcommand{\ee}{\end{equation}}
\newcommand{\bea}{\begin{eqnarray}}
\newcommand{\eea}{\end{eqnarray}}

\newcommand{\emptypage}
{
\thispagestyle{empty}
}

\fancypagestyle{plain}
{
\fancyhead[RO]{\slshape\footnotesize\leftmark}
\fancyhead[RE]{}
\fancyhead[LO]{}
\fancyhead[LE]{\slshape\footnotesize\leftmark}
\fancyfoot[C]{}
\fancyfoot[LE, RO]{\thepage}
}

\title{Highly Nonclassical Quantum States and Environment Induced Decoherence}
\author{P\'eter F\"oldi}
\date{2002}

\begin{document} 
\begin{spacing}{1.0}
\frontmatter
\emptypage
\begin{flushright}
\emptypage
\Large 
{\textsc{Highly Nonclassical Quantum States \\and Environment Induced Decoherence}}
\end{flushright}
\newpage
\begin{center}
\emptypage
\vspace* {0.5cm}
\Huge{\textsl{Highly Nonclassical Quantum States 
\\and Environment Induced Decoherence}}
\vskip 2cm
\Large{PhD Thesis} \\
\vskip 2.0cm
written by \\
\vskip 0.3cm
\LARGE{\textbf{P\'eter F\"oldi}}
\vskip 7.0cm
\Large
Supervisor: Dr.~Mih\'aly G.~Benedict\\
\vskip 2 cm
Department of Theoretical Physics\\
University of Szeged\\
Szeged, Hungary\\
2003.
\end{center}
\setcounter{page}{0}
\markboth{CONTENTS}{}
\tableofcontents
\listoffigures
\mainmatter
\addcontentsline{toc}{part}{PART I}
\newpage
\chapter*{Introduction \markboth{INTRODUCTION}{}}
\addcontentsline{toc}{chapter}{Introduction}
\emptypage
The discovery of the quantized nature of the electromagnetic 
radiation and atomic energies followed by the foundation of Quantum
Mechanics is one of the greatest achievements in physics.
Quantum theory is found to provide excellent description
of fields and elementary particles as well, and it became
a standard tool for investigating ``microscopic'' physical
objects. 

Dirac 
postulated the superposition principle to be
a fundamental concept even before the canonical Hilbert-space
formulation of the theory has been established. 
Although the predictive power of
quantum mechanics is based on this principle, 
it is counter-intuitive for a human mind that
experiences a world of classical mechanics since the beginning of
consciousness. 
The most famous example showing the incompatibility of 
the superposition principle with the usual way of thinking was
given by Schr\"odinger \cite{SCH35a,SCH35b}, where the fate of a cat
in a box is triggered by the decay of a radioactive atom. If the
duration of this {\emph {gedanken experiment}} equals to the half-life
of the atom, it is obvious that the survival probability 
of the cat is $\frac{1}{2}$. More surprisingly, the result 
will be an entangled state, the \emph{superposition} of a dead 
cat with a decayed atom \emph{and} a cat alive with an 
undecayed atom. The two states that
form the superposition -- as a matter of life and death --
are clearly distinct. Such a  superposition of two classically distinguishable 
states, which is usually called  
a Schr\"odinger-cat state, is  allowed 
in quantum systems, but never observed in everyday life.  
In fact, the term ``classically distinguishable states''
means states that can be interpreted in a classical world, where,
due to the lack of the superposition principle, their superposition
is not present. (Note that here and throughout this thesis
the attribute ``classical'' stands for the opposite of ``quantum''.)

In fact, most of the quantum mechanical states have properties that
are unusual from the classical point of view, and therefore, in some sense,
they are nonclassical. 
On the other hand, Schr\"odinger-cat states are in such a strong contradiction 
with the classical description of a physical system, that they can be
called \emph{highly} nonclassical quantum states without exaggeration.
The Wigner function of these states is negative over some regions
of its domain, which is the signature of (high) nonclassicality from our 
point of view.

An apparent implication of the quantum effects that are paradoxical 
from the classical point of view is that there exists a classical and a 
quantum realm in nature, with their respective laws. 
The quantum realm is usually identified with microscopic particles, 
but sometimes it is difficult to draw a non-flexible quantum-classical border.
E.g., considering a fullerene $(\mathrm{C}_\mathrm{60})$ molecule, 
there are experimental situations, such as  
the scattering of highly charged ions on $\mathrm{C}_\mathrm{60}$,
when a fully classical model provides agreement with the 
measured results \cite{H00}. 
On the other hand, this cage
of $60$ carbon atoms surrounded by $360$ electrons produces interference
fringes on a screen placed behind a grating 
\cite{A99}, that is, the molecule as a whole
exhibits genuinely quantum behavior. 

Alternatively, we can assume that quantum mechanics is universal, 
and so is the superposition principle. In this case, however, the
emergence of classical properties has to be explained in the framework
of quantum theory. There is a name for this fundamental problem: decoherence.
In other words, decoherence is the disappearance of the quantum 
superpositions that distinguishes a  Schr\"odinger-cat state from
the corresponding  classical mixture describing a system that is 
\emph{either} in one of the classical states \emph{or} in the other.
Models for decoherence usually describe it dynamically, that is,
decoherence is considered as a \emph{process} that is extremely fast
on everyday timescales.

Besides its fundamental importance, exploring the mechanisms of
decoherence can have practical applications as well. The recently born
and rapidly developing field of quantum information technology relies on the
quantum nature of the physical objects that store, carry and 
process information. This is the very origin of the classically 
unreachable computational power of quantum algorithms.
From this point of view, decoherence is the most serious obstacle 
still hindering the practical use of quantum computation (QC) \cite{NC00}.
Knowing the way in which decoherence destroys quantum superpositions
renders it possible to find promising decoherence-free states.
These states are exceptionally robust, they
keep their quantum properties for a time hopefully long enough
for implementing quantum algorithms.

In this work we consider the decoherence model that is based on the 
interaction of the investigated quantum system with its unavoidably
present (quantum) environment.
In Chap.~\ref{introchap} we summarize the basic concepts of this model, 
which is called environment induced decoherence.
Chap.~\ref{methodchap} is devoted to the description
of the methods that are useful in setting up and solving the
relevant dynamical equations, which, besides realizing the
conceptually important link between quantum and classical mechanics, 
provide a realistic description of open quantum systems.

In the second part of the thesis
we use these methods in order to analyze nonclassicality and 
investigate the effects of decoherence in concrete quantum systems. 
The presented results are based on the publications 
\cite{FBC98,FCB00,FCB01a,FBC01,FC02,FBC02b,FBC02a,F02}. 

In Chap.~\ref{oechap} we investigate the time evolution of wave 
packets in the anharmonic Morse potential, which can provide a realistic
model for a vibrating diatomic molecule. This chapter deals with
the situation when no environmental effects are present.
Using the Wigner function of the system, we show that for vibrations
with amplitudes exceeding the limits of the harmonic approximation,
spontaneous formation of Schr\"odinger-cat states occur. These highly 
nonclassical states are superpositions of two distinct states that are
localized both in position and momentum. 

As a result of the environmental influence, the Schr\"odinger-cat states
are expected to disappear rapidly, and it is known that when the potential is 
is approximated by a harmonic one, the result of the decoherence will
be the mixture of the constituent localized states.
Our analysis in Chap.~\ref{morsedecchap} shows that this is not the case
for the Morse oscillator. We introduce a master equation for a
general anharmonic system in interaction with a thermal bath of
harmonic oscillators. 
Using this equation we find that decoherence drives the system
into a density operator that can be interpreted as the mixture of
localized states equally distributed along the phase space orbit 
of the corresponding classical particle. That is, the information
related to the position along this orbit (``phase information'') 
is completely lost.
On the contrary, after the process of decoherence, the energy distribution 
of the system is still quite sharp, in fact the expectation value 
of the Hamiltonian is very close to its initial value. Because of the 
separation of the time scales of decoherence and dissipation, these processes
can be clearly distinguished. We define the characteristic
time of the decoherence as the time instant when the 
transition between the decoherence dominated and dissipation  dominated
time evolution takes place.

In Chap.~\ref{atomchap} the same definition is proven to be valid
for a system of two-level atoms interacting with the free radiational field.
This model offers the possibility of investigating the approach to the
macroscopic limit by increasing the number of atoms. We found that
the larger is this number, the more naturally ans sharply the time
evolution splits into two regimes.	
In this physical system the atomic coherent states \cite{ACGT72} can be given
a clear classical interpretation, they correspond to certain directions 
of the Bloch vector \cite{MS91}.
Therefore superpositions of different atomic coherent states are rightly called
atomic Schr\"odinger-cat states. We show by analytical short time calculations
that the coherent constituents of these
highly nonclassical states are robust against the effects of decoherence. 
Consequently, the decoherence
of the atomic Schr\"odinger-cat states is expected to lead to the 
classical mixture of the constituent atomic coherent states.
We obtain that this conjecture is true, unless decoherence is exceptionally
slow. In Chap.~\ref{atomchap} we give a scheme of decoherence that
remains valid also for the so-called symmetric Schr\"odinger-cat states,
which exhibit exceptionally slow decoherence.  

The basic object which is manipulated in QC algorithms 
is a \emph{qubit}, which is an abstract two-level quantum system. 
A system of two-level atoms can provide a physical realization
of a sequence of qubits. The usefulness of this realization
depends on the extent to which the difficulties related to the 
decoherence can be eliminated. In this context the
the possible  preparation of decoherence-free states is important,
this issue is discussed in Chap.~\ref{subprepchap}.
We consider the atoms to be in a cavity, and propose 
a method that can prepare decoherence-free states.
Besides the free time evolution in the cavity,
our scheme requires the possibility of changing the state of one of the
atoms on demand. The analysis of these requirement shows that our scheme
can be implemented with present day cavity QED setups.

\chapter{Environment induced decoherence}
\emptypage
\label{introchap}
\label{EIDsec}
The apparent lack of a superposition  of macroscopically distinct quantum
states (Schr\"{o}din-ger cats) has been an interesting and vivid problem since
Schr\"{o}dinger's famous papers \cite{SCH35a,SCH35b}. 
A successful approach, initiated by Zeh \cite{Z70} and 
developed by Zurek \cite{Z81}, obtains the loss of quantum 
coherence as the consequence of the inevitable interaction 
with the environment. Theoretical studies 
in this framework have investigated a variety of model systems usually coupled
to a collection of harmonic oscillators as an environment. 
Fundamental work has been done on this subject
in Refs.~\cite{Z82,CL83,UZ89,HPZ92,D81,SW85,D89,BJK99}, 
for reviews see \cite{Z93,GJK96}.
Important experiments have also been carried out during the last 
years \cite{BHD96,MKT00,FPC00}.

We note that whatever successful is the approach of the environment 
induced decoherence, it is not the only possible mechanism that can 
explain the phenomenon of decoherence.  
Spontaneous collapse models are conceptually different, they are based
on an appropriately modified Schr\"odinger equation, which automatically
leads to classical behavior for large systems. We shall not
consider these models here, a review can be found in 
Chap.~8 of Ref.~\cite{GJK96}.
The role of gravity is also often discussed in both of the two
main approaches, see \cite{K86,P86,D89,EMN89}.

\section{Formation of system-environment entanglement}
Apart from cosmology, we usually focus our interest on a specific part of the 
universe. This distinguished subsystem (our ``system'', $S$) is, 
however, unavoidably coupled to the ``rest of the world'', called
environment ($E$) in this context. Although the way in which we
single out our system can appear accidental or even artificial, it is clearly
necessary to obtain a useful, solvable model. Additionally, measurements 
performed on $S$ are in most of the cases ``local'', i.e., concern
the degrees of freedom of the system only. (In fact, the definition of the
``system'' in a theoretical model is closely related to the possible
measurements the outcomes of which are to be predicted.) Neglecting 
the $S$-$E$ interaction leads to results that are good approximations 
only for very well isolated systems and for short times. 
For a more realistic description of the necessarily open quantum system $S$,
the effects of the environment have to be taken into account.
The system-environment interaction builds up entanglement
(\emph{Verschr\"ankung}) between the 
the two quantum systems $S$ and $E$. In order to obtain results for the 
system only, we have to average over the unobservable environmental 
degrees of freedom. This process of ``tracing out the environment'' 
(see the next chapter) provides a density operator
of $S$ that usually describes a mixed state and contains all the information
that can be extracted by local measurements. 

In order to illustrate this concept, we consider a simple but expressive 
example \cite[pp.~41-42]{GJK96} with the interaction term
\be
V_{int}=\sum_n|n\rangle_S{}_S\langle n| \otimes B_E(n)
=\sum_n|n\rangle_S{}_S\langle n| B_E(n)
\label{EIDint}
\ee
connecting $S$ and $E$. The states $\{|n\rangle_S\}$ 
are assumed to form an orthogonal basis in the Hilbert space of the system, 
while $B_E(n)$ denote (Hermitian) environmental operators. The 
operators $|n\rangle_S{}_S\langle n|$ 
act in the Hilbert space of the system, and similarly $B_E(n)$ stands for
the tensorial product of the identity $\mathrm{id}_S$ with the
environmental operator $B_E(n)$. In what follows, when it is not necessary,
the tensorial product sign will be omitted in the notation.

Note that $V_{int}$ is special in the sense that
it does not contain cross terms like $|n\rangle_S{}_S\langle m|$, $n\neq m$, 
but it demonstrates the main effects well. Later on we shall consider more
general interactions as well. As a further approximation, we neglect
the self-Hamiltonians of $S$ and $E$ for the moment.  
Assuming an initially uncorrelated state 
\be
|\Psi(t=0)\rangle_{SE}=|\phi(0)\rangle_S|\Phi(0)\rangle_E=
\sum_n c_n |n\rangle_S\ |\Phi(0)\rangle_E,
\label{EIDprod}
\ee
the time evolution builds up $S$-$E$ correlations and leads to 
an entangled state:
\be
|\Psi(t)\rangle_{SE}=\sum_n c_n |n\rangle_S\ e^{\frac{-iB_E(n) t}{\hbar}}
|\Phi(0)\rangle_E =\sum_n c_n |n\rangle_S \ |\Phi\rangle_E^n,
\label{EIDent}
\ee
where $|\Phi\rangle_E^n=\exp(-iB_E(n) t/\hbar)|\Phi(0)\rangle_E$. This result
can be verified by Taylor expanding the time evolution operator 
$\exp(-iV_{int}t/\hbar)$.
The local or reduced density operator of the system is
\be
\rho_S=\mathrm{Tr}_E (\rho_{SE})
=\mathrm{Tr}_E \left[|\Psi\rangle_{SE}\ {}_{SE}\langle\Psi|\right],
\ee
where the operation $\mathrm{Tr}_E$ means trace over environmental degrees
of freedom. Initially
\be
\rho_S(0)=\sum_{n m}c^*_m c_n |n\rangle_S{}_S\langle m|,
\ee
and as Eq.~(\ref{EIDent}) shows, it evolves according to 
\be
\rho_S(t)=\sum_{n m}c^*_m c_n \ |n\rangle_S{}_S\langle m| \ 
{}_E^m\langle\Phi|\Phi\rangle_E^n.
\ee 
That is, in the basis defined by the interaction (\ref{EIDint}),
the off-diagonal elements of $\rho_S$ are multiplied by
the overlap of the corresponding (time dependent) environmental states,
while the diagonal elements remain unchanged. Depending on the form of
the operators $B_E(n)$, after a certain time the states $|\Phi\rangle_E^n$
can become orthogonal and hence the interaction diagonalizes the
reduced density operator of the system
\be
\rho_S(0)\rightarrow\sum_n \left|c_n \right|^2 |n\rangle_S{}_S\langle n|.
\label{EIDdec}
\ee
This phenomenon, the decoherence, can be expressed as the 
apparent collapse of the
state of the system: $\rho_{SE}$ at this time instant
still represents a pure state, but the phase relations of the states 
$\{|n\rangle_S\}$ are inaccessible for a local observer.
The RHS of Eq.~(\ref{EIDdec})
is formally identical with the density operator that would be
the result of a von Neumann-type measurement \cite{N32}
corresponding to the operator 
$\sum_n|n\rangle_S{}_S\langle n|$. The notion that the environment
continuously measures, or monitors the system, is understood in this
loose sense, without assuming the collapse of $|\Psi(t)\rangle_{SE}$. 

The essential reason for the disappearance of the interference terms of
$\rho_S(t)$ in the above example was the entanglement 
of the two systems $S$ and $E$. Similarly to the 
case of an EPR pair \cite{EPR35,B87}, 
where it is impossible to assign a pure state to one of the constituents of
the pair, $\rho_S$, which initially described a pure state, turns into
a mixture. This feature of the system-environment interaction is
present also in more sophisticated models where there is
an interplay between the interaction and internal dynamics of  
$S$ and $E$ governed by the self-Hamiltonians, see 
the second part of this work. 
Also in these more general situations the (by assumption pure) 
system + environment state $\left| \Psi \right\rangle_{SE}$
can be written in the Schmidt representation \cite{SCH07,KZ73,EK95,GJK96} 
at any time as 
\be 
\left| \Psi (t)\right\rangle_{SE} =\sum_{k}\sqrt{p_{k}(t)}\left| \varphi
_{k}(t)\right\rangle_S \left| \Phi _{k}(t)\right\rangle_E
\label {schmidt},
\ee
where the positive numbers $p_k$ add up to unity and 
$\left| \varphi _{k}\right\rangle_S $\ and $\left| \Phi
_{k}\right\rangle_E$ are elements of certain orthonormal 
bases (Schmidt bases) of the system and the environment, respectively.
A comparison shows that Eq.~(\ref{EIDent}) is a special case of
this generally valid representation, with 
$\left| \varphi _{k}\right\rangle_S=|k\rangle_S$ 
and $\left| \Phi _{k}(t)\right\rangle_E=|\Phi\rangle_E^k$, 
apart from a possible phase factor. 
Note that if the dimensionality of any of the involved Hilbert spaces is finite
then the number of nonzero terms in the sum (\ref{schmidt}) is necessarily
also finite. This holds even in the case when $E$ represents a continuum
\cite{SCH07,CLE02}. 

The participation ratio
\be
K=\frac{1}{\sum_k p_k^2},
\label{EIDpartic}
\ee
which is a real number ``counting'' the nonzero 
terms in Eq.~(\ref{schmidt}), can serve as a measure
of entanglement \cite{CLE02}.
For a summary of other approaches 
in quantifying entanglement, see Ref.~\cite{DHR01}. 
The participation ratio is related to the so-called Schmidt number \cite{P98},
which is the integer number of the nonzero coefficients $p_k$ in 
Eq.~(\ref{schmidt}).
However, $K$ is somewhat more practical, especially in numerical 
calculations when exact zeros are difficult to identify.
A product state like
the one given by Eq.~(\ref{EIDprod}), has a single term in its Schmidt sum, 
i.e., $p_{0}(0)=1$ and $p_{k}(0)=0$ for $k\neq 0$, and $K=1$ in this case.
Any interaction is clearly nonlocal (as it couples $S$ and $E$) 
and thus has the capacity of creating entanglement and consequently
increase the participation ratio.

Having a product state $|\varphi_0\rangle|\Phi_0\rangle$ at $t=0$,
the short-time dynamics of entanglement formation can be characterized 
by the decrease of the coefficient $p_0$ in the 
Schmidt decomposition (\ref{schmidt}). According to 
\cite{KZ73}, in leading order in time we can write
\be
p_{0}(t)=1-At^{2},
\ee with the rate of entanglement
\be A=\sum_{k\neq 0,l\neq 0}\left| {}_S\left\langle \varphi _{k}(0)\right|
{}_E\left\langle \Phi _{l}(0)\right| V \left| \varphi _{0}\right\rangle_S
\left|\Phi_0 \right\rangle_E \right| ^{2}.
\label{EIDentrate}
\ee
This quantity can be used to test the stability of a quantum state in
the presence of a given interaction Hamiltonian $V$:
small value of $A$ means that the initial system state $|\varphi_0\rangle$ 
becomes entangled slowly with the environment. 

We note that entanglement -- although it is peculiar from the
classical point of view -- is rather common in quantum systems.
The mere statistics of 2, 3, \dots partite random states shows that
the relative number of non-entangled states is rapidly disappearing
by increasing the number of the parties. More precisely, using an
appropriate measure, numerical evidence shows that the volume of
the separable states decreases exponentially as a function of the 
dimension of the composite system \cite{Z98,Z99}. 

\newpage
\section[Dynamical stability of quantum states]
{Dynamical stability of quantum states and 
\\the  direction of the decoherence}
\label{pointersec}
According to the previous section, the interaction of the
investigated quantum system $S$ and the environment $E$ builds up
$S$-$E$ entanglement. If the reduced density operator of
the system initially represented a pure state, it turns into 
a mixture as a consequence of the interaction. The direction of
the decoherence is related to the question how it is possible to determine
this mixture for a given initial system state.

Let start with a practical method that will be used in Chap.~\ref{atomchap},
where the dynamical equations are solved numerically.
We consider a single two-level atom,
which is clearly a microscopic quantum system. 
A general interaction with the 
environment has a twofold effect: It changes the energy of the atom,
and transforms an initially pure atomic state into a mixture.
A representative example can be the interaction of the atom with
the electromagnetic vacuum.
In this case the time scale of these processes, namely energy dissipation
and decoherence, is roughly the same, see Chap.~\ref{atomchap}. 
However, if we add more two-level
atoms and consider their ensemble as the investigated system,
usually it is possible to distinguish decoherence and dissipation dynamically,
because the characteristic time of the second process is much longer
than that of the first.
Then, soon after the fast decoherence, the reduced density operator
of the system is the  density operator into
which the decoherence has driven the atomic system.

It generally holds, that in ``macroscopic'' quantum systems
the ratio of the characteristic times related to dissipation and decoherence
is much larger than in ``microscopic'' cases.
(We note that depending on the initial state, this ratio $R$ can be
larger than unity even for microscopic objects:
For superpositions of microwave coherent states $R$ can be
controlled between 1 and 10, see Ref.~\cite{BHD96}. 
According to Ref.~\cite{MKT00}, in the case of a single 
${}^9\mathrm{Be}^+$ ion in a Paul trap, the value of 
$R$ can be as much as 25.) 

However, the characteristic time of the decoherence in a given
model (that is, $S$, $E$, and the interaction are specified)
depends on the initial state. The stable or robust states, 
for which this time is exceptionally long, are of special interest. 
These states are usually called pointer states, as they were
introduced in the context of a measurement process, where 
they correspond to the possible ``classical'' states of a measurement 
apparatus \cite{Z81}. Since the formulation of this concept,
pointer states have gained more general meaning, as the most
stable states of a quantum system, which does not need to be a
measurement apparatus. In this work the term ``pointer states''
is used in this extended sense.

Recalling Eq.~(\ref{EIDdec}), it can be seen that in the example 
of the previous section decoherence does not change the states
$\{|k\rangle_S\}$, therefore they are stable indeed.  
The reason for this fact is that every $|k\rangle_S$
is an eigenstate of the operator $\sum_n|n\rangle_S{}_S\langle n|$
that appears in the interaction Hamiltonian given by Eq.~(\ref{EIDint}).
More generally, when $H_S$, the self-Hamiltonian of the investigated
system can not be neglected, but it has common eigenstates with
the interaction term, like $aa^\dagger$ in the phase relaxation 
of the harmonic oscillator \cite{WM85}, the pointer states will be
these common eigenstates (that is, energy eigenstates).    

In more difficult situations, because of the interplay between the 
self-Hamiltonian and the interaction, it is not a trivial task
to identify the stable pure states. One can even construct artificial
models, where it is impossible to find pointer states.
However, e.g. the so-called predictability sieve \cite{ZHP93}, which is method
based on the relatively low entropy production of the pointer states,
works well for most of the physically relevant models.
This approach shows that the coherent states $|\alpha\rangle$ of a
harmonic oscillator are pointer states in different models
\cite[and see also \cite{D93}]{WM85,ZHP93}.
In Chap.~\ref{atomchap} we shall use
Eq.~(\ref{EIDentrate}) to find states for which the entanglement with
the environment builds up slowly. 
  
\vskip 12pt
Having determined the pointer states, an additional interesting question 
is the time evolution of their superpositions. 
We consider pointer states that can be labeled 
by a discrete index, but the possible answers are qualitatively the same 
in more general situations as well.
Recalling again the example of the previous section,
we can see that it is possible that the pointer states form an orthonormal
basis, the elements of which are distinguished by the environment. 
(Formally, this means that we have different environmental operators
$B_E(n)$ for each $n$ in the interaction term 
$V_{int}=\sum_n|n\rangle_S{}_S\langle n| B_E(n)$.)
That is, if $\{|n\rangle_S\}$ denotes the pointer basis, then,
according to the previous section,
we can calculate the result of the decoherence for any initial system state 
$|\psi\rangle$ in a particularly simple way:
\be
|\psi(t=0)\rangle=\sum_k c_k |k\rangle_S \rightarrow 
\rho=\sum_k |c_k|^2 |k\rangle_S{}_S\langle k|.
\label{EIDscheme}
\ee
Note that $|\psi\rangle$ could have been expanded in terms of any 
basis, but in this case the pointer states have the unique
property of satisfying the scheme (\ref{EIDscheme}). Thus, if the 
environment distinguishes the pointer states, then their superposition
rapidly transforms into a mixture. 

The robustness of the pointer states implies that they can survive
long enough to be observed. In fact, the known results show, that
these states have a clear classical interpretation \cite{WM85,ZHP93}.
Therefore the result (\ref{EIDscheme}) is in accordance
with the observation that there are no superpositions of classical
states in our macroscopic world. 

However, it is also possible that the interaction with the environment 
does not draw a distinction between 
some robust system states $\{|n\rangle_S\}_{n=1}^{N}$,
$N>1$. (This can be achieved by setting $B_E(1)=B_E(2)=\ldots=B_E(N)$ 
in the interaction term given by Eq.~(\ref{EIDint}).) 
Now any superposition of these
states are as stable as the pointer states themselves. In other words,
the states $\{|n\rangle_S\}_{n=1}^{N}$ span a 
\emph{decoherence-free subspace} (DFS). 
This possibility is of high importance when decoherence 
should be avoided, such as in a physical realization of quantum computational
methods. Clearly, there are physical systems, where we do not need to
find the pointer states in order to characterize a DFS, simply because
we have additional information that leads directly to the wanted DFS, see
Chap.~\ref{subprepchap}.

\vskip 12pt
The concept of the pointer states and methods 
that allow us to determine them, 
can provide explanations of emergence of classical properties
in an open quantum system. We note that  
superselection rules -- stating that certain quantum superpositions,
such as superpositions of different electric charge states, are not present 
in nature even in the microscopic level -- 
can also be investigated in the framework of environment 
induced decoherence \cite{Z82}.
In fact, the aim of the \textit{program of decoherence} \cite{Z70,Z82} is to
explain all superselection rules under the assumption of a universally
valid quantum theory.

\chapter[Interaction with the environment]
{Description of a quantum system interacting with its environment}
\emptypage
\label{methodchap}

In this chapter we give a brief overview of the usual mathematical tools 
capable to calculate the dynamics of open quantum systems. In these methods
the basic object -- the time evolution of which we are interested in -- 
can be the reduced density operator, or the state vector 
of the system, but it is also possible that a quasiprobability distribution
(QPD) \cite{WM85} of the system is to be calculated directly. 
\vskip 12pt

In the first case the reduced density operator of the system 
obeys non-unitary dynamics that can turn an initially pure
state into a mixture. Considering the system and its environment as 
a single, closed quantum system, the equation that governs the non-unitary
time evolution can be derived. 
One obtains in this way an integro-differential equation, called pre-master
equation that is nonlocal in time. In some cases 
it is possible to introduce approximations which
remove this nonlocality and lead to a differential equation
termed as master equation. In section \ref{MEsec} we illustrate
this process and analyze the role of the Born and Markov approximations
in a rather general example.   

It is also possible to transfer a given master equation 
into stochastic processes that involve the state vector of the system. 
Spontaneous collapse decoherence models (for a review see 
\cite[Chap.~8]{GJK96}) are often make use of the stochastic 
differential equations 
(SDE) \cite{G90} obtained in this way. 
If the reduced density operator of the system can 
be represented by a quasiprobability distribution (QPD), it can
be possible to transform a given master equation into a partial 
differential equation involving the respective QPD.
These methods will be discussed briefly in Sec.~\ref{miscmethsec}.

Note that sometimes not all the information contained by
the state of the system is needed to answer a specific question,
and it is possible to apply a technique that directly leads
to the required answer. E.g., in the case of spontaneous
emission \cite{WW30,WW31,A74} 
from a two-level atom, the quantity of interest is 
the population of the upper (or lower) atomic level and the 
off-diagonal elements of the $2\times2$ reduced density matrix are
in principle irrelevant (although in some models they can be
necessary in order to compute the populations). 
However, in the context of decoherence,
especially when our aim is to determine the pointer states
(Sec.~\ref{pointersec}), the complete state of the system itself is 
to be calculated. Therefore we shall not consider methods that
can be used to obtain the time evolution of a specific 
physical quantity and focus on more general
approaches. Heisenberg picture methods,
such as quantum Langevin equations \cite{MS91},
are not discussed here either.
   
\section{Master equations}
\label{MEsec}
According to the general situation outlined in Chap.~\ref{introchap}, 
we consider a quantum system ($S$) interacting
with its environment ($E$), which can be considered as a heat bath
or reservoir. This means that neither the energy, nor other
macroscopic parameters of the environment can change appreciably 
as a consequence of the system-environment coupling. The environment
as a reservoir is in most of the cases modeled by a large number of harmonic 
oscillators, standing for e.g.~the modes of the free electromagnetic field or
phonon modes in solids. A different, often used model describes
the reservoir as a set of atomic energy levels. 
We note that the logical steps followed in this section 
are not depending on the chosen reservoir model.

\vskip 12pt

Let the total (system plus environment) Hamiltonian be written
in the form
\be
H_{SE}=H_S+H_E+\epsilon\widehat{V},
\label{MEham}
\ee
where the parameter $\epsilon$ in the interaction Hamiltonian
$V=\epsilon\widehat{V}$ expresses the strength of the $S$-$E$
coupling. The starting point here is the von Neumann equation
for the total density operator:
\be
{{d}\over{dt}}\rho_{SE}=-{{i}\over{\hbar}}\left[H_{SE},\rho_{SE}\right],
\label{totalneumann}
\ee
and our aim is to clarify the role of the different approximations
applied in deriving a master equation for the reduced density
operator of the system, 
\be
\rho_{S}={\mathrm {Tr}}_E \  \rho_{SE}.
\ee
The rigorous way to proceed involves the projection techniques
of Nakajima \cite{N58} and Zwanzig \cite{Z60a, Z60b}, where one splits
the information contained in $\rho_{SE}$ into a ``relevant'' and
``irrelevant'' part. In our case, if the system and the environment
are initially uncorrelated, i.e., $\rho_{SE}(t=0)=\rho_{S}(0)
 \rho_{E}(0)$, the relevant part would be $\rho_{S}(t) 
\rho_{E}(0)$. (Note that while $\mathcal{P}\rho_{SE}(t)=\rho_{S}(t) 
\rho_{E}(0)$ defines a proper projection, the map $\rho_{SE}(t)
\rightarrow\rho_{S}(t)$ does not. Besides $\rho_{S}(t)$ a ``reference state'',
that is, an environmental density operator is needed as a result of 
a projection. 
In the above mentioned initially uncorrelated case the 
reference state acquires physical significance as a part 
of $\rho_{SE}(t=0)$.) 

However, the physical meaning of the master equation approach is seen
more clearly by choosing a more transparent method.
In the following we consider a rather general example in a way
similar to the derivation in Ref.~\cite{WM94}, but having performed 
the Born and Markov approximations 
the final equation  will be the same as
if it were calculated using the projection method.
In the current chapter we
concentrate on the generality of the discussion, we point out
what the necessary approximations are when obtaining a master
equation. Later on, in Chap.~\ref{morsedecchap}
this method will be used to treat the specific problem of decoherence
of wave packets in the anharmonic Morse potential. It will be 
also shown that if we assume that $H_S$ has equidistant spectrum
(which is clearly not the case in a general anharmonic system) 
a simpler master equation is obtained that can
describe a system of two-level atoms interacting with the
environment of a thermal photon bath, see Chap.~\ref{atomchap}.

The von Neumann equation (\ref{totalneumann}) in an interaction picture 
reads
\be
{{d}\over{dt}}\rho_{SE}^i(t)
=-{{i}\over{\hbar}}\left[V^i(t),\rho_{SE}^i(t)\right],
\label{interneumann}
\ee
where the interaction picture operators are defined in the following way
\be
\rho_{SE}^i(t)=U^{\dagger}(t)\rho_{SE}U(t), \ \ \ \  
V^i(t)=U^{\dagger}(t)VU(t),
\label{MEintpdef}
\ee
using the unitary operator 
\be
U(t)=e^{-{{i(H_S+H_E)t}\over{\hbar}}}.
\ee
Integrating the 
equation of motion (\ref{interneumann}), we obtain
\be
\rho_{SE}^i(t)=\rho_{SE}^i(0)-{{i}\over{\hbar}}\int_0^t dt_1
\left[V^i(t_1),\rho_{SE}^i(t_1)\right].
\label{MEgeneralstart}
\ee
Iterating this solution and performing the trace over reservoir variables
we find
\bea
\rho_{S}^i(t)=\rho_{S}^i(0)+
\nonumber
\sum_{k=0}^{\infty}\left( -{{i}\over{\hbar}}
\right)^k \int_0^t dt_1 \int_0^{t_1} dt_2 \cdots
\\
\times 
\int_0^{t_{k-1}} dt_{k}
{\mathrm {Tr}}_E \left[ V^i(t_1), \left[ V^i(t_2),\ldots\left[ V^i(t_k),
\rho_{SE}^i(0)\right]\right]\right].
\label{MEsum}
\eea
Now the $k$-th term in the sum is proportional to $\epsilon^k$, 
see Eq.~(\ref{MEham}). If we consider a weak interaction, it is sufficient
to take into account only the first two terms with $k=1$ and $2$.
This is analogous to the usual approach of time dependent perturbation 
theory, and also to the Born expansion of the scattering 
amplitude \cite{CTDL77}.
Therefore the restriction of the interaction to at most second
order is a kind of Born approximation. Sometimes a different approximation,
which will be described later, is also called Born approximation, 
therefore the neglection of higher order terms in Eq.~(\ref{MEsum})
can be termed as the first part of the Born approximation, yielding
\be
{{d}\over{dt}}\rho_{S}^i(t)= -{{i}\over{\hbar}} {\mathrm {Tr}}_E \left[ V^i(t),
\rho_{SE}^i(0)\right]-{{1}\over{\hbar^2}}\int_0^t dt_1 
{\mathrm {Tr}}_E \left[ V^i(t),\left[ V^i(t_1),
\rho_{SE}^i(0)\right]\right].
\label{MEalmost}
\ee 
Assuming that the system and the environment is initially uncorrelated
$\rho_{SE}(t=0)=\rho_{S}(0)\rho_{E}(0)$, $\rho_{E}(0)$
corresponds to thermal equilibrium and
$V$ has no diagonal matrix elements in the eigenbasis of $H_E$, 
the first term vanishes on the RHS of Eq.~(\ref{MEalmost}). 
With these realistic assumptions we have
\be
{{d}\over{dt}}\rho_{S}^i(t)= -{{1}\over{\hbar^2}}\int_0^t dt_1 
{\mathrm {Tr}}_E \left[ V^i(t),\left[ V^i(t_1),
\rho_{SE}^i(0)\right]\right].
\label{MEalmost2}
\ee 

The only approximation made so far was the step from Eq.~(\ref{MEsum})
to Eq.~(\ref{MEalmost}), which was justified by the weakness of the 
perturbation induced by the interaction Hamiltonian.
Clearly, this approximation (as a perturbative result) 
introduces a limit of the applicability
of Eq.~(\ref{MEalmost2}), because for a time $t$ too long, the neglected
terms in Eq.~(\ref{MEsum}) could change the time evolution significantly.
In principle this difficulty could be circumvented by 
dividing the time interval $[0,t]$ into $N$ smaller subintervals
with sufficiently short duration of $\Delta t=t/N$ and applying 
Eq.~(\ref{MEalmost2}) successively. 
Within one of these  short time intervals $[(n-1)\Delta t, n \Delta t]$, 
the replacement of $\rho_{SE}^i((n-1)\Delta t)$ with $\rho_{SE}^i
(n\Delta t)$ in the integrand does not affect that property of the equation of
motion that it is correct up to second order in the interaction.
In this way we introduced a natural coarse graining of the time evolution,
so that ${{d}\over{dt}}\rho_{S}^i(n\Delta t)$ does
not depend on the density operators $\rho_{SE}^i$ that belong to
earlier times. That is, the equation
\be
{{d}\over{dt}}\rho_{S}^i(n\Delta t)= -{{1}\over{\hbar^2}}\int_{(n-1)\Delta t}
^{n\Delta t} dt_1 
{\mathrm {Tr}}_E \left[ V^i(n\Delta t),\left[ V^i(t_1),
\rho_{SE}^i(n\Delta t)\right]\right].
\label{MEalmost3}
\ee    
defines a Markovian sequence of density operators 
$\{\rho_{S}^i(n\Delta t)\}_{n=0}^N$. This step is the Markov approximation.

However, it is difficult to calculate the elements
of this Markovian chain according to Eq.~(\ref{MEalmost3}), 
because in order to be able to perform the trace over the reservoir, 
we have to know the total $\rho_{SE}$ at the starting point of
each short time interval.
The final approximation follows from  the assumption that the state
of the reservoir does not change appreciably due to the interaction.
More precisely, we assume $\Delta t$ to be long compared to the
relaxation time of the environment. Consequently, on the time
scale defined by $\Delta t$, the system-environment 
correlation that builds up due to the interaction affects only the
system. Formally, this second part of the Born approximation is performed
by replacing $\rho_{SE}^i(n\Delta t)$ with 
$\rho_{S}^i(n\Delta t)\rho^i_E(0)$ in Eq.~(\ref{MEalmost3}).

By setting $n\Delta t=0$ and $(n+1)\Delta t=\tau$ the equation of motion 
in the Born-Markov approximation reads:
\be
{{d}\over{dt}}\rho_{S}^i(\tau)= -{{1}\over{\hbar^2}}\int_0^{\tau} dt_1 
{\mathrm {Tr}}_E \left[ V^i(\tau),\left[ V^i(t_1),
\rho_{S}^i(\tau)\rho^i_E(0)\right]\right].
\label{MEthatsit}
\ee     

In summary, the validity of the Born-Markov approximation is 
based on the possibility of the separation of 
the environmental and system time scales: If there are
time intervals which are short enough to allow the
cutoff of the interaction at the second order terms, and, simultaneously,
long enough for the relaxation in the environment to take place,
then the Born-Markov approximation can be used.
We note that the considerations that led from Eq.~(\ref{MEgeneralstart})
to Eq.~(\ref{MEthatsit}) are rather general, 
the only assumption concerning the interaction Hamiltonian
was that it has no diagonal matrix elements in the eigenbasis of $H_E$. 
In Chap.~\ref{morsedecchap} the interaction Hamiltonian $V$ as well as $H_S$
will be specified and the integration in
Eq.~(\ref{MEthatsit}) will be performed to obtain a master equation
that describes a vibrating diatomic molecule in interaction with
the environment of thermal photon modes. 
\newpage
\section{Other methods}
\label{miscmethsec}
\vskip 24 pt
As an alternative of the method summarized in the
previous section, it is possible to ``unravel'' 
\cite{SB96} the master equation 
into stochastic processes that involve the state vector
of the system. Solving the stochastic differential equation (SDE) 
\cite{G90} several times, an ensemble of pure states, i.e., rank $1$ 
density operators is obtained.
Properly renormalizing and summing up these projectors we arrive
at a density operator that describes the ensemble.
The notion unraveling means that in the limit of infinite number of
ensemble elements  the corresponding density operator will
be identical to the solution of the master equation.

In this sense individual outcomes of the stochastic process
have no physical interpretation, but this not the only possible
point of view.
Indeed, in spontaneous collapse models (for a short review see
Ref.~\cite[Chap.~8]{GJK96}), the stochastic equation
replaces the usual Schr\"{o}dinger equation, i.e., the former one
is postulated to be the fundamental equation describing the time
evolution. This interpretation leads
to spontaneous collapse of the wave function of the system of
interest without referring to any disturbance due to the 
environment. The parameters in these models are chosen such
as to permit the same dynamics to be valid for both microscopic
and macroscopic systems but leading to different observable behavior 
in the two cases.  However, the approach of this thesis is based
on the universality of the {\em Schr\"{o}dinger equation}
and describes the appearance of classical properties in
quantum systems as a consequence of inevitable interaction with the
environment. 
Therefore  
we shall not adopt the idea that physical interpretation can
be associated to individual outcomes of stochastic processes
being the unraveling of a master equation. However,
these stochastic equations undoubtedly must be  considered as very useful
tools to obtain approximate solutions of the underlying
master equation. 

Additionally, if the reduced density operator of the system can 
be represented by a quasiprobability distribution (QPD), it can
be possible to transform a given master equation into a partial 
differential equation involving the respective QPD. 
The resulting partial differential equation is often
turns out to have the form of a Fokker-Planck equation.
After a brief overview of the quasiprobability distributions
(Sec.~\ref{QPD}), a typical example will be shown in
Sec.~\ref{fokkersec}.


\subsection{Wigner functions}
\label{QPD}
Quasiprobability distributions (QPDs) are used extensively in quantum 
physics for various problems, and are specially instructive in 
visualizing the process of decoherence.
These distributions map the state of a quantum system on
a continuous parameter space that can be identified with the phase space of 
the system. From a more mathematical point of view, this 
continuous parameter space can be considered as a coadjoint orbit of the
underlying Lie group \cite{AMO98}.

In the case of an oscillator, the phase space is a plane that
is traditionally parametrized by two real numbers, $x$ and $p$.
A system of $N$ two-level atoms (see Chap.~\ref{atomchap}), 
if they are invariant with respect
to permutations, is identical to the subspace
characterized by the $j=N/2$ eigenvalue of the usual 
angular momentum operator $J^2$. In this system the relevant symmetry group
is $SU(2)$, and the phase space is the surface of a 2-sphere.
The usual coordinates on this Bloch-sphere 
are the azimuthal and polar angles, $\theta$ and $\phi$.

In the following the construction of the Wigner functions $W(x,p)$ and 
$W(\theta, \phi)$ will be given in a way that points out the 
similarities. Note that Wigner functions are not the only possible QPDs
in either systems, but as more general quasidistributions
will not appear later in this work, it sufficient to concentrate
on $W(x,p)$ and $W(\theta, \phi)$. The construction of additional
QPDs in the above systems can be found in Refs.~\cite{CG69ab} and 
\cite{A81}, and the relation of these methods is discussed in 
Ref.~\cite{FBC98}. 

Given a density operator of the system, $\rho$, the corresponding Wigner
functions are defined as the expectation value of the
respective kernel operators
\be
W(x,p)=\mathrm{Tr}\left[\rho\Delta(x,p)\right],
\label{usualwfunct}
\ee
\be
W(\theta,\phi)={\rm Tr}\left[ \rho \Delta(\theta,\phi) \right],
\label{wfunct}
\ee
where, according to \cite{S57,A81}
\be
\Delta(x,p)=\frac{1}{\pi^2}\int_{-\infty}^\infty
\int_{-\infty}^\infty\
\!\!\!\! d u \ d v \ \ e^{i(vX-uP)} e^{i(up-vx)}
\label{swkernel}
\ee
and 
\be
\Delta(\theta,\phi)=\sum_{K=0}^N 
\sum_{Q=-K}^K T^\dagger_{KQ} Y_{KQ}(\theta,\phi).
\label{tkqkernel}
\ee
The operators $X$ and $P$ are the dimensionless 
position and momentum operators ($[X,P]=i$), 
while $Y_{KQ}$ denote the spherical harmonics \cite{BL81}
and $T_{KQ}$ stand for the spherical multipole operators \cite{A81}. 
Since the kernels given by Eqs.~(\ref{swkernel}) and (\ref{tkqkernel})
are Hermitian, both the spherical (\ref{wfunct}) and the ``planar''
(\ref{usualwfunct}) Wigner functions are real. These functions are normalized
with respect to the appropriate (invariant) measures
\be
\int_{-\infty}^\infty\int_{-\infty}^\infty \!\! 
dx \ dp \ W(x,p)=1, \ \ \ \ 
\int_{0}^\pi\int_0^{2 \pi} \!\! \sin{\theta} 
d\theta\  d\phi \ W(\theta,\phi)=1.
\label{wignorms}
\ee
However, the value of these Wigner functions
can be negative in certain domains of the phase space, 
that is why they are called \emph{quasi}\-distri\-butions. 
This is a manifestation of the 
fact that quantum mechanics is not equivalent to a classical
statistical theory. Conversely, a state with non-negative
Wigner function is rightly considered as classical. 
Thus, for a given density operator $\rho$, the degree of nonclassicality
can be characterized by the aid of the corresponding Wigner function.
The quantity
\cite{BC99} 
\be
M_{nc}(\rho)=1-{{I_+(\rho)-I_-(\rho)}\over{I_+(\rho)+I_-(\rho)}},
\label{QPDnoncl}
\ee
is found to be an appropriate measure of nonclassicality \cite{BC99,FC02}.
Here $I_+(\rho)$ and $I_-(\rho)$ are the 
moduli of the integrals of the Wigner function
over those domains of the phase space where it is positive and negative, 
respectively. On using Eqs.~(\ref{wignorms}), we obtain that $0\leq M_{nc}<1$.
The disappearance of nonclassicality is of course closely related 
to the decoherence: as we shall see later in several examples,
decoherence drives the system into 
a state with positive Wigner function, implying $M_{nc}=0$.
  
\subsection{Partial differential equations}
\label{fokkersec}
In the case of a time dependent density operator $\rho(t)$, 
Eqs.~(\ref{wfunct}) and (\ref{usualwfunct})
assign a Wigner function to $\rho(t)$ at any time instant.
However, sometimes it is favorable (and more instructive) to calculate the
time dependent Wigner function directly. In this section we consider 
the example of the amplitude damped harmonic oscillator (HO) which
is the special case of the model described in Sec.~\ref{MEsec},
with $H_S$ representing 
a distinguished oscillator with angular frequency 
$\omega_S$ (our ``system'') that is coupled to a set of environmental 
oscillators via its destruction operator, $a$. 
The environment is assumed to be in thermal equilibrium at a given 
temperature $T$. 
The calculations that will be performed later in Sec.~\ref{mastersec} can be
adapted to this case, yielding the interaction picture master equation 
\be
\frac{\partial \rho}{\partial t}=\frac{\gamma}{2}(\overline{n}+1)
\left(
2a\rho a^\dagger-a^\dagger a \rho -\rho a^\dagger a
\right)
+ \frac{\gamma}{2}\overline{n}
\left(
2a^\dagger\rho a-a a^\dagger \rho -\rho a a^\dagger 
\right),
\label{ampdamp}
\ee
where $\overline{n}=1/(\exp(\frac{\hbar\omega_S}{kT})-1)$, 
$\rho$ is the interaction picture reduced density operator of the
system and $\gamma$ denotes the damping rate \cite{WM85}. 

Combining Eqs.~(\ref{usualwfunct}) and (\ref{ampdamp}) we can express
$\partial W(x,p,t)/\partial t$ in terms of the operators $a$, $a^\dagger$,
$\Delta(x,p)$ and $\rho$. At this point it worth introducing the
complex variable $\alpha=(x+ip)/2$. Then the identities
\begin{eqnarray} 
a^\dagger \exp(\alpha a^\dagger-\alpha^*a)&=&\left(\frac{\partial}
{\partial \alpha}+\frac{\alpha^*}{2} \right)\exp(\alpha a^\dagger-\alpha^*a),
\nonumber
\\
a \exp(\alpha a^\dagger-\alpha^*a)&=&\left(\frac{\alpha}{2}-\frac{\partial}
{\partial \alpha^*} \right)\exp(\alpha a^\dagger-\alpha^*a)
\end{eqnarray}
and their adjoints inserted into the definition (\ref{usualwfunct})   
lead to
\be
\frac{\partial W(\alpha,t)}{\partial t}= \frac{\gamma}{2}
\left(\frac{\partial}{\partial \alpha}\alpha + 
\frac{\partial}{\partial \alpha^*}\alpha^*\right)W(\alpha,t)
+ \gamma\left(\overline{n}+\frac{1}{2}\right)\frac{\partial^2}
{\partial \alpha \partial \alpha^*}W(\alpha,t).
\label{wigfockpl}
\ee
This partial differential equation has the form of a  
Fokker-Plank equation \cite{G90}. (Note that this is not a general 
consequence of the procedure outlined above, there are situations
when the resulting equation is not so well-behaved as Eq.~\ref{wigfockpl}.)
Considering a Wigner function with a single peak, the qualitative
behavior of $W(\alpha,t)$ can be seen even intuitively. 
There are regions on the complex plane  $\alpha$, where the first term
in Eq.~(\ref{wigfockpl}), which contains only first derivatives, has  
opposite sign. This causes $W(\alpha,t)$ to increase (decrease) 
where the sign is positive (negative), resulting in the overall
motion of the peak. Therefore this first term is called the drift term.
On the other hand, the second (diffusion) term broadens the distribution and
-- due to the normalization -- decreases the peak value.

\begin{figure}[htb]
\begin{center}
\psfrag{x}[tl][bl]{$\mathrm{Re}\  \alpha$}
\psfrag{p}[tl][bl]{$\mathrm{Im}\  \alpha$}
\psfrag{W(x,p)}[tl][tl]{$W$}
\includegraphics*[bb=80 70 1350 510 , width=15.0cm]{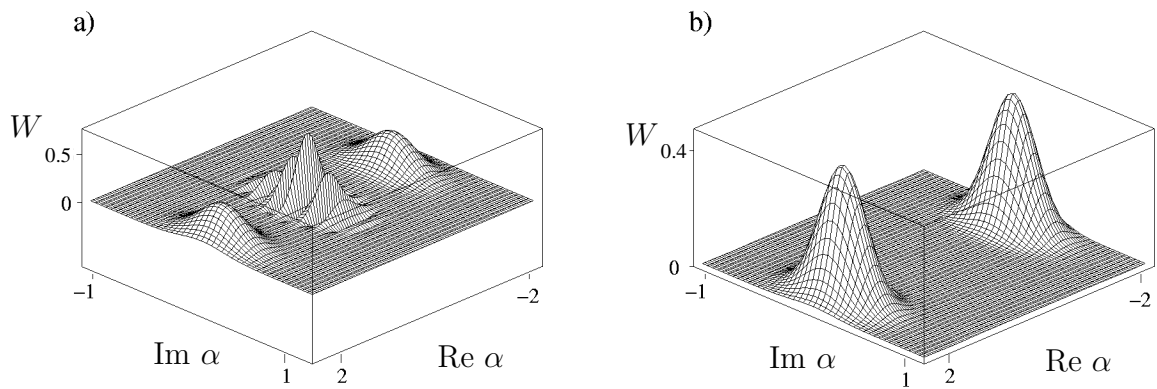}
\caption[Amplitude damping of the harmonic oscillator: Wigner view]
{
Schematic time evolution of a Wigner function
according to Eq.~(\ref{wigfockpl}). 
The most important consequence of the amplitude damping 
is the disappearance of the quantum interference between the positive
hills that correspond to the initial oscillator coherent states
$|\alpha=2\rangle$ and $|\alpha=-2\rangle$.
\label{osccatwig}
}
\end{center}
\end{figure}

As an important application from the viewpoint of decoherence, 
we consider the initial Wigner function that corresponds to
the superposition of two oscillator coherent states \cite{G63a,G63b}
$|\Phi\rangle=1/\sqrt 2 (|\alpha=2\rangle+|\alpha=-2\rangle)$, 
see Fig.~\ref{osccatwig} a). 
The positive hills represents the two coherent states, while the strong
oscillations between the hills are signatures of the quantum coherence
of $|\alpha=2\rangle$ and $|\alpha=-2\rangle$. These coherent states
have clear classical interpretation, and therefore their superposition 
can be called a Schr\"{o}dinger-cat state, see 
Ref.~\cite[and references therein]{JDA93,B90}. 
Fig.~\ref{osccatwig} a) is a typical Wigner function for these nonclassical 
states. The effect of the amplitude damping 
is shown in Fig.~\ref{osccatwig} b), it leads to the disappearance
of the quantum interference represented by the oscillations. 
This is what we expect according
to the Fokker-Planck equation (\ref{wigfockpl}), 
because $W(\alpha,t)$ changes rapidly in the regions where it 
oscillates, implying very fast diffusion that smears out the oscillations.
On the level of the master equation (\ref{ampdamp}), this result is the
manifestation of the fact that coherent states of the HO are pointer 
states (see Sec.~\ref{pointersec}) to a very good approximation \cite{WM85} 
in the case of the amplitude damping interaction.

\begin{figure}[htb]
\begin{center}
\psfrag{x}[tl][bl]{$\mathrm{Re}\  \alpha$}
\psfrag{p}[tl][bl]{$\mathrm{Im}\  \alpha$}
\psfrag{W(x,p)}[Bl][tl]{$W$}
\includegraphics*[bb=80 70 1330 510 , width=15.0cm]{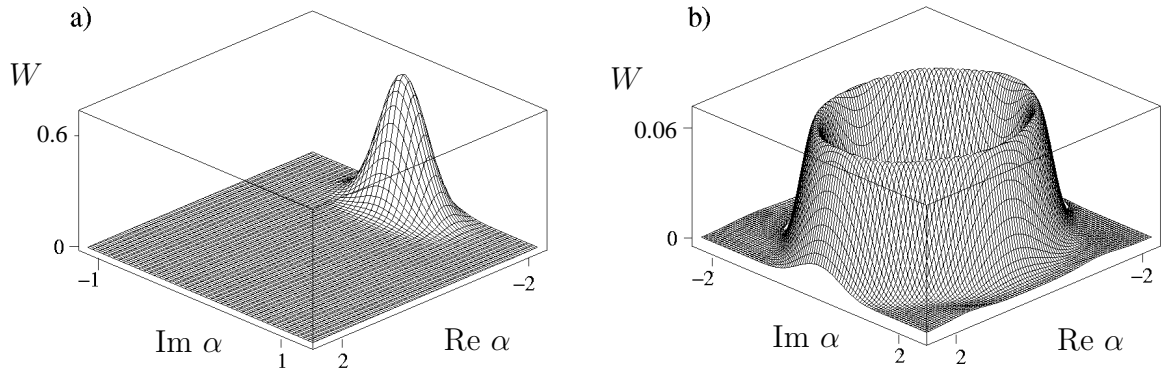}
\caption[Wigner representation of the phase relaxation in harmonic oscillator]
{
Wigner functions visualizing phase relaxation. The initial state 
shown in a) is a coherent state $|\alpha=-2\rangle$. Part b) of the figure
corresponds to the result of the phase relaxation.
\label{wigrelax}
}
\end{center}
\end{figure}

A qualitatively different decoherence mechanism related to
the HO is the so-called phase relaxation \cite{WM85}. 
Since our results in the anharmonic Morse system has similarities
with this process, it is worth summarizing here the
phase relaxation as well. Now the relevant master equation is
\be
\frac{\partial \rho}{\partial t}=\frac{\gamma}{2}
\left(
2a^\dagger a\rho a^\dagger a -a^\dagger a a^\dagger a \rho 
-\rho a^\dagger a a^\dagger a
\right),
\label{phrelax}
\ee
and, as it has been already mentioned (Sec.~\ref{pointersec}), the
eigenstates of the HO Hamiltonian are pointer states in this case.
This means that according to the general scheme given by 
Eq.~(\ref{EIDscheme}), the result of the decoherence will be the
a mixture of energy eigenstates with the weights defined by the initial
state. That is, the energy of the system remains unchanged during
the process of decoherence, but the phase information is completely
destroyed: The distance between the origin and the highest values of the
Wigner function shown in Fig.~\ref{wigrelax} b) is the same as it was 
initially (Fig.~\ref{wigrelax} a)), but $W(\alpha)$ is cylindrically 
symmetric now.


\addtocontents{toc}{\par \vspace{0.5cm} \hspace{6.5cm} *** 
\vspace{-1.0cm} \par}
\newpage
\addcontentsline{toc}{part}{PART II}

\chapter{Molecular wave packets in the Morse potential}
\emptypage
\label{oechap}
Peculiar quantum effects of wave packet motion in anharmonic  
potentials have been predicted in several model systems 
\cite{PS86,AS00,VE94}.  
We are going to investigate the role of anharmonicity in the 
case of the Morse potential. This model potential is often used to
describe  a vibrating diatomic molecule having a finite number 
of bound eigenstates together with a dissociation continuum. 
Our initial wave packets will be Morse coherent states \cite{BM99},
and in the current chapter we consider the case 
when the environment does not influence
the dynamics of the system \cite{FC02}.
We show that the Wigner functions of the system
exhibit spontaneous formation of Schr\"{o}dinger-cat states 
at certain stages of the time evolution. 
These highly nonclassical states are coherent 
superpositions of two localized states corresponding to 
two different positions of the center of mass. 
The degree of nonclassicality is also analyzed
as the function of time for different initial states. 
Our numerical calculations are based on a novel, essentially
algebraic treatment of the Morse potential \cite{MF02}.

The same system in the case when the environmental effects 
are present will be analyzed in Chap.~\ref{morsedecchap}.
\section{The Morse oscillator as a model of a vibrating diatomic molecule}
\label{morsemodelsec}
Our description of molecular vibrations is based on the Morse 
Hamiltonian \cite{HH79}, that can be written in the  
following dimensionless form
\begin{equation}
H=P^2+(s+1/2)^2[\exp(-2X) -2\exp(-X)],
\label{ham}
\end{equation}
where the shape parameter, $s$, is related to the dissociation
energy $D$, the reduced mass of the molecule $m$, and the range parameter of
the potential $\alpha$ via $s={{\sqrt{2mD}}\over{\hbar \alpha}}-1/2.$
The dimensionless operator $X$ in Eq.~(\ref{ham}) corresponds
to the displacement of the center of mass of the diatomic system
from the equilibrium position, and the canonical commutation relation
$[X,P]=i$ also holds.

The Hamiltonian (\ref {ham}) has
$[s]+1$ normalizable eigenstates (bound states), plus the continuous energy
spectrum with positive energies. 
The wave functions of the bound eigenstates of 
$H$ are $\psi_n(y)=\sqrt{[n!(2s-2n)]/[(2s-n)!]}\ 
y^{s-n}e^{-y/2}L_{n}^{2s-2n}(y)$,
where $y=(2s+1)e^{-x}$ is the rescaled position variable, and 
$L_n^{2s-1}(y)$ is a generalized Laguerre polynomial.
The corresponding eigenvalues are
$E_{m}(s)=-(s-m)^{2},$ $m=0,1,\ldots \lbrack s]$,
where $[s]$ denotes the largest integer that is 
smaller than $s$.

In the following we solve the  Schr\"{o}dinger equation 
\be
{{d}\over{d t}}|\phi\rangle=-i {{2\pi}\over{2s+1}} 
H|\phi\rangle,
\label{schr}
\ee 
where time is measured in units of $t_0=2\pi/\omega_0$, with 
$\omega_{0}=\alpha \sqrt{\frac{2D}{m}}$
being the circular frequency of the small oscillations in the potential.

The initial states of our analysis will be Morse coherent 
states \cite{BM99,MBB01} associated with the wave functions
\begin{equation}
\langle y|\beta \rangle= {\frac{\left( 1-|\beta|^2\right)^s}
{\sqrt{\Gamma(2s)}(1-\beta)^{2s}}}
y^s \exp\left(-{{y}\over{2}}{{1+\beta}\over{1-\beta}}\right).
\end{equation}
We expand these states in terms of a suitable finite basis:
\begin{eqnarray}
|\beta\rangle&=&\sum_{n=0}^{N} c_n |\psi_n\rangle = 
\sum_{n=0}^{[s]}\Big[\sqrt{\frac{(2s-2n)\Gamma(2s-n+1)}{n!\Gamma(2s)}}
\frac{\Gamma(2s-n)}{\Gamma(2s-2n+1)}
\frac{(1-|\beta|^2)^{s}}{(1-\beta)^n} \nonumber \\
&\times& {}_2{\mathrm F}_1(-n,2s-n;2s-2n+1;1-\beta)\  |\psi_n\rangle\Big]
+\sum_{n=[s]+1}^{N} c_n |\psi_n\rangle,
\label{cohwf}
\end{eqnarray}
where  ${}_2{\mathrm F}_1$ is the hypergeometric function of the
variable $1-\beta$. The first $[s]+1$ elements of the basis 
$\{|\psi_n\rangle\}_{n=0}^N$ are the bound states, 
and the continuous part of the
spectrum is represented by a set of orthonormal states which
give zero overlap with the bound states.
The energies of the states
$|\psi_n\rangle$, $n>[s]$ follow
densely each other, approximating satisfactorily the 
continuous energy spectrum \cite{MF02}.

We note that the states in Eq.~(\ref{cohwf}) are ``single mode''
coherent states in contrast to those of \cite{BA99}, where
the dynamics of two-mode coherent states were investigated
for various symmetry groups, including SU$(1,1)$, 
which is in a close relation to the relevant 
symmetry group of the Morse potential \cite{BIA01}.

The label  $\beta$ in Eq.~(\ref{cohwf}) 
is in one to one correspondence 
with the expectation values 
\be
\langle X\rangle_{\beta}=\ln \left ({\mathrm {Re}}{{1+\beta}\over{1-\beta}} \right),
\ \ 
\langle P\rangle_{\beta}=s {{{\mathrm {Im}}{[(1+\beta)/(1-\beta)]}}\over{\mathrm {Re}}
{[(1+\beta)/(1-\beta)}]},
\ee
therefore we can use the notation $|x_0,p_0\rangle$ 
for the state $|\beta\rangle$ that gives
$\langle X\rangle=x_0$ and $\langle P\rangle=p_0$. 
The localized wave packet corresponding to $|x_0,p_0\rangle$
is centered at $x_0$ ($p_0$) in the coordinate (momentum) representation.
\begin{figure}[htbp]
\begin{center}
\includegraphics*[bb=70 40 780 520 ,width=10cm]{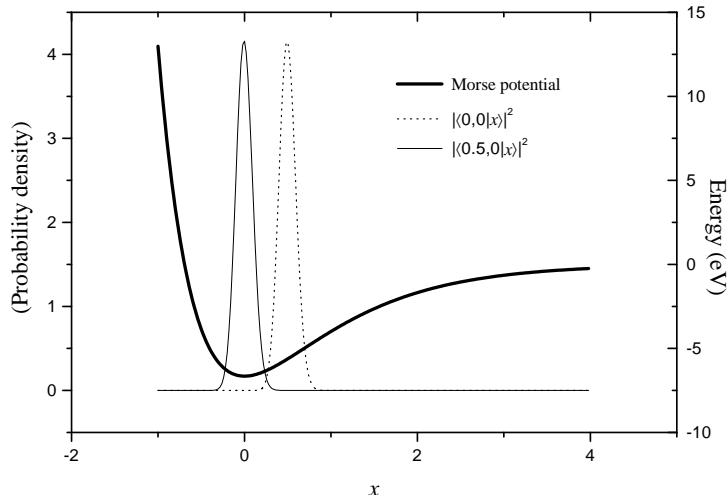}
\caption[Wave functions of two Morse coherent states]
{
The absolute square of the wave functions corresponding to the Morse
coherent states $|x_0,p_0=0\rangle$, with
$x_0=0.0$ (ground state) and $x_0=0.5$. 
These plots correspond to the case of the 
NO molecule, where $s=54.54$. (We have generated a movie file showing 
the time evolution of the wave function plotted with dotted line, it can
be found in Ref.~\cite{FC02}.)
\label{potwf}
}
\end{center}
\end{figure}
 
In our calculation
we have chosen the NO molecule as our model, where $m=7.46\ {\mathrm{a.u.}}$, 
$D=6.497\ {\mathrm {eV}}$ and  $\alpha=27.68\ {\mathrm{nm^{-1}}}$ \cite{HH79},
yielding $s=54.54$. That is, this molecule has $55$ bound states, 
and we found that a basis of dimension $N+1=150$ is sufficiently
large to handle the problem. 
The absolute square of the wave 
functions $|\langle x|0,0\rangle|^2$ and  $|\langle x| 0.5,0\rangle|^2$
is depicted in Fig.~\ref{potwf}, where $V(x)$ is also shown. Fig.~\ref{potwf} 
indicates that initial displacements, $x_0$, having the order of
magnitude of unity will not lead to  ``\emph{small} oscillations''.

The Morse coherent states \cite{BM99,MBB01} 
can be prepared by an appropriate
electromagnetic pulse that drives the vibrational state of the molecule
starting from the ground state into an approximate  coherent state. 
An example can be found in \cite{MBF01}, where the effect of an 
external sinusoidal field is considered.


\section{Behavior of expectation values as a function of time}
Starting from  $|\phi(t=0)\rangle=|x_0,p_0=0\rangle$ as initial states,
first we consider the dependence of the
$\langle X\rangle(t)$ curve on $x_0$.  
\begin{figure}[htb]
\begin{center}
\psfrag{taxis}{$t/t_0$}
\psfrag{xlabel}{$\langle X \rangle$}
\includegraphics*[bb=45 50 780 520 , width=10cm]{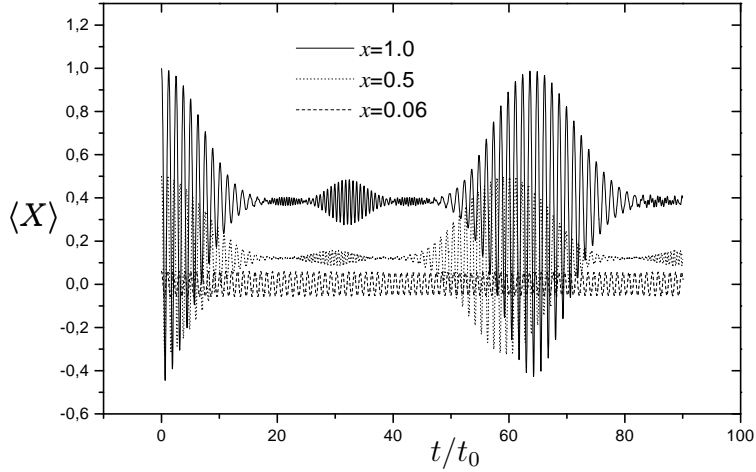}
\caption[The expectation value of the position operator 
in the Morse potential as a function of time (without environment)]
{
The expectation value of the dimensionless position operator as a function
of time. The initial states were  $|\phi(t=0)\rangle=|x_0,p_0=0\rangle$, with
$x_0=1.0$, $x_0=0.5$ and $x_0=0.06$.
\label{xexp}
}
\end{center}
\end{figure}
It is not surprising that for small values of $x_0$ ($\leq 0.06$) these 
curves show similar oscillatory behavior as in the case of the harmonic
oscillator, see Fig.~\ref{xexp}.
However, when anharmonic effects become important, a different
phenomenon can be observed: the amplitude of the oscillations
decreases almost to zero, then faster oscillations with
small amplitude appear but later we re-obtain almost exactly 
$\langle X\rangle(0)$, and the whole
process starts again. 
Fig.~\ref{xexp} is similar to the 
collapse and revival in the Jaynes-Cumings (JC) model \cite{JC63,ENSM80}, 
but in our case the non-equidistant spectrum of the Morse
Hamiltonian is responsible for the effect. 
There are important situations when revivals and fractional revivals 
\cite{PS86,AP89,LAS96,DK00} of the wave packet can be 
described analytically
\cite{AS00}, but in a realistic model for a diatomic molecule the
difficulties introduced by the
presence of the continuous spectrum implies choosing an appropriate 
numerical solution.

The expansion of the initial state in our finite basis
$|x_0,0\rangle=\sum_n c_n(x_0) |\psi_n\rangle$  shows that for values 
of $x_0$ shown in Fig.~\ref{xexp}  the maximal $|c_n(x_0)|$ belongs to
$n<[s]$. That is, the expectation value
\be
\langle X\rangle(t)=\sum_{n,k=0}^{N} c_n(x)  c_k^*(x) 
\langle \psi_k|X|\psi_n\rangle 
\exp\left[it{{2\pi}\over{2s+1}}(E_k(s)-E_n(s))\right]
\label{xpexp}
\ee
is dominated by the bound part of the spectrum.
Damping of the amplitude of the oscillations is
due to the destructive interference between the various Bohr frequencies and
we observe revival when the exponential terms rephase again.

\begin{figure}[htb]
\begin{center}
\includegraphics*[bb=40 15 770 530 , width=10cm]{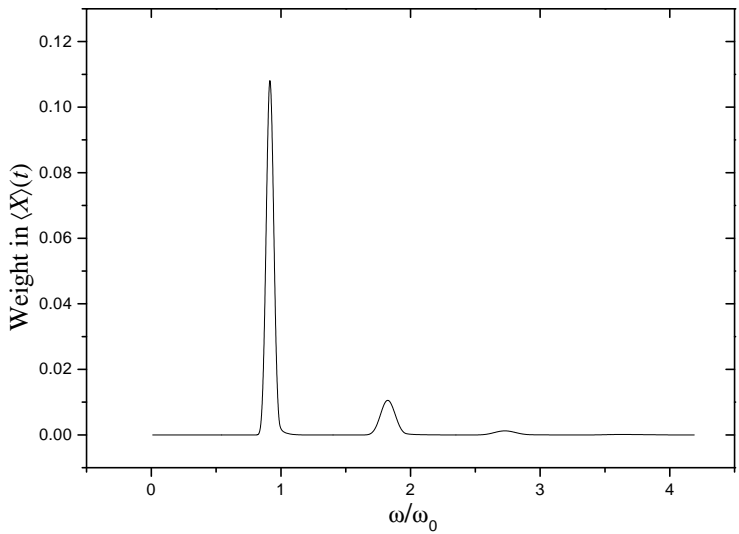}
\caption[The dominant frequencies responsible for the collapse-revival
phenomenon in the expectation value of the position operator]
{
The weight of the dominant frequencies responsible for the collapse-revival
phenomenon in the expectation value of the position operator.
The first peak is related to the matrix elements 
$\langle\psi_n|X|\psi_{n+1}\rangle$, while the second family of
nonzero weights corresponds to the matrix elements 
$\langle\psi_n|X|\psi_{n+2}\rangle$.
\label{freqs}
}
\end{center}
\end{figure}

Quantitatively, we have determined the dominant frequencies in 
Eq.~(\ref{xpexp}) for $x_0=0.5$ and
found that they fall into two families, see Fig.~\ref{freqs}.
The first family is related to
the matrix elements $\langle\psi_n|X|\psi_{n+1}\rangle$ and a has a sharp
distribution around $\omega_1=0.9 \omega_0$. The contribution 
of the second family to the sum in Eq.~(\ref {xexp}) is much weaker, 
these frequencies around $\omega_1=1.81 \omega_0$
correspond 
to the matrix elements $\langle\psi_n|X|\psi_{n+2}\rangle$. 
The width the first distribution $\Delta \omega_1=0.1\omega_0$ allows 
us to estimate the
revival time as $2\pi/\Delta \omega_1=62.8 t_0$, while
$\Delta \omega_2=0.17\omega_0$ is responsible for the  partial
revival at $t/t_0\approx 30$, see Fig \ref{xexp}. 
Following Refs. \cite{PS86,AP89}, we denote by
$t_{rev}$ the time when
the anharmonic terms in the spectrum induce no phase shifts,
that is, the initial wave packet is reconstructed.
At $t_{rev}/2\approx 60t_0$ all these phase factors are $-1$, while 
$t/t_0=30$ corresponds to a quarter-revival, i.e., to time
$t_{rev}/4$.


\section{Time evolution of the Wigner function of the system}
In order to gain more insight concerning the physical process
leading to the collapse-revival phenomenon seen in Fig.~\ref{xexp}, 
one can look at the coordinate representation of the wave function
$\phi(x,t)=\langle x|\phi(t)\rangle$. 
In the representative case of $|\phi(t=0)\rangle=|x_0,0\rangle$,
the wave function is an initially
well localized wave packet that gradually falls apart into several packets and 
then conglomerates again, see Ref.~\cite{FC02}.

\begin{figure}[htb]
\begin{center}
\includegraphics*[bb=90 40 1330 510 , width=15.0cm]{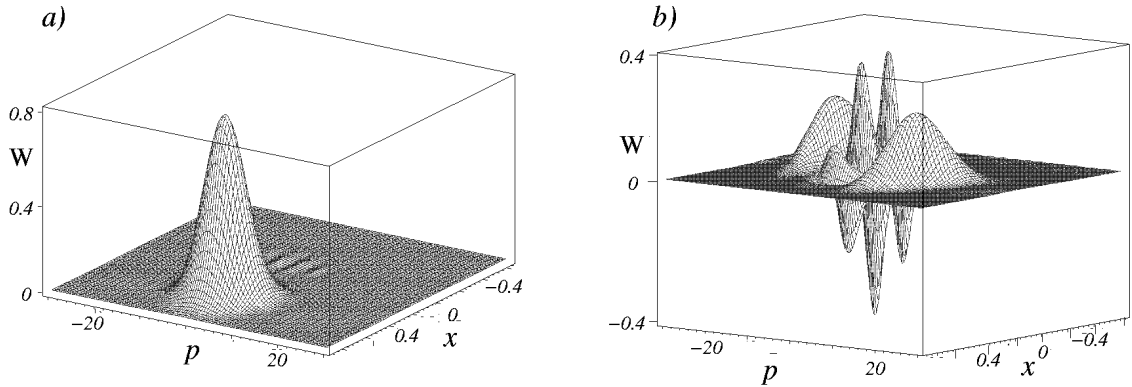}
\caption[Wigner view of Schr\"odinger-cat state formation in 
the Morse potential]
{
Wigner functions of the Morse system at 
the initial stage of the time evolution and the formation of 
a Schr\"{o}dinger-cat state.
The plots a) and b) correspond to $t/t_0=0$ and $t/t_0=30$, respectively.
The initial state was $|\phi(t=0)\rangle=|x_0,p_0=0\rangle$, with
$x_0=0.5$. (The movie file showing the time evolution
of $W$ can be found in Ref.~\cite{FC02}.)
\label{wigs}
}
\end{center}
\end{figure}

Starting from the same initial state it is more instructive to visualize 
the time evolution
by the aid of the Wigner function $W(x,p,t)$ 
that reflects the state of the  system in the phase space, see 
Sec.~\ref{QPD}. The definition given by Eq.~(\ref{usualwfunct})
can be reformulated for a pure state that is represented by its
wave function $\phi(x,t)$, yielding
\be
W(x,p,t)={{1}\over{2\pi}}\int_{-\infty}^{\infty} \phi^*(x+u/2,t) \phi(x-u/2,t) 
e^{i up} d u.
\ee 
Fig.~\ref{wigs} a) shows the initial stage of the time evolution, 
while Fig.~\ref{wigs} b) corresponds to $t/t_0=30.$ 
This second Wigner function is typical for
Schr\"{o}dinger-cat states, compare with Fig.~\ref{osccatwig}.
$W(x,p)$ in Fig.~\ref{wigs} b)
corresponds to a superposition of two states that are well-localized 
in both momentum and coordinate, and represented by the two positive hills 
centered at $x_1=-0.1$, $p_1=-18.0$ and $x_2=0.3$, $p_2=12.0$. 
The strong oscillations between them shows the quantum
interference of these states. 

According to the our calculations, there are a few periods
around $t/t_0=30$, while the state of the system can be 
considered to be a phase space Schr\"{o}dinger-cat state.
During this time the Wigner function is similar to the one shown
in Fig.~\ref{wigs} b), and it rotates around 
the equilibrium position.
Similar behavior of the Wigner function was found in \cite{ER91}
for the JC model. 
This effect is responsible for the partial revival around $t/t_0=30$
shown in Fig.~\ref{xexp}, where the frequency of the oscillations 
is twice that of the oscillations
around $t=0$: 
in the neighborhood of $t/t_0=30$
there are two wave packets moving approximately the same 
way as the coherent state soon after $t=0$.


\section{Measuring nonclassicality}
\label{nonclsec}
According to Sec.~\ref{QPD},
the Wigner function of a state $|\phi\rangle$ can be used to determine
the nonclassicality of $|\phi\rangle$. Having calculated $W(x,p)$,
it is straightforward to obtain the quantity $0\leq M_{nc}<1$ (defined 
by Eq.~(\ref{QPDnoncl})), which is an appropriate measure of the 
nonclassicality \cite{BC99}.
%
\begin{figure}[htbp]
\begin{center}
\psfrag{tlabel}{$t/t_0$}
\psfrag{mlabel}{$M_{nc}$}
\includegraphics*[bb=50 50 780 520 ,width=10cm]{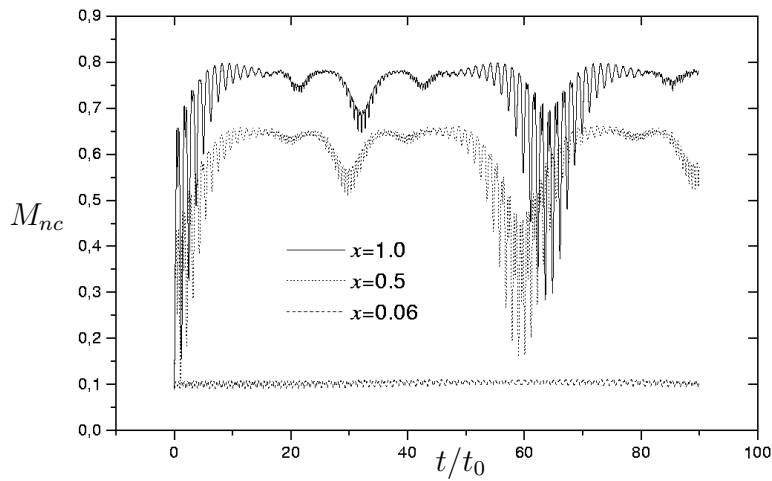}
\caption[Nonclassicality during the process of Schr\"odinger-cat 
state formation]
{
Nonclassicality as a function
of time. The initial state was  $|\phi(t=0)\rangle=|x_0,p_0=0\rangle$, with
$x_0=1.0$, $x_0=0.5$ and $x_0=0.06$.
\label{noncl}
}
\end{center}
\end{figure}
Fig.~\ref{noncl} shows $M_{nc}$ as a function of time 
for the same initial states as 
in Fig.~\ref{xexp}. For the small initial displacement 
of $x_0=0.06$, we see that the Wigner function is positive 
almost everywhere, the state
can be considered as a classical one during the whole time evolution.

For larger initial displacements we can easily identify two time scales. 
The shorter one is the period of the wave packet in the potential,
while the second time scale can be identified with the revival time.
Looking at the initial part of the curve $M_{nc}(t)$, we observe that
the state of the system is the most classical at those turning points
where $\langle X\rangle>0$, see Fig.~\ref{potwf}.
On the other time scale, the collapse of
the oscillations in $\langle X \rangle$ presents itself as the increase of 
$M_{nc}$, and the revival turns the state into a more classical one.
When the state of the system can be considered as a Schr\"{o}dinger-cat state,
$M_{nc}(t)$ has a small local minimum, but it still has significant values
indicating strong nonclassicality.      


\section{Conclusions}
We have found that in  
the potential of the NO molecule,  
when anharmonic
effects are important, the time evolution naturally leads to the formation
of Schr\"{o}dinger-cat states at certain stages of the time evolution.
These highly nonclassical states correspond to the superposition of two
molecular states which are well localized in the phase space. 

\newpage
\chapter{Decoherence of molecular wave packets}
\emptypage
\label{morsedecchap}
The correspondence between classical and quantum dynamics of anharmonic 
systems has gained significant attention in the
past few years \cite{PS86,AP89,VE94,AS00}.
A short laser pulse impinging on an atom or a molecule excites 
a superposition of several stationary states, and the resulting 
wave packet follows the orbit of the corresponding classical 
particle in the initial stage of the time evolution.
However, the nonequidistant nature of the involved energy 
spectra causes peculiar quantum effects, broadening of the initially
well localized wave packets, revivals and partial revivals 
\cite{AP89,PS86,VE94,AS00,LAS96,DK00}. 
As we saw in the previous chapter, partial revivals are in close connection
with the formation of Schr\"odinger-cat states, which, in this context, 
are coherent superpositions of two spatially separated, 
well localized wave packets \cite{JV94}. 
Phase space description of vibrational 
Schr\"odinger-cat state formation using animated Wigner
functions can be found in \cite{FC02}.
According to Chap.~\ref{introchap}, 
these highly nonclassical states are expected to be
particularly sensitive to decoherence. The aim of this
chapter is to analyze the process of decoherence for the spontaneously
formed Schr\"odinger-cat states in the anharmonic Morse potential.

In the following we introduce 
a master equation that takes into account the fact that in a general
anharmonic system the relaxation rate of each energy eigenstate is different.
This master equation is applied to the case of wave packet motion in 
the Morse potential that is often used to describe a 
vibrating diatomic molecule. 
Considering the  phase space description of decoherence,
we show how the phase portrait of the system reflects the damping of 
revivals in the expectation values of the position and momentum operators due
to the effect of the environment. 
We also calculate and plot 
the time evolution of the Wigner function corresponding
to the reduced density operator of the Morse system. 
The Wigner function picture visualizes the fact that although
our master equation reduces to the 
amplitude damping equation (\ref{ampdamp}) in 
the harmonic limit, the anharmonic
effects lead to a decoherence scheme which is similar to the
phase relaxation (see Sec.~\ref{fokkersec} and also Ref.~\cite{WM85}) 
of the harmonic oscillator (HO).
It is found that the time scale of decoherence is much
shorter than that of dissipation, 
and gives rise to density operators which are mixtures of
localized states along the phase space orbit of the corresponding 
classical particle.
We illustrate the generality of this
decoherence scheme by presenting the time evolution of an energy
eigenstate as well.  
We also calculate the decoherence time for various 
initial wave packets. We show that decoherence is faster for wave packets
that correspond to a classical particle with a phase space
orbit of larger diameter.

\section{A master equation describing decoherence in the Morse system}
\label{mastersec}
We consider a vibrating diatomic molecule and recall the Morse Hamiltonian
\begin{equation}
H_S=P^2+(s+1/2)^2[\exp(-2X) -2\exp(-X)],
\label{haminmastersec}
\end{equation}
which is often used to describe this system, see Sec.~\ref{morsemodelsec}. 
The initial wave packets of our analysis -- similarly to the previous
chapter -- will be Morse coherent 
states \cite{BM99}, $|x_0,p_0\rangle$, which are localized on 
the phase space around the point $(x_0, p_0)$, see 
Fig.~\ref{wigs}.
Although the construction given in \cite{MF02}
would allow us to use arbitrary initial states, 
for our current purpose it suffices to consider
states $|x_0,p_0\rangle$ with negligible
dissociation probability, i.e., coherent states that practically
can be expanded in terms of the bound states
$|\phi_n\rangle$,  $n=0,1,\ldots \lbrack s]$. This means that
the relevant part of the spectrum of $H_S$ is nondegenerate 
and discrete.

The environment is assumed to consist of the
modes of the free electromagnetic field 
\be
H_E=\sum_k \hbar\omega_k (a^{\dagger}_k a_k+1/2).
\label{MEenvH}
\ee
We assume the following interaction Hamiltonian
\be
V=\hbar \mathcal{X}^{\dagger}\sum_k g_k a_k+\hbar \mathcal{X}\sum_k g_k a^{\dagger}_k,
\label{MEinterH}
\ee
where, for the sake of simplicity, the coupling constants $g_k$ 
were taken to be real. Wishing to keep the derivation as general as it 
is possible, the only necessary restriction on the operator $\mathcal{X}$ 
is that it must have a strictly upper triangular matrix
in the eigenbasis $\{|\phi_n\rangle\}$, i.e., $\mathcal{X}$ transforms each
eigenstate of $H_S$ into a superposition of different eigenstates 
corresponding to \emph{lower} energy values.
$\mathcal{X}^{\dagger}$ is the Hermitian conjugate of $\mathcal{X}$. 
This is the application of the rotating wave approximation (RWA)
to an anharmonic, multilevel system.
Well-known examples imply that from the viewpoint of decoherence RWA
is a permissible approximation.
According to Ref.~\cite{A74}, in the case of spontaneous emission,
RWA on the initial Hamiltonian modifies the level shifts induced by
the environment. This could be expected, because the total Hamiltonian
with and without RWA has usually different spectra. However, the damping
term that provides the time scale of the spontaneous emission is
practically unaffected by keeping the counter rotating terms in the 
interaction Hamiltonian. A similar result was found for the case of
a spin-${{1}\over{2}}$ system in external magnetic field \cite{BS40}
and also for a single two-level atom in electric field \cite{AT55}.
 
Note that in the particular case of a vibrating diatomic molecule, 
the operators $\mathcal{X}$ and $\mathcal{X}^\dagger$
gain a clear interpretation: in the eigenbasis of Morse Hamiltonian $H_S$, 
they are the upper and lower 
triangular parts of the molecular dipole moment operator, $\hat\mu$.
We will assume that $\hat\mu$ is linear \cite{YL98}, that is, proportional 
to the displacement $X$ of the center of mass of the diatomic system
from the equilibrium position. 
Although in this chapter the resulting master equation will be 
applied to describe
the decoherence of a vibrating diatomic molecule, we do not
perform the $X\rightarrow$ $\mathcal{X}$ substitution during the derivation
in order to indicate the generality of our approach.

Using the specific operators above, we can return to Eq.~(\ref{MEthatsit})
\be
{{d}\over{dt}}\rho_{S}^i(\tau)= -{{1}\over{\hbar^2}}\int_0^{\tau} dt_1 
{\mathrm {Tr}}_E \left[ V^i(\tau),\left[ V^i(t_1),
\rho_{S}^i(\tau)\rho^i_E(0)\right]\right],
\label{MEthatsitinmastersec}
\ee   
and perform the integration in order to obtain
a differential equation that describes the time evolution of
the reduced density matrix of our system, $\rho_{S}$. The 
interaction picture operator $V^i$
is defined by Eq.~(\ref{MEintpdef}), and its expansion in the eigenbasis 
$\{|\phi_n\rangle\}$ of the system Hamiltonian has the form 
\be
V^i(t)=
\sum_{m>n}e^{{{i(E_m-E_n)t}\over{\hbar}}}\mathcal{X}^{\dagger}_{m n} 
|\phi_m\rangle\langle\phi_n|\ 
\sum_k e^{{-i\omega_k t}}g_k a_k+h.c.,
\label{MEinterdetail}
\ee
where we invoked that $\mathcal{X}_{m n}^{\dagger}=0$ if $m\leq n$. 
$E_0$ denotes the ground state energy, and the eigenvalues of $H_S$
follow each other in increasing order: $E_m>E_n$, whenever $m>n$.
Therefore the integrand in Eq.~(\ref{MEthatsitinmastersec}) 
contains $16$ terms. However, by assuming that the environment 
is in thermal equilibrium at a given temperature $T$, we can deduce
that the terms containing $\sum_{k,l} 
 g_k g_l \ a^{\dagger}_k a^{\dagger}_l$ and its adjoint
give no contribution, because 
${\mathrm {Tr}}_E [a^{\dagger}_k a^{\dagger}_l\rho_E]=0$. 
(We note that for some specially prepared reservoirs, such as 
squeezed reservoirs, this quantity need not be zero, see \cite{WM94}.)
Moreover, since 
\be
{\mathrm {Tr}}_E \left( a^{\dagger}_k a_l \ \rho_E \right)
=\delta_{k l} \langle a^{\dagger}_k a_k\rangle_{\rho_E}
\ \ \ {\textnormal {and}} \ \ \ \ 
{\mathrm {Tr}}_E \left( a_k a^{\dagger}_l \ \rho_E \right)
=\delta_{k l} \langle a_k a_k^{\dagger}\rangle_{\rho_E},
\ee 
we have 
\be
{\mathrm {Tr}}_E \sum_{k,l} 
 g_k g_l \  a^{\dagger}_k a_l \ \rho_E =
\sum_{k}(g_k)^2 \overline{n}_k,
\ \ \ \ \ \textrm{and}\ \ \ \ \ 
{\mathrm {Tr}}_E \sum_{k,l} 
 g_k g_l \ a_k a_l^{\dagger} \ \rho_E =
\sum_{k}(g_k)^2 \left(\overline{n}_k +1\right),
\label{MEtrE}
\ee  
where $\overline{n}_m=\langle a^{\dagger}_m a_m 
\rangle_{\rho_E}=1/(\exp(\frac{\hbar\omega_m}{kT})-1)$ is the average
number of quanta in the $m$-th mode of the environment.
According to 
the assumption that the environment consists of a {\emph {large}} number
of harmonic oscillators, we can convert the sum over modes to frequency-space 
integral $\int_0^{\infty} d\omega D(\omega)\ldots$, where $D(\omega)$ denotes
the density of states which is proportional to 
$\omega^2$ in our case. The continuous version of Eq.~(\ref{MEtrE}) 
combined with Eq.~(\ref{MEthatsitinmastersec}) yields to
\bea
\lefteqn{
{{d}\over{dt}}\rho_{S}(\tau)= -
\int_0^{\tau} dt_1 \int_0^{\infty} d\omega D(\omega) g^2(\omega)}
\nonumber
\\
&\times&
\Big[ 
\mathcal{X}(\tau)\mathcal{X}^{\dagger}(t_1)\rho_{S}(\tau) \overline{n}(\omega) 
e^{{-i(t_1-\tau)\omega}}
+ \mathcal{X}^{\dagger}(\tau)\mathcal{X}(t_1)\rho_{S}(\tau) \left(\overline{n}(\omega)+1
\right) e^{{i(t_1-\tau)\omega}}
\nonumber
\\
&-&\mathcal{X}^{\dagger}(\tau)\rho_{S}(\tau)\mathcal{X}(t_1)\left( \overline{n}(\omega)+1
\right) e^{{i(t_1-\tau)\omega}} 
- \mathcal{X}(\tau)\rho_{S}(\tau)\mathcal{X}^{\dagger}(t_1)  
\overline{n}(\omega) e^{{-i(t_1-\tau)\omega}}
+\mathrm{h.c.}\Big], 
\nonumber
\\
\label{MEbeforeint}
\eea    
where the superscript $i$ referring to the interaction picture was omitted.
Choosing the first term as a representative example, 
the application of Eq.~(\ref{MEinterdetail}) leads to
\bea 
\lefteqn{
I_1=\int_0^{\tau}\!\! dt_1 \!\!\int_0^{\infty}\!\!
 d\omega D(\omega) g^2(\omega)
\mathcal{X}(\tau)\mathcal{X}^{\dagger}(t_1)\rho_S(\tau) \overline{n}(\omega) 
e^{{-i(t_1-\tau)\omega}}}
\nonumber
\\
&=&\!\!\!\!\!\! \sum_{l, m<l, n<l}\!\!\!\!
\mathcal{X}_{m l} 
\mathcal{X}^{\dagger}_{ln} e^{{{i(E_m-E_n)\tau}\over{\hbar}}}|\phi_m\rangle\langle\phi_n|\ \rho_S(\tau)  
\int_0^{\tau}\!\!\!\! dt_1 \!\!\int_0^{\infty} \!\!
d\omega D(\omega) g^2(\omega)
\overline{n}(\omega)  e^{{-i(t_1-\tau)(\omega-\omega_l+\omega_n)}},  
\nonumber
\eea
\be
\label{empty}
\ee
where $\omega_m=E_m/\hbar$ and $\omega_l=E_l/\hbar$. Interchanging the
order of the time and frequency integral and introducing the variables
$t_2=\tau-t_1$, $\omega_{l  n}=\omega_l-\omega_n>0$ we obtain 
\be
I_1=\sum_{l, m<l, n<l}\mathcal{X}_{m l} 
\mathcal{X}^{\dagger}_{ln}  e^{{{i(E_m-E_n)\tau}\over{\hbar}}}
|\phi_m\rangle\langle\phi_n|\   \rho(\tau)  
\int_0^{\infty} d\omega\int^{\tau}_0 dt_2 D(\omega) g^2(\omega)
\overline{n}(\omega) e^{{it_2(\omega-\omega_{l  n})}}.
\label{MEbefdelta}
\ee
At this point it is worth recalling that $\tau$ is the duration
of a time interval which is short from the system's point of view,
i.e., the \emph {interaction picture} reduced density operator 
changes a little during $\tau$. However, since the interaction is
assumed to be weak, the relation $\tau\gg 1/\omega_{l  n}$ 
also holds. This allows us
to extend the upper limit of the time integration to infinity in 
Eq.~(\ref{MEbefdelta}). Then the identity
\be
\int_0^{\infty}du \ e^{\pm iwu}=\pi \delta(w) \pm i {\mathrm {Pv}} \left(
{{1}\over{w}}\right),
\label{deltaid}
\ee  
where the Dirac-$\delta$ and Cauchy principal value distributions appear on
the RHS, allows us to evaluate the integral in $I_1$. Eq.~(\ref{deltaid})
shows that the effect of the environment is twofold: first it slightly modifies
the energy spectrum of the system. This effect is related to the 
imaginary term in Eq.~(\ref{deltaid}).
If our aim is not the calculation of the level shifts themselves, 
then they can be neglected, provided the interaction is not 
too strong \cite{A74}.
The second effect of the environment (related to the first term in  
Eq.~(\ref{deltaid})) is to induce transitions between
the (shifted) system energy levels, and this kind of environmental
influence is responsible for the decoherence. 
Therefore we can use the approximation 
\bea
I_1&\approx&\sum_{l, m<l, n<l}\mathcal{X}_{m l} 
\mathcal{X}^{\dagger}_{ln}  e^{{{i(E_m-E_n)\tau}\over{\hbar}}}
|\phi_m\rangle\langle\phi_n|\   \rho(\tau) 
\int_0^{\infty} D(\omega) g^2(\omega)\overline{n}(\omega)
\pi \delta(\omega-\omega_{l  n})
\nonumber
\\
&=&U^{\dagger}(\tau)\mathcal{X} 
\mathcal{X}^{\dagger}_{a}\rho_{S}U(\tau),
\label{MEonepart}
\eea 
where we returned to the explicit notation of the interaction picture,
and the matrix elements of the operator 
$\mathcal{X}^{\dagger}_{a}$ are defined by
\be
\langle \phi_m |\mathcal{X}^{\dagger}_{a}|\phi_n\rangle=
\langle \phi_m |\mathcal{X}^{\dagger}|\phi_n\rangle \ \ \pi  
D(\omega_{n m}) g^2(\omega_{n m}) \overline{n}(\omega_{n m}).
\label{MEXelem1}
\ee

The master equation in the Schr\"{o}dinger picture,  neglecting the
terms inducing level shifts, reads:
\bea
{{d}\over{dt}}\rho_{S}(\tau)&=&
-{{i}\over{\hbar}}\left[H_S,\rho_{S}(\tau)\right]
-\mathcal{X}^{\dagger}\mathcal{X}_{e}\rho_{S}(\tau)-\mathcal{X}\mathcal{X}_{a}^{\dagger}\rho_{S}(\tau)-
\rho_{S}(\tau)\mathcal{X}_{e}^{\dagger}\mathcal{X}-\rho_{S}(\tau)\mathcal{X}_{a}\mathcal{X}^{\dagger}
\nonumber
\\
&+&\mathcal{X}_{a}^{\dagger}\rho_{S}(\tau)\mathcal{X}
+\mathcal{X}_{e}\rho_{S}(\tau)\mathcal{X}^{\dagger}
+\mathcal{X}^{\dagger}\rho_{S}(\tau)\mathcal{X}_{a}
+\mathcal{X}\rho_{S}(\tau)\mathcal{X}^{\dagger}_{e},
\label{ME}
\eea
where each term following the unitary one (the commutator with $H_S$)
is calculated similarly to $I_1$, and
\be
\langle \phi_m |\mathcal{X}_{e}|\phi_n\rangle=
\langle \phi_m |\mathcal{X}|\phi_n\rangle \ \ \pi  D(\omega_{n m}) g^2(\omega_{n m})
\left(\overline{n}(\omega_{n m})+1\right).
\label{MEXelem2}
\ee
The subscript $e$ here and 
and $a$ in Eq.~(\ref{MEXelem1}) refers to emission 
and absorption, respectively.
As we can see, the matrix elements (\ref{MEXelem1}) and (\ref{MEXelem2}) 
of the operators that induce the transitions 
depend on the Bohr frequency of the involved transition,
which is a genuine anharmonic feature.  
In the special case of the HO, when
$H_S$ has equidistant spectrum, and
$\mathcal{X}$ is identified with the usual annihilation operator $a$, 
both $\mathcal{X}_{a}$ and $\mathcal{X}_{e}$ are
proportional to $\mathcal{X}\equiv a$, and Eq.~(\ref{ME}) reduces to
the amplitude damping master equation (\ref{ampdamp}) 
at a finite temperature.
\vskip 12pt 

In certain cases one can further simplify Eq.~(\ref{ME}). When
the environment induced relaxation rates are much lower than the relevant
Bohr frequencies, the system Hamiltonian induces oscillations
that are very fast even on the time scale of decoherence and vanish
on the average. Ignoring these fast oscillations we arrive at the
interaction picture master equation 
\begin{equation}
\frac{d}{dt}\langle\phi_i|\rho_{S}|\phi_j\rangle=
\delta_{i,j}\sum_{k\neq i}\gamma_{ik}\langle\phi_k|\rho_{S}|\phi_k\rangle
-\Gamma_{ji}^c \langle\phi_i|\rho_{S}|\phi_j\rangle,
\label{ME2}
\end{equation}
that has already been obtained in Refs.~\cite{L66, A73}
in order to treat the spontaneous emission
of a multilevel atom. 
In Eq.~(\ref{ME2}), $\gamma_{ik}$ denotes a relaxation rate, that is 
the probability of the $|\phi_k\rangle\rightarrow|\phi_i\rangle$ 
transition per unit time, while $\Gamma_{ji}^c=1/2\sum_k(\gamma_{ik}+
\gamma_{jk})$, where
\begin{equation}
\gamma_{ik}=\left\{
\begin{aligned}
&2\  \langle \phi_i |\mathcal{X}_{e}|\phi_k\rangle
\langle \phi_i |\mathcal{X}|\phi_k\rangle
&& {\textnormal{if}}\  i<k,
\\
&0 && {\textnormal{if}}\  i=k,
\\
&2\  \langle \phi_i |\mathcal{X}_{a}|\phi_k\rangle
\langle \phi_i |\mathcal{X}|\phi_k\rangle
&& {\textnormal{if}}\  i>k.
\label{MEpoptrprob}
\end{aligned}
\right.
\end{equation}

However, due to the elimination of 
the fast oscillations related to $H_S$, Eq.~(\ref{ME2}) is not suitable 
for investigating the wave packet motion and decoherence simultaneously,
therefore we propose to use Eq.~(\ref{ME}). On the other hand we note that 
Eq.~(\ref{ME2}) radically reduces the computational costs of calculating
the time evolution for long times, which might be necessary when
the system-environment coupling is very weak.

Supposing that our knowledge is limited to
the populations $P_n=\langle \phi_n|\rho_s|\phi_n\rangle$,
both Eq.~(\ref{ME}) and Eq.~(\ref{ME2}) leads to the Pauli type equation
\begin{equation}
\frac{d}{dt}P_n=\sum_k \left( \gamma_{nk} P_k-\gamma_{kn}P_n \right).
\label{pauli}
\end{equation}
Requiring the condition of detailed 
balance \cite{R65} in Eq.~(\ref{pauli}) leads to the steady-state 
thermal distribution at the temperature of the environment.
\vskip 12pt

In the case of a diatomic molecule in the free electromagnetic field, 
$g^2(\omega)D(\omega)\propto \omega^3$, and we assume that $\mathcal{X}=X$,
thus the nonzero matrix elements in Eqs.~(\ref{MEXelem1}) and (\ref{MEXelem2})
are
\begin{equation}
\begin{aligned}
\langle \phi_m |\mathcal{X}_{a}|\phi_n\rangle&=\lambda\ 
\langle \phi_m |X|\phi_n\rangle \ {\omega_{n m}}^3\ 
\overline{n}(\omega_{n m}),&&\ n<m \\
\langle \phi_m |\mathcal{X}_{e}|\phi_n\rangle&=\lambda\ 
\langle \phi_m |X|\phi_n\rangle \  {\omega_{n m}}^3 
\left(\overline{n}(\omega_{n m})+1\right),&&\ n>m 
\label{MorseX}
\end{aligned}
\end{equation}
where the matrix elements of $X$ can be calculated using 
the algebraic method summarized in \cite{BM99}, and
$\lambda=\pi g^2(\omega)D(\omega)/\omega^3$
is an overall, frequency independent coupling constant. 
For the sake of definiteness we have chosen the NO molecule as our
model.

In order to get insight into the 
interplay between wave packet motion and decoherence, it
is worth considering a stronger molecule-environment interaction
than the electromagnetic field modes can provide.
Keeping the structure of Eqs.~(\ref{MorseX}), this can be done by
increasing the value of $\lambda$.
Here we present calculations with two different
coupling constants, $\lambda_1$  and  $\lambda_2$ which are chosen 
so that at zero temperature  
$\omega_{01}/\gamma_{01}\approx10^5$ and $4\times 10^3$ 
for $\lambda=\lambda_1$ and $\lambda_2$, respectively. 
This model allows for the numerical integration of  
the master equation (\ref{ME}) (that provides more details of the 
dynamics than Eq.~(\ref{ME2}))
in a time interval that is long enough to identify the
effects of decoherence. These effects can be summarized
in a decoherence scheme (see Sec.~\ref{wignersec}) that has 
a clear physical interpretation, and which is valid also in the weak
molecule-environment interaction, when (\ref{ME2}) is more efficient to
calculate the time evolution.

\section{Time evolution of the expectation values}
\label{expectsec}
Starting from  $|x_0,p_0=0\rangle$ as initial states,
we saw in Chap.~\ref{oechap} that the qualitative behavior of the 
expectation value 
$\langle X\rangle(t)=\langle\psi(t)| X|\psi(t)\rangle$ draws the limit
of small oscillations. In the absence of environmental
coupling (i.e., $\lambda=0$), for $x_0\leq 0.06$, $\langle X\rangle(t)$
(as well as $\langle P\rangle(t)$) exhibits sinusoidal oscillations. 
\begin{figure}[htb]
\begin{center}
\psfrag{plabel}[tl][tl]{$\langle P \rangle$}
\psfrag{xlabel}{$\langle X \rangle$}
\psfrag{tlabel}{$t/t_0$}
\includegraphics*[bb=80 50 670 470 , width=10cm]{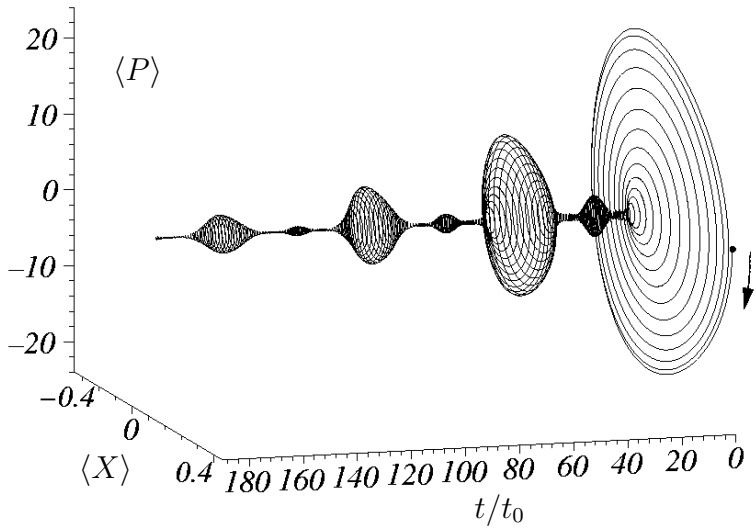}
\caption[Phase portrait of an anharmonic wave packet subject to decoherence]
{
Phase portrait corresponding to the time evolution of the initial state
$|\psi(t=0)\rangle=|x_0,p_0=0\rangle$, with $x_0=0.5$. 
The parameters are $\lambda=\lambda_1$, $T=5 \ \hbar\omega_{01}/k$, and
$t_0$ is the period of the small oscillations in the 
potential. The initial point $\langle X \rangle=0.5$, $\langle P \rangle=0$
together with the starting direction is also indicated.  
\label{portrait}
}
\end{center}
\end{figure}
For larger initial displacements from
the equilibrium position, the anharmonic effects become apparent.
The collapse and revival in 
$\langle X\rangle(t)$ and $\langle P\rangle(t)$ 
can be explained by referring
to the various Bohr frequencies that determine their time dependence:
dephasing of these frequencies leads to the collapse of the expectation 
value, and we observe revival when they rephase again. 

For the initial state of $|x_0,0\rangle$, with $x_0=0.5$,
the original phase of the 
eigenstates is restored \cite{PS86,AP89} 
around the full revival time $t_{rev}=110\ t_0$, where $t_0$ is the period 
of the small oscillations in the potential.   
At $t/t_0=55$ and $t/t_0=27.5$
half and quarter revivals \cite{PS86,AP89} can be observed.
Fig.~\ref{portrait} shows the damping of the revivals both in
$\langle X\rangle(t)$ and $\langle P\rangle(t)$
when interaction with the environment is turned on. 
Note that the phase portrait of the corresponding classical particle
would be a helix with monotonically decreasing diameter,
revivals are of quantum nature. However, Fig.~\ref{portrait}
does not provide a complete description of the time evolution in
the phase space, this can be given by
using Wigner functions, see Sec.~\ref{wignersec}.


\section{Decoherence times}
\label{dectimesec}
Our master equation (\ref{ME}) describes decoherence as well
as dissipation. However, the time scale of these processes is generally
very different, providing  a useful tool to distinguish the stages of the
time evolution that are dominated either by decoherence or 
dissipation \cite{FCB01a}.
\begin{figure}[htb]
\begin{center}
\psfrag{elabel}[tl][tl][0.8]{$S(t)$}
\psfrag{trlabel}[bl][bl][0.8]{$\mathrm{Tr}\left[\rho_S^2(t)\right]$}
\psfrag{eaxis}{$S$}
\psfrag{traxis}[br][bl]{$\mathrm{Tr}\rho_S^2$}
\psfrag{taxis}{$t/t_0$}
\includegraphics*[bb=65 45 775 515 , width=10cm]{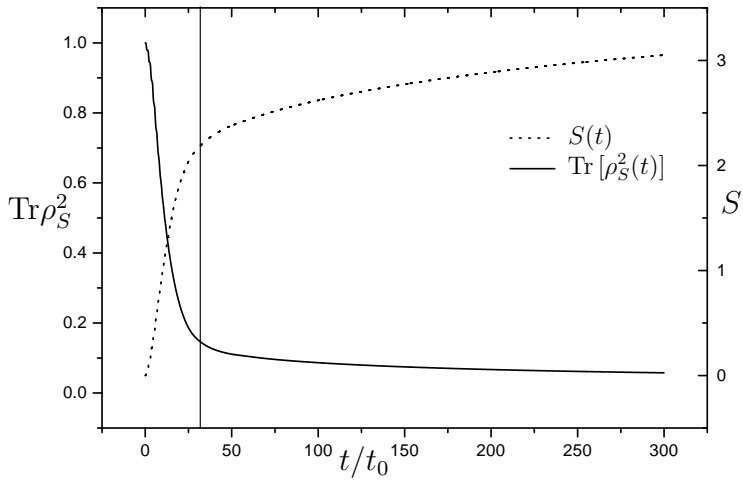}
\caption[Determination of the decoherence 
time in the Morse system by the aid of timescale separation]
{
The entropy and the purity of the reduced density matrix of the Morse system
as a function of time, calculated using Eq.~(\ref{ME}). The coupling parameter
(see Sec.~\ref{mastersec}) is $\lambda=\lambda_1$ and
$T=10 \ \hbar\omega_{01}/k$. The initial state was  
$|\psi(t=0)\rangle=|x_0,0\rangle$, with $x_0=2.0$.
\label{dtdef}
}
\end{center}
\end{figure}
In Fig.~\ref{dtdef} an example is depicted showing how the method of 
time scale separation works.
We have calculated the entropy 
\begin{equation}
S=-\mathrm{Tr}\left[\rho_S \ln(\rho_S)\right],
\end{equation}
as well as the quantity $\mathrm{Tr}[\rho_S^2]$, 
which measures the purity of the reduced density operator. Note that
the $\mathrm{Tr}$ operation without subscript refers to the
trace in the system's Hilbert space.
Decoherence time $t_d$ is defined as the time instant 
that divides the time axis into two parts
where the character of the physical process is clearly different. 
Initially both $S(t)$ and $\mathrm{Tr}[\rho_S^2(t)]$ change rapidly 
but having passed  $t_d$ (emphasized by a vertical line in 
Fig.~\ref{dtdef}), the moduli of their derivative
significantly decrease. After $t_d$ the entropy and the purity 
change on the time scale which is characteristic of 
the dissipation of the system's energy during the whole process. 
The time dependence of the participation 
ratio $K$ given by Eq.~(\ref{EIDpartic}) is found 
to be similar to that of the entropy and purity. We note that the typical 
value of $K$ at the decoherence time was around $5$, that is, just
a few modes of the environment were active. The same surprising 
result was found in Ref.~\cite{CLE02}, in the context of spontaneous 
emission from a two-level atom.
  
In summary, decoherence dominated time evolution
turns into dissipation dominated dynamics around $t_d$.
In the next section we shall determine the density operators 
into which the process of decoherence drives the system. 
In connection with these results we have verified that 
the states around the decoherence time 
do not change appreciably in a time interval
that covers the possible errors in determining $t_d$.

An interesting question is the dependence of the decoherence time on the 
initial state of the time evolution. 
We calculated $t_d$ as a function of the initial displacement
for the case of displaced 
ground states (that is, coherent states with zero momentum, 
$|x_0,0\rangle$) as initial states.
It was found that for all values of $\lambda$ and $T$, the
decoherence time is longer for smaller initial displacements. 
Additionally, for fixed $\lambda$ and $T$ the function
$t_d(x_0)$ can be well approximated by an exponential curve
$t_d(x_0)=t_d(0) \exp(-\kappa x_0)$. E. g., for $\lambda=\lambda_1$,
$T=10 \ \hbar\omega_{01}/k$ and $0<x_0\leq 2$ the parameters take 
the values $t_d(0)=93\  t_0$ and $\kappa=0.97$.

It is known (see Chap.~\ref{oechap} and Ref.~\cite{AP89}) that 
quarter revivals in an anharmonic potential
lead to the formation of Schr\"{o}dinger-cat states, i.e., states that are
superpositions of two distinct states localized in space 
\cite{AP89} as well as in momentum \cite{FC02,KP95}. 
On the other hand,   
smaller initial displacements correspond to classical phase space orbits
with smaller diameter. Consequently the quantum interference related to 
nonclassical states that are formed during the course of time 
cover a smaller area in the phase space in this case. 
This means that our result is a manifestation of the general
feature of decoherence that increasing the ``parameter of nonclassicality'',
which is the diameter of the corresponding classical orbit in our case,
causes faster decoherence \cite{GJK96}. 
A similar result was found in \cite{FCB01a} for the case of 
decoherence in a system of two-level atoms \cite{BC99,BBH00}. 


\section{Wigner function description of the decoherence}
\label{wignersec}
In order to visualize the time evolution of the reduced density matrix of the
Morse system we have chosen the Wigner function picture, which has been 
summarized in Sec.~\ref{QPD}.
This description allows us to investigate the correspondence 
between classical and quantum dynamics. 
\begin{figure*}[h]
\begin{center}
\includegraphics*[bb=60 60 1370 1070 , width=14.5cm]{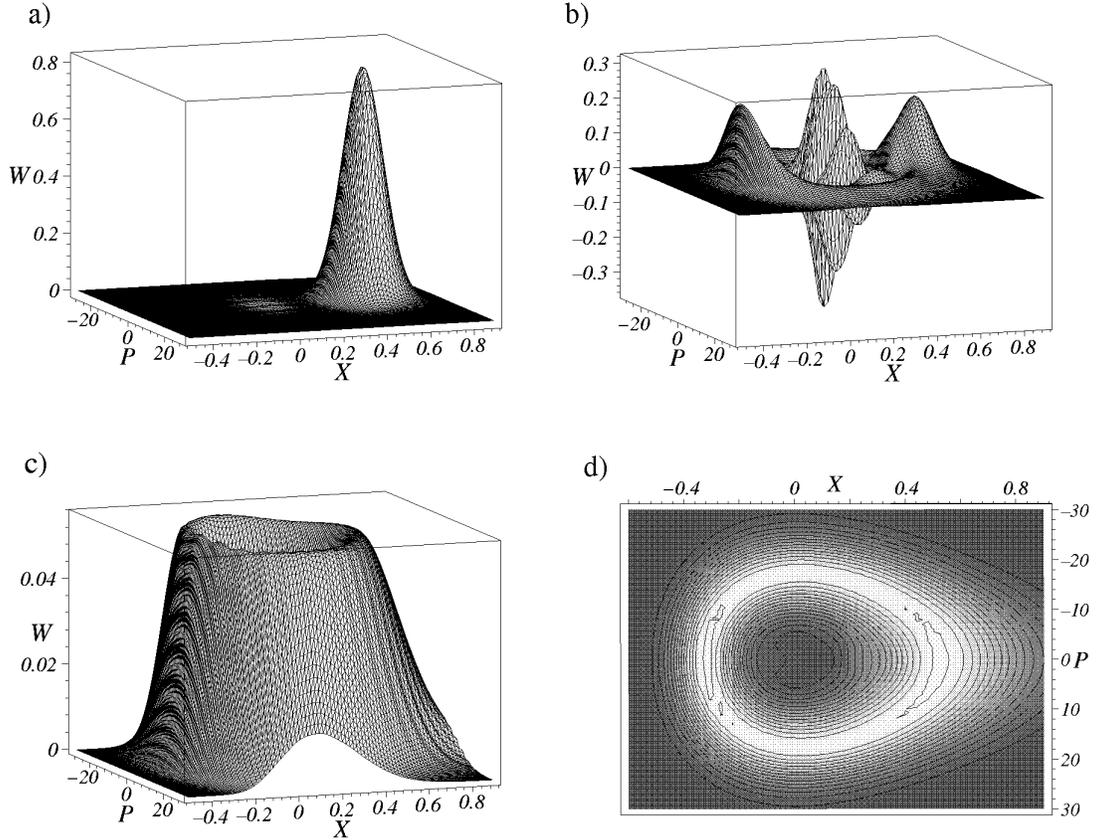}
\caption[Wigner functions visualizing the decoherence 
of a wave packet in the Morse potential]
{
Time evolution of the Wigner function corresponding to the initial state
$|\psi(t=0)\rangle=|x_0,0\rangle$, with $x_0=0.5$. The coupling parameter is
(see Sec.~\ref{mastersec}) $\lambda=\lambda_2$ and
$T=0.3 \ \hbar\omega_{01}/k$. The plots a) and b) 
correspond to time instants $t/t_0=0\  \text{and}\  27.5$, while
both c) and the contour plot d) are snapshots taken at $t/t_0=137.5$.  
\label{wig1}
}
\end{center}
\end{figure*}
 
First we recall the ideal case without environment.
Then, in the initial stage of the time evolution, the positive hill 
corresponding to the wave packet $|x_0,p_0\rangle$
follows the orbit of the classical particle that has started 
from $(x_0,p_0)$ at $t=0$. However, due to the uncertainty relation,
the Wigner function as a quasiprobability distribution has a finite
width, and this fact combined with the form of the Morse potential 
implies the stretching of the Wigner function along the classical 
orbit in the course of time. (See Ref.~\cite{KP95} for similar results
with the Husimi $Q$ function.) After a certain time the
increasingly broadened wave packet
becomes able to interfere with itself,
and around the quarter revival time one can observe two positive hills
chasing each other at the opposite sides of the classical orbit.
The strong oscillations of $W$ between the hills represent the quantum 
correlation of the constituents of this molecular Schr\"{o}dinger-cat 
state \cite{JV94}.
Later on the initial Wigner function is restored almost exactly and
Schr\"{o}dinger-cat state formation starts again. Detailed 
Wigner function description of
these processes that are related to the free time evolution can be found
in \cite{FC02}. 

In the case when environmental effects are present,
we found that decoherence follows a general scheme. A representative series 
of Wigner functions is shown in Fig.~\ref{wig1}. The snapshots correspond
to the initial state and time instants when the first and third
Schr\"{o}dinger-cat state formation  would occur in the absence of 
the environment. Consequently, the Wigner function in Fig.~\ref{wig1} b) 
corresponds almost to a Schr\"{o}dinger-cat state, but this state is 
already a mixture. However, there are still negative parts of the function
in between the positive 
hills centered at $x_1=0.51,\  p_1=0$ and  $x_2=-0.34,\  p_2=0$.
The ``ridge'' that connects these hills
along the classical orbit is absent in a pure Schr\"{o}dinger-cat state,
see Fig.~\ref{wigs} b). 
Later on this ridge becomes more and more pronounced 
and at the decoherence time 
we arrive at the positive (that is, classical in the sense of
Sec.~\ref{QPD}) Wigner function of  Fig.~\ref{wig1} c) and d).
According to the contour
plot  Fig.~\ref{wig1} d), the highest values of this function trace out
the phase space orbit of the corresponding classical particle.
That is, $\rho_S^{dec}$, the reduced density matrix 
that arises as a result of decoherence, 
can be interpreted as a mixture of localized states that are
equally distributed along the
orbit of the corresponding classical phase space orbit. 
\begin{figure*}[htb]
\begin{center}
\includegraphics*[bb=50 60 1350 1100 , width=14.5cm]{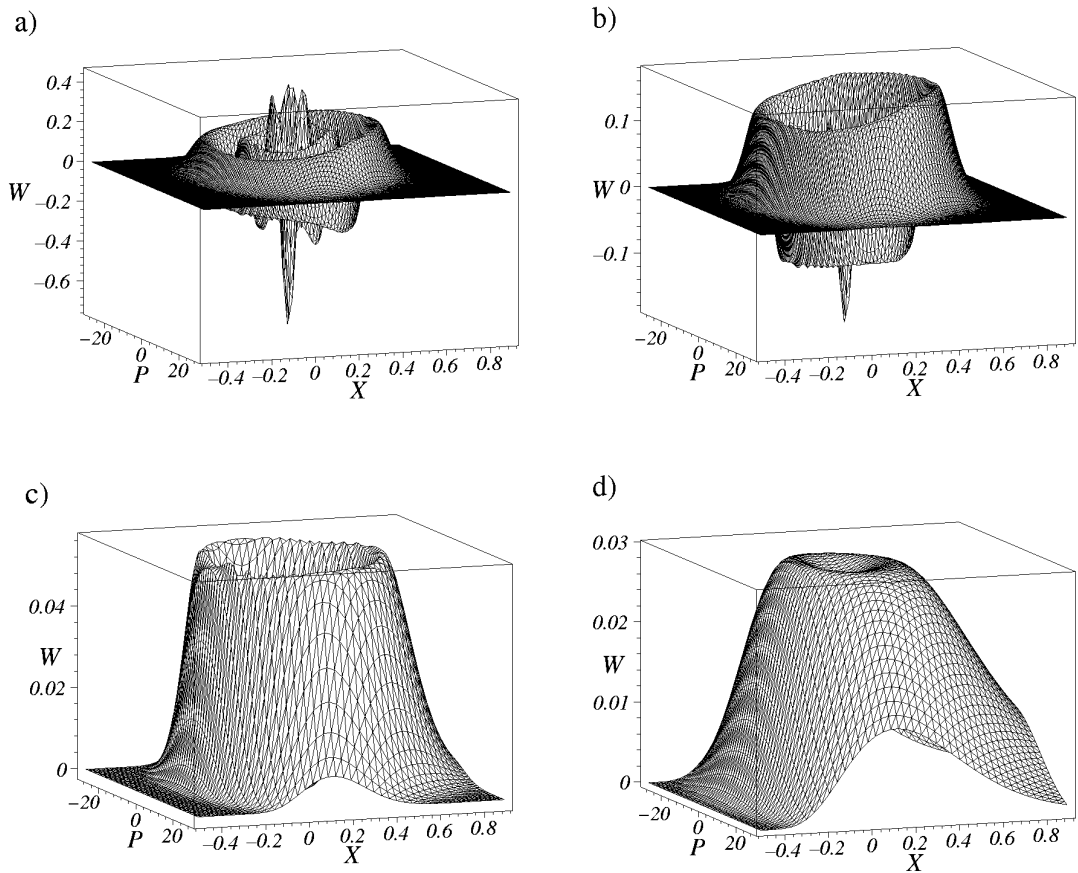}
\caption[Decoherence of an eigenstate of the 
Morse potential: Wigner functions]
{
Time evolution of the Wigner function corresponding to the fifth bound
state as initial state.
The coupling parameter
(see Sec.~\ref{mastersec}) is  $\lambda=\lambda_1$ and
$T=10 \ \hbar\omega_{01}/k$. The plots a), b), c) and d)
correspond to time instants $t/t_0=0,\   27.5, \ 330$  and $1000$, 
respectively.
\label{wig2}
}
\end{center}
\end{figure*}

It is worth comparing this result with the case of the HO, when 
the master equation (\ref{ME}) reduces to the
amplitude damping equation (\ref{ampdamp}), see Sec.~\ref{fokkersec}.
It is known that harmonic oscillator coherent states are
robust against the decoherence described by the 
amplitude damping master equation (as well as against the Caldeira-Leggett
\cite{CL83} master equation \cite{ZHP93}), 
the initial superposition of coherent 
states turns into the statistical mixture of essentially the
same states. This is a consequence of the facts that these
states are eigenstates of the destruction operator $a$, and
the operators in the nonunitary terms of Eq.~(\ref{ME}) 
are proportional to $a$ and $a^{\dagger}$ in the harmonic case. 
None of these statements can be transferred to the anharmonic system, 
where the Morse coherent states do not remain localized during the
course of time, even without environment.
Therefore the scheme of the decoherence is qualitatively different for the
harmonic and anharmonic oscillators: Our results in the anharmonic system are
similar to the phase relaxation in the harmonic case \cite{WM85}, where the
energy of the system remains unchanged, but the phase information is 
completely destroyed, see Sec.~\ref{fokkersec}. 
We note that a similar result was obtained in Ref.~\cite{B01},
where the rotational degrees of freedom were considered as a 
reservoir for the harmonic vibration of hot alkaline dimers.

Our decoherence scheme is universal to a large extent.
In the investigated domain of the coupling constants $\lambda_1\leq \lambda
\leq\lambda_2$ and temperatures ranging from $T=0$ 
to $T=15 \ \hbar\omega_{01}/k$,
it is found to be valid for all initial states, not only for coherent states.
Fig.~\ref{wig2} shows an example when the initial state is not a wave packet,
it is the fifth bound state, corresponding to $E_5$, which
is very close to $\langle0.5,0|H_S|0.5,0\rangle$, so direct comparison
with Fig.~\ref{wig1} is possible. As we can see, although the two Wigner
functions are initially obviously very different, they follow different routes
(that takes different times) to the \textit{same} 
state: Fig.~\ref{wig1} c) and 
Fig.~\ref{wig2} c) are practically identical. The final plot 
in Fig.~\ref{wig2} indicates how the Wigner function represents the 
long way to thermal equilibrium with the environment: the distribution
becomes wider and the hole in the middle disappears.    

It is expected that the loss of phase information has observable
consequences. According to the Franck-Condon principle,
the absorption spectrum of a molecule around
the frequency corresponding to an electronic transition between 
two electronic surfaces depends on the vibrational state. 
The time dependence of the spectrum should exhibit the 
differences between the pure state of an oscillating wave packet and
the state $\rho_S^{dec}$ and the thermal state.
More sophisticated experimental methods based on the
detection of fluorescence \cite{KZ90} or fluorescence
intensity fluctuations \cite{WT00}, surely have the capacity
of observing the dephasing phenomenon considered in this chapter. 
  
\section{Conclusions}
We investigated the decoherence of wave packets in the Morse potential.
The decoherence time for various initial states was calculated and it
was found that the larger is the diameter of the phase space orbit described 
by a wave packet, the faster is the decoherence. We obtained a general
decoherence scheme, which has a clear physical interpretation: The reduced
density operator that is the result of the decoherence is a mixture of
states localized along the corresponding classical phase space orbit.

\addtocontents{toc}{\newpage}
\chapter[Two-level atoms and decoherence]
{
A system of two-level atoms in interaction with the environment
}
\emptypage
\label{atomchap}
Two-level atoms are essential objects in quantum optics,
several important models rely on this notion \cite{JC63,ENSM80,A74,WW30,WW31}.
Clearly, most atoms have much more than two energy levels, i.e.,
considering only two of them is a simplification.
However, in usual (experimental) situations the initial conditions
and the frequency of the external 
electromagnetic field or the long lifetime of the lower level
supports the two-level view of the atomic system.
Additionally, a two-level atom provides a physical
realization of a \emph{qubit}, which is the basic entity in 
quantum computation (QC)\cite{NC00,PHBK99,BD00}.

In the present chapter we investigate a system which is a candidate for the 
experimental study of decoherence and possibly also for practical 
applications. The model consists of  
several identical two-level atoms (the system) interacting with a  large 
number of photon modes in a thermal state (the environment). 
It has the advantage that it is simple to make the correct transition from a 
microscopic system to a macroscopic one by increasing the number of atoms.
We point out how the master equation (\ref{ME}) reduces to the equation 
appropriate in this case \cite{BSCHH71,A74,HR85}, and use it to 
analyze the evolution of the reduced density matrix of the atomic system.

By analytical short-time calculations we show 
that the atomic coherent states \cite{ACGT72} of our system are robust against 
decoherence caused by the realistic interaction we consider. 
The possibility of classical interpretation and this behavior 
justifies that the superpositions
of atomic coherent states are relevant with respect to the original 
problem of Schr\"{o}dinger, and such a highly nonclassical 
superposition is rightly called 
an atomic Schr\"{o}dinger-cat state \cite{BCB97,BC99,JV90}.
We also note that there are several proposals for the experimental preparation
of these type of states \cite{APS97,GG97,GG98}.

We present the decoherence and dissipation properties 
of  atomic Schr\"{o}dinger-cat states 
based on numerical computations of their time evolution.
It will be seen that similarly to the case of the Morse 
system (see Chap.~\ref{morsedecchap}), although 
the one and the same solution of the master equation 
describes both decoherence and dissipation, 
the time scales of these processes differ by orders of magnitude.
Using this fact, we show how one can make a clear distinction between 
these two processes despite of the interplay between them, 
and define the decoherence time. 
This decoherence time strongly depends on the initial conditions,
notably, it is particularly large for a special 
set of initial cat states \cite{BC99,BBH00}.
This will be termed as slow decoherence in contrast with the general case which
will be referred to as rapid decoherence.

The interplay between decoherence and energy dissipation 
is the most appreciable in connection with the concept of 
pointer states that has been summarized in Sec.~\ref{pointersec}. 
It will be shown that when the decoherence is rapid, 
then the constituent coherent states of the initial state are pointer states 
to a very good approximation. However, when there is enough time for 
dissipation, i.e., when decoherence is slow, then the initial atomic coherent 
states themselves evolve into mixtures, and  
therefore a refined scheme of decoherence 
holds.

In order to underline the contrast between 
rapid and slow 
decoherence we superpose four atomic coherent states
corresponding to the vertices of a suitably oriented tetrahedron.
The time evolution of this four component cat state will be studied 
by the aid of the spherical Wigner function (Sec.~\ref{QPD}). 
As it is expected, the interaction with the environment 
selects that pair from the initial superposition 
which constitutes a long-lived cat state.

\section{Description of the model}
\label{model}
We consider 
a system of identical two-level atoms interacting 
with the environment of macroscopic number of photon modes. With dipole 
interaction and in the rotating wave approximation
the total system is
described by the following model Hamiltonian:
\begin{equation}
H=H_S+H_E+V=\hbar \omega_{\mathrm{a}}J_{z}+ \sum_{k} \hbar \omega _{k}a_{k}^{\dagger
}a_{k}+ \sum_{k} \hbar g_{k}\left( a_{k}^{\dagger }J_{-}+a_{k}J_{+}\right) ,
\label{model_ham}
\end{equation}
where $\omega_{\mathrm{a}}$ is the transition frequency between the two atomic 
energy levels,  $\omega_k$ denote the frequencies of the modes of the 
environment and  $g_k$ are 
coupling constants. The operators $J_+$, $J_-$ and $J_{z}$ in the 
interaction term are
dimensionless collective atomic operators obeying the usual angular
momentum commutation relations \cite{D54}.
On replacing $\mathcal{X}^{\dagger}$ and 
$\mathcal{X}$ by $J_+$ and $J_-$ respectively,
the process we followed in Sec.~\ref{mastersec} leads to the 
interaction picture master equation \cite{BSCHH71,A74}
\begin{eqnarray}
\frac{\rm{d}\rho (t) }{\rm{d}t} & = & -\frac{\gamma }{2}
\ (\overline{n}+1)\ \left(J_{+}J_{-}\rho (t)
+\rho (t) J_{+}J_{-}-2J_{-}\rho (t) J_{+}\right)
\nonumber \\
& & -\frac{\gamma }{2}
\ \overline{n}\ \left(J_{-}J_{+}\rho (t)
+\rho (t) J_{-}J_{+}-2J_{+}\rho (t) J_{-}\right).  \label{mastereq}
\end{eqnarray}
Here $\overline{n}$ is the mean number of photons in the environment,
$\gamma=\pi D(\omega_a) g^2(\omega_a)$ denotes the damping rate, 
and for the sake of simplicity
the subscript $S$ of the reduced density operator of atomic system
has been dropped. 
Note that the same master equation can be obtained 
by considering a low-Q cavity containing Rydberg atoms \cite{HR85}.

If the state of the atomic system was initially invariant
with respect to the permutations of the atoms, i.e., it was a superposition 
of the totally symmetric Dicke states \cite{D54}, the dipole 
interaction described by 
$V= \sum_{k} \hbar g_{k}\left( a_{k}^{\dagger }J_{-}+a_{k}J_{+}\right)$ 
in (\ref{model_ham}) would not destroy this symmetry.
Therefore we may restrict
our investigation to the totally symmetric $N+1$
dimensional subspace of the whole Hilbert-space of the atomic system. 
This subspace corresponds to the first column in Fig.~\ref{Dladder} 
and it is isomorphic to an angular momentum eigensubspace labeled by $j=N/2$.  
This model has been proven to be valid in
cavity QED experiments with many atoms, as reviewed 
in \cite{HR85}.

The environment as a static reservoir (represented by the thermal photon modes)  
continuously interacts with the atomic system influencing its dynamics.
As it is obvious, the dissipation of the energy leads to thermal equilibrium in the system,
corresponding to the stationary solution of the master equation (\ref {mastereq}). 
However, as it will be shown here, the same master equation describes also 
a much more interesting process. 
The continuous "monitoring" \cite{Z81} of the 
atomic system by the environment results in the total loss of the 
coherence of the quantum superpositions in the system.
This decoherence process is generally extremely fast compared to the dissipation, except for special
initial conditions  
which will be discussed in section \ref{times}.


\section {The initial stage of the time evolution}
\label{robust}
In this section we apply the
general concepts introduced in Sec.~\ref{EIDsec}
to our system in order to find the initial states 
for the master equation (\ref {mastereq}) which
are relevant to the original problem of Schr\"{o}dinger \cite{SCH35a,SCH35b}
concerning the unobservability of macroscopic superpositions.

First we consider the short-time behavior of the total
system. 
At zero temperature the photon field of the present model is  in its pure 
vacuum state $\left| 0\right\rangle$, therefore the initial state factorizes as
\be
\left| \Psi (0)\right\rangle =\left| \varphi _{0}(0)\right\rangle
\left| 0\right\rangle,
\ee
i.e., there is a single term in the Schmidt decomposition 
(see Sec.~\ref{EIDsec}) of the compound state.  Due to the
interaction, this product state evolves into
a more general Schmidt sum like Eq.~(\ref{schmidt}), or in other words it 
turns into an entangled state. According to the summary in Sec.~\ref{EIDsec}, 
the rate of entanglement can be obtained as
\be A=\sum_{k\neq 0,l\neq 0}\left| \left\langle \varphi _{k}(0)\right|
\left\langle \Phi _{l}(0)\right| V\left| \varphi _{0}(0)\right\rangle
\left| 0\right\rangle \right| ^{2}.
\ee
Using the 
explicit form of the interaction Hamiltonian $V$ in Eq.~(\ref{model_ham}),
a straightforward calculation leads to
\be
A={\cal C}\left( J_{+},J_{-}\right) :=\left\langle J_{+}J_{-}\right\rangle
-\left\langle J_{+}\right\rangle \left\langle J_{-}\right\rangle,
\label{normcorr} 
\ee
i.e., in our system the rate of entanglement is the normally
ordered correlation function of the operators $J_{+}$\ and $J_{-}$. 

\begin{figure}[htbp]
\begin{center}
\includegraphics*[bb=35 75 270 275 , width=8cm]{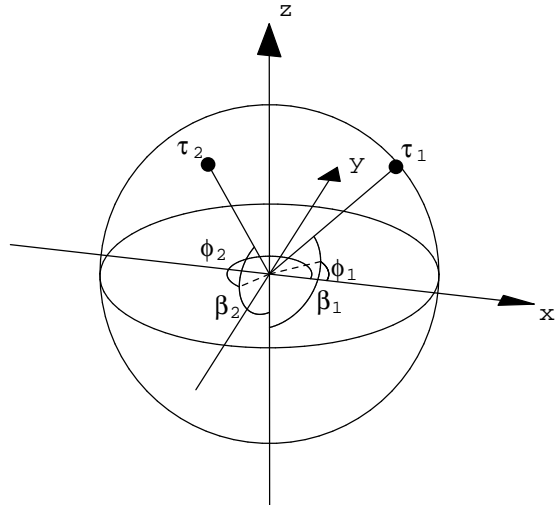}
\end{center}
\caption[Scheme of an atomic Schr\"{o}dinger cat state]{
Scheme of an atomic Schr\"{o}dinger cat state defined by Eq.~(\ref{cats}).
The points labeled by $\tau_1$ and $\tau_2$ represent the 
corresponding atomic coherent
states on the surface of the Bloch-sphere. The angles defining the  
parameters $\tau_1$ and $\tau_2$ are also shown.
}
\label{sphere}
\end{figure}

Let us turn to the 
case of finite temperatures,
when the total system has to be represented by a 
mixed state 
even at $t=0$. The linear entropy, defined as
\be
S_{\mathrm{lin}}={\rm Tr}(\rho - \rho^2), \label {linent}
\ee
can be regarded as a relevant measure of decoherence \cite{GJK96,Z93}.
Restricting ourselves again to the initial regime of the time evolution, 
we can make use of the master equation (\ref {mastereq}) and calculate the 
time derivative of the linear entropy at $t=0$
\be
\left( \frac{\partial S_{\mathrm{lin}}}{\partial t}\right) _{t=0}=\gamma \left(
\left\langle n\right\rangle {\cal C}\left( J_{-},J_{+}\right) +\left( \left\langle
n\right\rangle +1\right) {\cal C}\left( J_{+},J_{-}\right) \right).
\label{sat0}
\ee
The normally (antinormally) ordered correlation function
${\cal C}\left( J_{+},J_{-}\right)$ (${\cal C}\left( J_{-},J_{+}\right)$), 
disappears in the eigenstate $\left| j, m=-j \right\rangle$  
($\left|  j, m=j \right\rangle$) of  $J_{-}$\ ($J_{+}$).
However, the collective atomic operators $J_{-}$ and $J_{+}$ have no simultaneous eigenstates 
which would annullate the right hand side of Eq.~(\ref {sat0}).
Nevertheless, we are going to show that if the number of atoms $N=j/2$
is large enough, then the 
correlation functions in Eq.~(\ref{sat0}) are negligible in a class of 
states called atomic coherent states \cite{ACGT72}. 
These states are 
labeled by a complex parameter 
$\tau=\tan(\beta/2)\exp(-i\phi)$ (for the angles $\beta$ and $\phi$ see Fig.~\ref{sphere}) 
and can be expanded in terms of 
the eigenstates of the operator $J_z$ (Dicke states) as
\be  |\tau \rangle=\sum^j_{m=-j} \left( 
\begin {array}{cc} 2j \\ j+m \end {array} \right)^{1 \over 2} {\tau^{j+m} \over
{(1+|\tau|^2)^j}} |j,m\rangle \label{cohs}. \ee

For large $j$, the atomic coherent states are approximate
eigenstates of the operators $J_{-}$ and  $J_{+}$ \cite{ACGT72,BBH00}.
This statement is understood in the sense that the square of the cosine of 
the angle $\alpha$ between $\left| \tau \right\rangle$ and 
$J_- \left| \tau \right\rangle$
\be
\cos^2 \alpha={{\left| \left\langle \tau \left| J_- \right| \tau \right\rangle \right|^2}
\over
{\left\langle \tau \left| \right. \tau \right\rangle
\left\langle \tau \left| J_+ J_- \right| \tau \right\rangle}}
\ee
differs from unity by a factor which scales as $(j\tau^2)^{-1}$. 
Thus $\alpha$ becomes negligible in the $j \to \infty$ limit for finite $\tau$ 
\cite {BBH00}. The same 
statement holds for the operator $J_+$, therefore both
correlation functions 
in Eq.~(\ref {sat0}) are indeed negligible in the atomic 
coherent states (\ref{cohs}).

This suggests that the atomic
coherent states are rather stable against the decoherence induced by
the photon modes, i.e., they can serve as a model of 
classical-like
macroscopic quantum states. This result is in analogy with the stability of the oscillator
coherent states obtained in \cite{Z93}.

Two such states, $\left| \tau_1 \right\rangle$ and
$\left| \tau_2 \right\rangle$ can be considered as macroscopically distinct, 
whenever
the distance between the parameters $\tau_1$ and $\tau_2$ is
sufficiently large on the complex plane. This implies that the
coherent superposition of these states yields an appropriate 
model of the original paradox of Schr\"{o}dinger.

Based on these results, the superpositions 
\begin{equation}
|\Psi _{12}\rangle ={\frac{{|\tau_1\rangle +|\tau_2\rangle }}
{\sqrt{2(1+{\rm{Re}}\; \langle \tau_1|\tau_2\rangle )}}}
\label{cats}
\end{equation}
will be called atomic Schr\"{o}dinger cat states \cite{BCB97,BC99,FCB00},
see Fig.~\ref{sphere}. 
Now we are going to present our results on the decoherence and 
dissipation dynamics of these type of  states.

\section{Time scales}
\label{times}
\begin{figure}[htbp]
\begin{center}
\includegraphics*[bb=60 10 805 520 , width=10cm]{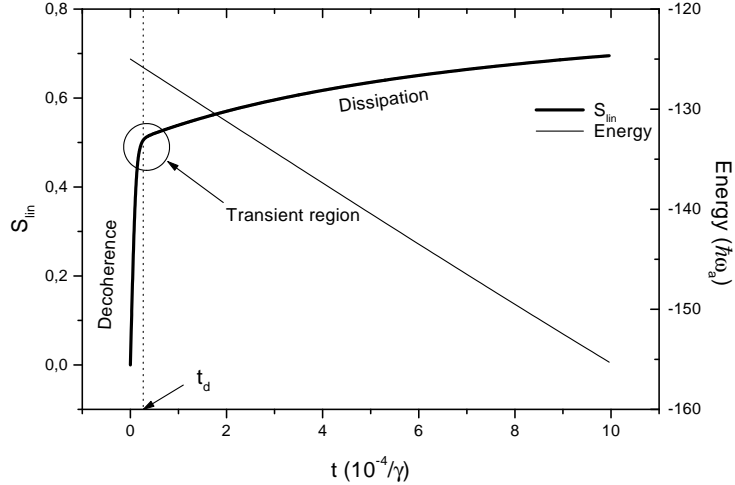}
\end{center}
\caption[Definition of the decoherence time in a system of two-level atoms]
{The two regimes of the time evolution.
(Initially: $\tau_1=\tan\pi/4$, $\tau_2=0$.)
 The number of atoms is $N=500$ and the average number of photons is $\overline{n}=1$, $\td \approx 6 \times 10^{-5}/\gamma$}. 
\label{scales}
\end{figure}

A typical result of the  numerical integration of Eq.~(\ref {mastereq}) is
that the time evolution of the 
states given by Eq.~(\ref{cats}) 
can be characterized by two different time scales, as illustrated by 
Fig.~\ref{scales}, where the linear entropy and the energy of the 
atomic system is plotted versus time. 
As we can see, 
there exists a time instant $\td$ (marked with an arrow in Fig.~\ref{scales}) 
when the character of the physical process changes radically. (It is worth
comparing this figure with Fig.~\ref{dtdef}, which was obtained in the case 
of the Morse system.) 
Initially $S_{\mathrm{lin}}(t)$ increases rapidly 
while 
the dissipated energy of the atoms is just a small  fraction of that part of the energy which 
will  eventually be transferred to the environment.  
On the other hand, for longer times $t\gg \td$  both curves 
change on the same time scale. 
The energy of the atomic system 
decays exponentially as a function
of time allowing for identifying the characteristic time of the dissipation, 
$\tdiss$, with the inverse of the exponent.
(We note that in Fig.~\ref{scales} the plotted time interval is much
shorter than 
$\tdiss$, thus the exponential behavior is not seen.)
More detailed calculations have shown that 
for high temperatures 
the energy and the linear entropy exhibit similar 
exponential behavior in the second 
regime of the time evolution. Their exponents coincide 
with 2-3\% relative error.
This implies that the 
initial stage of the time  
evolution is dominated by decoherence 
while after $\td$ the dissipation determines the dynamics. 
Accordingly, we define the characteristic time of the decoherence -- by the 
same token as we did in Sec.~\ref{dectimesec} --
as the instant when the slope of the curve $S_{\mathrm{lin}}(t)$  
decreases appreciably. We note that 
$\td$ defined in this way is in accordance with the decoherence time defined 
previously in \cite{BC99} for
a specific initial state.

It is remarkable that although a few hundred atoms do not really constitute a
macroscopic
system, the difference of the time scales is obviously seen 
in Fig.~\ref{scales}.
It is generally true that the larger $N$ is, the more naturally and sharply 
the time evolution splits into two regimes. 

Now we turn to the investigation of the dependence of the 
decoherence time $\td$ on the initial conditions.  
\begin{figure}[htbp]
\begin{center}
\includegraphics*[bb=25 40 715 510 , width=10cm]{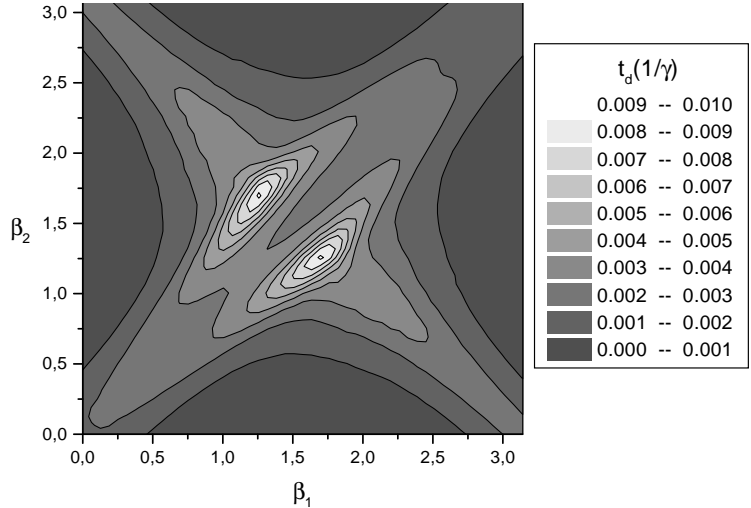}
\end{center}
\caption[Decoherence time as a function of the parameters of the initial
atomic Schr\"{o}dinger cat state]
{The dependence of the characteristic time of the decoherence on
the parameters of the initial Schr\"{o}dinger cat state: $\tau_1=\tan\beta_1/2$, 
$\tau_2=\tan\beta_2/2$. 
The number of atoms is $N=50$ and the average number of photons is $\overline{n}=3$.}
\label{dectimes}
\end{figure}
Fig.~\ref{dectimes}
shows the contour plot of the decoherence time versus the parameters $\beta_1$ 
and $\beta_2$ (see Fig.~\ref{sphere}) of the initial atomic Schr\"{o}dinger cat 
state (\ref{cats}). We have set $\phi_1=\phi_2=0$ for simplicity.
As we can see, the effect of decoherence is remarkably slower when $\beta_1\approx
\beta_2$ which was expected since 
in this case the overlap of the two initial coherent states is not negligible,
so these states can not be considered as ``macroscopically distinct''.
Much more surprising is the fact that cat states which were initially 
symmetric with respect to the 
$(x,y)$ plane
(i.~e.~$\beta_1\approx\pi-\beta_2$ ) also decohere slower 
\cite{BC99,FCB00}, but it is in accordance with the analytical 
estimations of Braun et.al.~\cite{BBH00}. 
In the following sections we shall refer to these states as {\it symmetric}
ones.

\section {The direction of the decoherence}
\label {pointer}
We saw in the previous section that the interplay between decoherence
and dissipation is reflected in the time evolution of the 
superpositions given by
Eq.~(\ref {cats}). In this section we shall focus on 
the direction of the process resulting from the dynamics governed by the master
equation (\ref {mastereq}).

According to Sec.~\ref{EIDsec},
the interaction with a large number of degrees of freedom 
selects naturally 
the so-called pointer basis \cite{Z81} in the Hilbert-space of the 
system subject to decoherence.
This process favors the constituent states of the 
pointer basis in the sense that 
the system  is driven towards a classical statistical mixture of these states. 
Thus, from the present point of view $\rho(\td)$ is the relevant quantity to be examined.

Recalling the analytical results of sec. \ref {robust}, it seems 
plausible to expect 
that the atomic coherent states (\ref {cohs}) will be pointer states.

By introducing the 
density matrix which corresponds to the classical statistical mixture of 
the initial coherent states:
\be
\rho_{\mathrm{cl}}(\tau_1,\tau_2)={1 \over 2} \left( |\tau_1 \rangle 
\langle \tau_1| +
|\tau_2\rangle  \langle \tau_2| \right),
\label{rhocl}
\ee  
the expected scheme of the decoherence reads:
\be
|\Psi _{12}\rangle \langle \Psi _{12}| \rightarrow \rho_{\mathrm{cl}}(\tau_1,\tau_2).
\label{scheme}
\ee
We shall refer to $\rho_{\mathrm{cl}}$ as the classical density matrix. 

\begin{figure}[htbp]
\begin{center}
\includegraphics*[bb=25 10 795 520 , width=10cm]{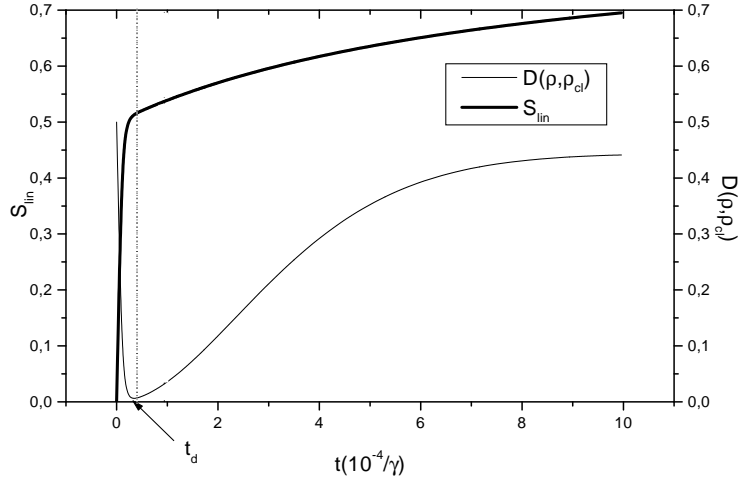}
\end{center}
\caption[A new  measure of decoherence in the system of two-level atoms]
{The linear entropy $S_{{\mathrm{lin}}}$ and the distance $D$ between $\rho$ and
$\rho_{\mathrm{cl}}$ (defined by Eq. \protect{(\ref {distance}})) in the case of a rapidly decohering 
Schr\"{o}dinger cat state ($\tau_1=\tan\pi/4$, $\tau_2=0$). 
The number of atoms is $N=500$ and the average number of photons is $\overline{n}=1$.}
\label{fastdec}
\end{figure}
The distance between the 
actual density matrix $\rho(t)$
and $\rho_{\mathrm{cl}}$, defined with
\be
D \left(\rho(t),\rho_{\mathrm{cl}} \right)={\rm Tr}\left[ \left( \rho(t)-\rho_{\mathrm{cl}} \right)^2 \right],
\label{distance}
\ee
is always decreasing fast.
Except for the case of slowly decohering cat states which will be discussed below,  
$D \left(\rho(t),\rho_{\mathrm{cl}} \right)$ reaches its minimal value 
at the {\it decoherence time},  see  Fig.~\ref {fastdec}.
This minimal value is very close to zero implying that the 
density matrix of the system at this instant is nearly the same as the classical density
matrix (\ref{rhocl}).
This fact 
justifies the definition of the characteristic time of the
decoherence in section \ref {times}, 
and it is in excellent agreement with the decoherence
scheme (\ref{scheme}). 

Due to the exceptionally slow decoherence, we have to modify 
this picture if the initial  state is a symmetric superposition. 
In this case the decoherence time is so long that the atomic coherent 
constituents of the initial state are also appreciably affected 
by the time evolution until the decoherence time, $t_d$.  
The state of the atomic system at  $t_d$ will be a mixture, 
which is the same as if the system had started from  
$\rho_{\mathrm{cl}}$ at t=0. In other words the evolution 
follows the modified scheme: 
\be
|\Psi _{12}\rangle \langle \Psi _{12}| \rightarrow 
\tilde{\rho}_{\mathrm{cl}}(\tau_1, \tau_2, t)
\label{scheme2}
\ee
where the time dependent classical density matrix
$\tilde{\rho}_{\mathrm{cl}}(\tau_1, \tau_2, t)$ is the one 
which would evolve from the statistical mixture
(\ref{rhocl}) 
$\rho_{\mathrm{cl}}(\tau_1,\tau_2)=\tilde{\rho}_{\mathrm{cl}}(\tau_1, \tau_2, t=0)$ 
according to the same master equation (\ref {mastereq})
as the actual atomic density matrix. 
The distance between the time dependent classical density matrix, $\tilde{\rho}_{\mathrm{cl}}$ 
and  $\rho(t)$ becomes negligible at $\td$, and asymptotically reaches zero for long times
in the case of all the initial conditions. 

\section {Wigner functions of four component Schr\"{o}dinger cat states}
\label {wigner}
\begin{figure}[htbp]
\begin{center}
\includegraphics*[bb=35 75 270 275 , width=8cm]{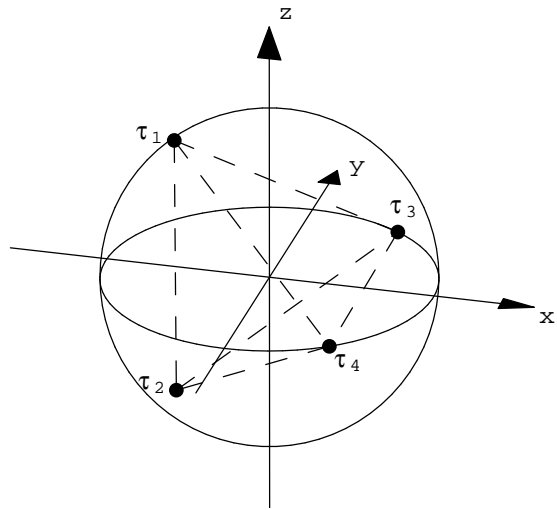}
\end{center}
\caption[Phase space scheme of a 4 component atomic cat state]
{Phase space scheme of the 4 component cat state. The atomic coherent states 
constituting the superposition (\ref{4cats}) are represented by the points labeled by
$\tau_1,\ \ldots,\ \tau_4$.
They are arranged to form the vertices of a tetrahedron
as shown.}
\label{tetrahedron}
\end{figure}

The results of the previous section have shown that both 
the characteristic time and the direction of the decoherence
strongly depend on the initial conditions. 
Now we illustrate this fact by tracking
the decoherence of the superposition of four atomic coherent states
\be
\left| \Psi_{1234} \right \rangle = {\frac{{|\tau_1\rangle +|\tau_2\rangle+|\tau_3\rangle+|\tau_4\rangle }}
{\sqrt{2(2+{\rm{Re}}\; \sum_{i>k}\langle \tau_i|\tau_k\rangle )}}}. 
\label{4cats}
\ee
Since four points on the surface of a sphere are not distinguished with respect to each other
if and only if they are the vertices of a regular 
tetrahedron inscribed in the sphere, we set the
components of $\left| \Psi_{1234} \right \rangle$ according to this pattern.
On the other hand,
the $z$ axis is distinguished in the present model because of the form of the
Hamilton operator (\ref{model_ham}), therefore we orient the tetrahedron 
with one edge 
parallel to the $z$ axis and the opposite edge parallel to the $y$ axis, 
see Fig.~\ref{tetrahedron}.
Although we have in principle 
two 
substantially different ways 
of considering the state represented by Fig.~\ref{tetrahedron} as the superposition of
{\it two} atomic Schr\"odinger-cat states,
according to the results of the previous section
one expects that the environment naturally selects one of these possibilities via the 
different time evolutions:
the quantum coherence between the {components} of the symmetric pair
 $\left| \Psi_{12} \right \rangle \propto \left| \tau_{1} \right \rangle + \left| \tau_{2} \right \rangle$    
disappears slowly, 
while all the other pairs are rapidly decohering superpositions.

We are going to
visualize the decoherence process of $\left| \Psi_{1234} \right \rangle$ 
by the aid of the spherical Wigner function. 
It is a real function over the unit sphere 
(which is the appropriate phase space in the present case)
being in a linear one-to-one correspondence with the density matrix of 
the atomic system, see Sec.~\ref{QPD}.
For previous applications of this 
function see \cite{BCB97,BC99,DASCH94,CB96,BM98,CFW99}.

\begin{figure}[htbp]
\begin{center}
\includegraphics*[width=15cm, height=17cm]{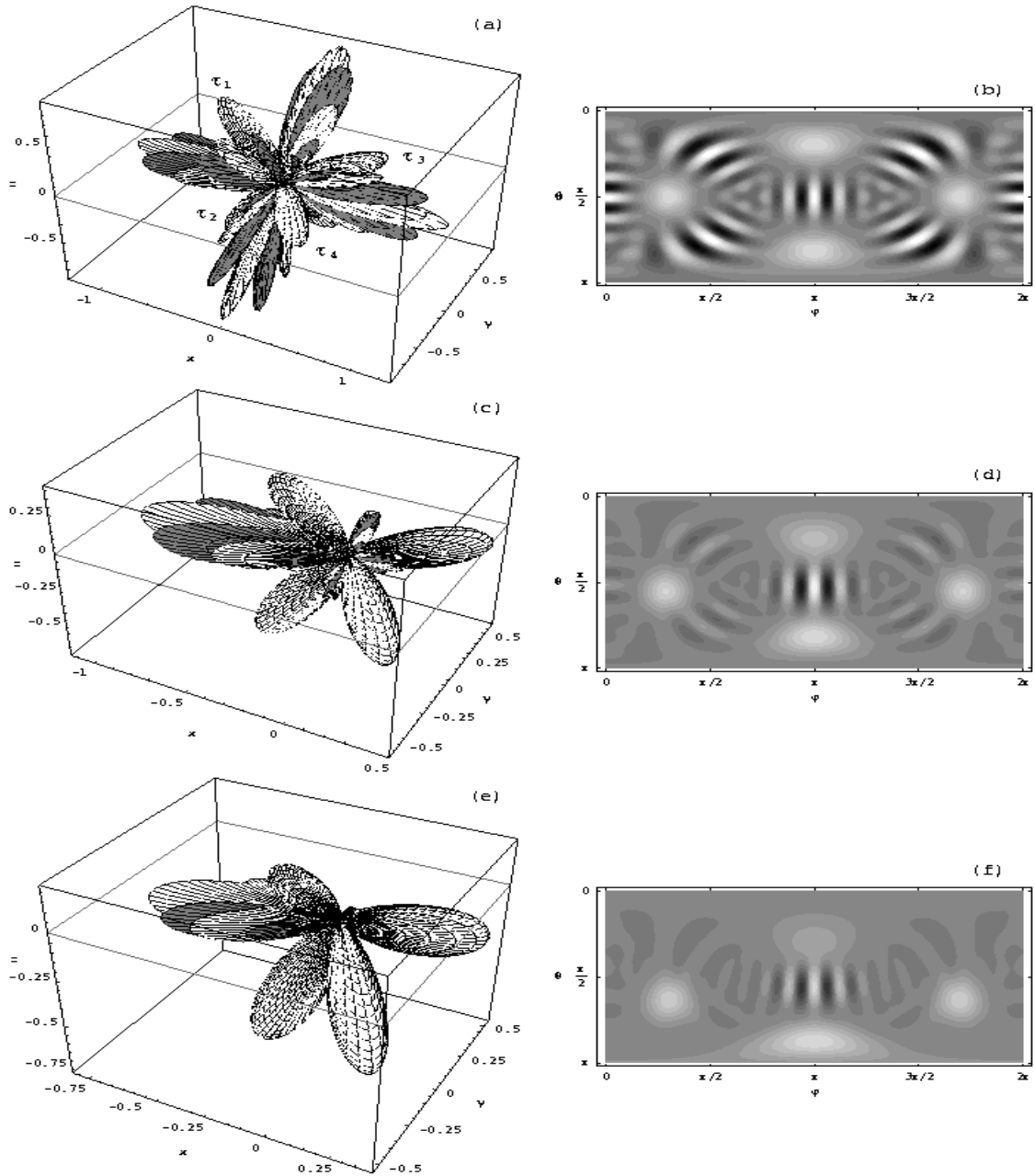}
\end{center}
\caption[Wigner view of the decoherence of the 4 component atomic cat state]
{
Wigner view of the decoherence of the 4 component cat state (\ref{4cats}).
We plot the spherical Wigner function (\ref{wfunct}) both as a polar
plot [a), c), e)] and as a contour plot [b), d), f)]. A polar plot is obtained
by measuring the absolute value of the function in the corresponding
direction, 
and the resulting surface is shown in light where the 
values of the Wigner function are positive, and
in dark where they are negative. Similarly, light shades of the contour
plot correspond to positive function values. 
Plots a) and  b) show the initial state and in a)
the lobes corresponding
to the initial 
coherent constituents are also labeled according to  Fig.~\ref{tetrahedron}.
Plots c), d) and e), f) show the spherical Wigner function
at $t=0.015/\gamma$ and $t=0.04/\gamma$, respectively.
}
\label{wigfig}
\end{figure}

The Wigner function  given by Eq.~(\ref{wfunct}) suggestively
maps the time evolution of the state (\ref {4cats}) 
onto the unit sphere, as shown 
in Fig.~\ref{wigfig}. We plot the Wigner functions
of the atomic system at three time instants, both as a polar 
plot and as a contour plot. Dark shades mean negative,
light shades mean positive function values.
The four positive lobes, pointing from the center to the vertices of the tetrahedron shown 
in Fig.~\ref{tetrahedron},  
correspond  to the four atomic coherent states in (\ref {4cats}). 
Due to the dissipation all these lobes will move slowly downwards.
The initial interference pattern (Figs.~\ref{wigfig} a) and b)) 
has the regularity of the 
tetrahedron, there are equally pronounced oscillations along 
all the edges, representing the quantum 
coherence between the coherent states.

Figs.~\ref{wigfig} c) and d) depict the situation after a time  
which is short in the sense that the 
the shapes of the lobes of the coherent states are not appreciably affected 
(no dissipation),  but the interference 
is already negligible between them, except for the single pair 
along the vertical edge of 
the tetrahedron. As it is seen from Fig.~\ref{tetrahedron} 
this is the pair which 
represented initially the {\it symmetric} atomic Schr\"odinger-cat state
$\propto\left| \tau_{1} \right \rangle + \left| \tau_{2} 
\right \rangle$ in (\ref {4cats}).
The coherence between 
the components of this pair of states  is nearly unaffected as 
shown by the strong oscillations. 

A qualitatively different stage of the time evolution 
is shown in Figs.~\ref{wigfig} e) and f) at a later time. 
The coherent constituents are already affected by the dissipation, 
the uppermost one rather strongly,
but the quantum coherence between the 
components of the symmetric pair is still present. 
On the contrary, the interference between all the other components 
has already disappeared.

In view of the present results, if the initial state of the
atomic system is a superposition of coherent states so that there are
symmetric pairs of coherent states in the expansion, 
then the coherence between the components of these
symmetric pairs will survive much longer than between any other terms.

\section{Conclusions}

We have investigated 
the decoherence of superpositions of macroscopically distinct quantum states in a
system of two-level atoms embedded in the environment of thermal photon modes.
Utilizing the Schmidt decomposition and the linear entropy, we have shown that atomic
coherent states are robust against decoherence, both at zero
and non-zero temperatures. This result is in analogy with the harmonic oscillator
case and justifies the definition of atomic 
Schr\"{o}dinger cat states as superpositions of atomic coherent states.

By solving the master equation (\ref{mastereq}) we have identified 
two different regimes of the time evolution with the help of the linear entropy. The first one is
dominated by decoherence while the second one is governed by dissipation.
Based on several computational runs focusing on the characteristic times it was found 
that $\td$ decreases much faster than $\tdiss$ as the function of the number 
of atoms, $N$. Consequently $\td$ becomes many orders of magnitude smaller than the 
characteristic time of dissipation for macroscopic samples, and 
even for e.g. $N=500$ atoms and an average photon number of 
$\overline{n} = 1$ the ratio $\tdiss/\td$ is around a few hundred depending 
on the initial conditions. 
However, there are very important exceptional cases, called slow decoherence, 
when the atomic coherent 
states constituting the initial atomic Schr\"{o}dinger cat state are symmetric with
respect to the equator of the Bloch sphere. 

Using a new measure $D$, we have shown that at the characteristic time of decoherence the  system is always
very close to the state described by the time dependent classical density matrix. Apart from
the exceptional case of 
slow decoherence, the coherent states appearing in (\ref {cats}) 
are approximate pointer states. When due to its symmetry the initial cat state is a 
long-lived superposition, also its constituent coherent states have time to transform 
into mixtures until $\td$. We have given a modified scheme of decoherence which is 
valid also for slow decoherence. 

We have demonstrated the important difference between rapid and slow  decoherence by
tracking the time evolution of a four component superposition with the help
of the spherical Wigner function. The initial interference pattern having the
symmetry of a tetrahedron rapidly disappears except for the single slowly decohering
pair.


\newpage
\chapter[Preparing decoherence-free states]
{Preparation of decoherence-free, \\subradiant states in a cavity}
\label{subprepchap}
\emptypage
Decoherence is a difficulty to overcome in the context of QC.
Quantum error correction codes \cite{S95} offer a possibility for this
purpose. A somewhat different approach relies on a specific 
decoherence-free subspace (DFS) 
\cite{PHBK99,LCW98,BBTK00}. This idea is essentially a passive error
correction scheme: quantum operations are restricted to the DFS in 
which all the quantum superpositions are much less fragile 
than in subspaces strongly coupled to the unavoidably present environment.

Subradiant states of a system of two-level atoms 
\cite{D54,SRAD,GH82,P85,DB96} has
recently gained wide attention because of their exceptionally
slow decoherence.  
This stability of quantum superpositions inside the subradiant
subspaces originates from the low probability of photon emission, 
which means very weak interaction between the
atoms and their environment.    
Hence the subradiant states span a
DFS in the atomic Hilbert-space.
We note that there are proposals that intend to perform QC within 
this DFS \cite{BBTK00}.

Here we propose a scheme that can be used to prepare subradiant states 
in a cavity. Our method is based on second order perturbation
theory but the exact results verify the
validity of the perturbative approach \cite{FBC02b}.
We also investigate to what extent our scheme
is independent of the state of the cavity field.
The analysis of the conditions shows that this scheme is 
feasible with present day techniques achieved in atom cavity 
interaction experiments \cite{WVHW99,BHD96,H97,R99}.


\section{Description of the system}
We investigate a system of $N$ identical two-level atoms in a single mode
cavity. Each individual atom is equivalent to a spin-$1/2$ system, and
the whole atomic ensemble can be described by the aid of 
collective atomic operators  $J_+$, $J_-$ and $J_z$
obeying the same algebra as the usual angular momentum 
operators\cite{D54}.
We consider the following model Hamiltonian:
\begin{equation}
H=H_0+V=\hbar \omega_{\mathrm{a}}J_{z}+ \hbar \omega _{c}a^{\dagger
}a+ \hbar g \left( a^{\dagger }J_{-}+aJ_{+} \right),
\label{ModHam}
\end{equation}
where $a$ and $a^\dagger$ are the annihilation and creation operators 
of the cavity mode, $\omega_{\mathrm{a}}$ is the transition frequency 
between the two atomic 
energy levels,  $\omega_c$ denotes the frequency of the cavity 
mode, different from 
$\omega_{\mathrm{a}}$, and $g$ is the 
coupling constant. 
We note that the Hamiltonian (\ref{ModHam}) 
is written in the framework of Dicke's theory, i.e.,
with the assumption 
that all the atoms are subjected to the same field, which is a good 
approximation when the size of the atomic sample is small compared to
the wavelength of the cavity mode. As discussed later in detail, 
there are experimental situations where this
requirement is fulfilled.
Our proposed scheme for preparing subradiant states
involves a detuned cavity. We shall assume that the detuning is much larger
than the resonant Rabi frequency:
\begin{equation}
\omega_c-\omega_a= \Delta \gg g.
\end{equation}

Now any state  of the atomic system and the cavity field
can be expanded as a linear combination of the eigenstates of $H_0$.  
These are tensorial products
of collective atomic states and number states of the field: 
$|j,m,\lambda\rangle \otimes |n\rangle$, where the indices 
$j,m$ and $\lambda$ label the 
atomic state (also called Dicke state \cite{ACGT72}), 
while $n$ refers to the $n$-th Fock state of the
mode. 
\begin{figure}[htbp]
\begin{center}
\includegraphics*[bb=0 0 836 256 , width=10cm]{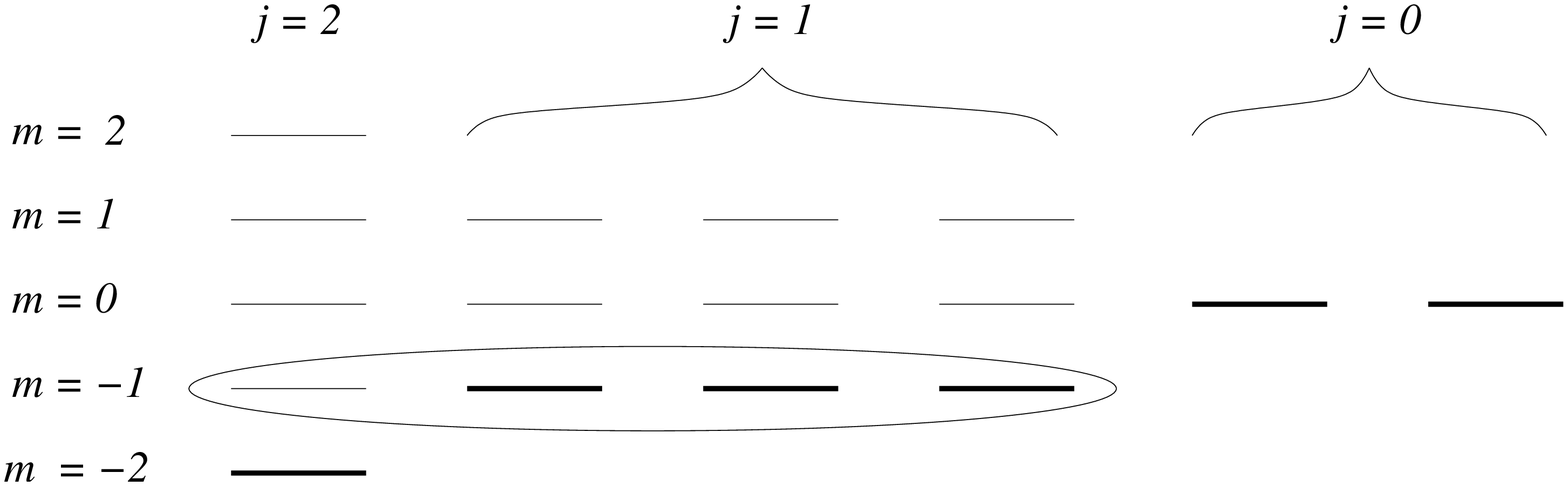}
\end{center}
\caption[Dicke ladders]
{Dicke ladders for $N=4$ atoms. Each 
line corresponds to a collective atomic state, the subradiant states
are emphasized by thick lines. The circle denotes the 4-fold degenerate
first excited level. The first column corresponds to the completely 
symmetric subspace.
}
\label{Dladder}
\end{figure}
The quantum number $j$ corresponds to the eigenvalues of the
operator $J^2=J_z^2+(J_+J_-+J_-J_+)/2$. This index is in one-to-one
correspondence with the Young diagram
\cite {H62} that describes the permutation 
symmetry of the state. The possible values of $j$ are $N/2, N/2-1,\dots $,
the smallest value being  $0$ if $N$ is even and $1/2$ if $N$ is odd.
The index $m$ of the $|j,m,\lambda\rangle$
Dicke state labels the eigenstates of the collective atomic operator 
$J_z$, that is essentially proportional to the 
energy of the atomic subsystem.
This is the index that is decreased (increased) by one under 
the action of the operator $J_-$ ($J_+$):
\be
J_-|j,m,\lambda\rangle=\sqrt{j(j+1)-m(m-1)}|j,m-1,\lambda\rangle,
\label{jpaction}
\ee
including the case when $m=-j$, when the result is the zero vector.
The states with $m=-j$ are the lowest ones of the Dicke ladders
\cite{D54}, they are called subradiant, because they have 
no dipole coupling to other lower lying states, see Fig.~\ref{Dladder}
for the case of $N=4$.    
Finally the index $\lambda$ distinguishes states with the
same $j$ and $m$. For more details see 
Refs. \cite{D54,FCB01a,H62,ACGT72,CP87,KSS92,BC99}.

Besides the collective atomic states $|j,m,\lambda\rangle$, we shall 
also use the natural basis that assigns a well defined state to each  
individual atom. These vectors will be labeled by a string of $0$-s and $1$-s
corresponding to the ground and excited sates, respectively.
E. g., the ground state of the atomic subsystem
is written in this basis as $|\stackrel{1}{0} \stackrel{2}{0}
\ldots \stackrel{N}{0}\rangle$; this state (as well as the fully
excited one) is also 
an element of the Dicke basis, $|00\ldots 0\rangle=
|j=N/2,m=-N/2,\lambda=1\rangle$.


\section{Preparation of subradiant states}
\label{subradprepsec}
The form of the Hamiltonian (\ref{ModHam}) implies that
the time evolution of the system shall exhibit two time 
scales: The first characteristic time is due to the self-Hamiltonian $H_0$ and
is approximately $2\pi/\omega_a$ (or $2\pi/\omega_c$) and
the second is proportional to $2\pi/g$. Generally $g \ll 
\omega_a \approx \omega_c $ 
and the faster process induced by $H_0$ can be eliminated by going into
an interaction picture. However, if the frequency difference $\Delta$ is large
enough, then the energy transfer between two adjacent eigenstates of $H_0$, 
differing in only one photon number, becomes negligible. 
This means that the amplitude of
the corresponding collective Rabi oscillations will be very small, that is,
the  process 
on the second time scale will be unnoticeable and even slower mechanisms will 
become apparent. In this section we are going to show that this situation,
which is similar to the one considered in Refs.~\cite{SM99}
and \cite{ZG00}, gives rise to the preparation of subradiant states.

Hereafter we shall focus on the solution of the Schr\"{o}dinger equation
in the case when just a single atom is excited at $t=0$.
This initial state can be prepared by  
starting from the state $|00\ldots0\rangle$,
and exciting one well defined control atom.
This excitation can be achieved via a third much higher lying level,
so that the wavelength of the addressing pulse allows to focus it on the
desired target atom \cite{N99}.
For the sake of simplicity we always consider the control atom as being the 
first, hence the initial state will be written as 
\be
|\phi(0)\rangle=|100\ldots 0 \rangle \otimes|n-1\rangle.
\label{veryfirst}
\ee 

In order to find the complete analytical solution of the 
Schr\"{o}dinger equation
induced by the Hamiltonian (\ref{ModHam}), in principle one should 
calculate all the eigenvalues and the corresponding eigenstates of $H$. 
Although this problem can be solved analytically \cite{TC68}, 
more insight is given by a simple perturbative
approach. 
The exact nonperturbative numerical solution of the Schr\"{o}dinger equation
verifies that results obtained via perturbation theory yield  excellent  
approximations, see section \ref{compsec}.  

The state 
\begin{eqnarray} 
|1\rangle \equiv \left({{1}\over{\sqrt{N}}}
\sum_{k=1}^N |0\ldots 0\stackrel{k}{1}0 
\ldots 0\rangle  \right)\otimes|n-1\rangle= \nonumber \\
= |j=N/2,m=-N/2+1,\lambda=1\rangle\otimes|n-1\rangle ,
\label{symmetric}
\end{eqnarray}
which is in the completely symmetric subspace, and the 
subradiant states: 
\begin{equation} 
|i\rangle \equiv |j=N/2-1,m=-N/2+1,\lambda=i-1\rangle \otimes|n-1\rangle
\label{kn}
\end{equation}
 with $i=2,3...N$,  
have the same unperturbed energy, they span the $N$-fold degenerate 
eigensubspace 
of $H_0$ corresponding to the eigenvalue 
$E^0(n)=\hbar(n \omega_c-N\omega_a/2)-\hbar\Delta$. For the case of
$N=4$, the atomic part
of the states $|i\rangle$ is denoted by a circle in Fig.~\ref{Dladder}. 

It can be seen that first order degenerate perturbation theory is 
not giving any correction to the energy, because all the matrix elements of
$V$ between the states above vanish: The action 
of $V$ on vectors $|j,m,\lambda\rangle\otimes|n-1\rangle$ 
gives a linear combination of $|j,m-1,\lambda\rangle\otimes|n\rangle$ and 
$|j,m+1,\lambda\rangle\otimes|n-2\rangle$ that are orthogonal
to the states (\ref{symmetric}) and (\ref{kn}).
In order to obtain nonzero energy corrections we have to perform a 
second order degenerate perturbation calculation\cite{LL65},  
and find the eigenvalues of the matrix: 
\be
\sum_m {{\langle i|V|m\rangle\langle m|V|k\rangle}
\over{E^0(n)-E_m^0}},
\label{secular2}
\ee
where the sum runs over all eigenstates of $H_0$ with eigenvalue
$E_m^0\neq E^0(n)$.
The only nonvanishing energy corrections in second order are the following:
\begin{eqnarray}
\delta E_1&=&\hbar {{g^2}\over{\Delta}}(Nn-2N-2n+2),\nonumber \\ 
\delta E_i&=&\delta E_1+\hbar N{{g^2}\over{\Delta}}, \ \  i=2,3\ldots N.
\label{corrections}
\end{eqnarray}

At this point we can formulate the requirements that assure the
validity of the perturbation theory: the magnitude of 
$\delta E_1$ and $\delta E_i$ must be much smaller than 
$\hbar |\Delta|$, the minimum of the
difference between $E^0(n)$ and all other unperturbed energy levels. 

The most important consequence of Eqs.~(\ref{corrections}) 
is that the Bohr frequencies that determine the 
time dependences
of the subradiant and non-subradiant states are different.
 
Now we expand the initial state (\ref{veryfirst}) 
as the linear combination of the fully 
symmetric (non-subradiant)
state $|1\rangle$,
and an appropriate subradiant state:
\begin{eqnarray}
|2\rangle={{1}\over{\sqrt{N(N-1)}}}\Big[(N-1)|100\ldots 0 \rangle- 
\nonumber \\ 
\sum_{k=2}^N |0\ldots 0\stackrel{k}{1}0 
\ldots 0\rangle  \Big]
\otimes|n-1\rangle.
\label{subradiant}
\end{eqnarray}
By assigning the symbol $|2\rangle$ 
to the state in Eq.~(\ref{subradiant}), we have utilized the freedom
of choosing a basis in the subradiant subspace. 
Now the initial state reads 
\be
|\phi(0)\rangle={{1}\over{\sqrt{N}}}|1\rangle+\sqrt{{N-1}\over{N}}|2\rangle.
\label{expansion}
\ee 
By the aid of this expansion and using the Bohr frequencies 
resulting from (\ref{corrections}), it is easy to calculate the 
time evolution of the state (\ref{expansion}).
Discarding an overall phase factor, this time dependent state has the form
\be
|\phi(t)\rangle={{1}\over{\sqrt{N}}}\exp\left(i N {{g^2}\over{\Delta}} 
t \right)|1\rangle
+\sqrt{{N-1}\over{N}}|2\rangle,
\label{expevol}
\ee
or, on using Eqs.~(\ref{symmetric}) and  (\ref {subradiant}):
\begin{eqnarray}
|\phi(t)\rangle= 
\Big[ \left( N \cos(\alpha t)-i(N-2)\sin(\alpha t)\right)|100\ldots 0 \rangle
\nonumber \\
+2 i\sin(\alpha t) \sum_{k=2} |0\ldots 010 \ldots 0\rangle \Big] 
\otimes|n-1\rangle /N.
\label{updownevol}
\end {eqnarray}
Here we introduced the parameter
\be 
\alpha={{N g^2}\over{2\Delta}},
\label{gamma}
\ee 
which is {\it independent} of $n$.
Because of this latter fact, from now on 
the state of the cavity field will be omitted in the notation.
We also note that the characteristic time 
of the time evolution, ${{2\pi}/{\alpha}}$, is much longer
than that of the free evolution due to $H_0$, being the consequence 
of the fact that the evolution described in Eq. 
(\ref{expevol}) is induced by a weak, nonresonant
interaction.

Eq.~(\ref {updownevol})  reveals that in $|\phi(t)\rangle$ the weight of the 
state  $|100\ldots 0 \rangle$ and those of the states with the 
first atom unexcited 
changes during the course of time. As we can see, the moduli 
of the corresponding coefficients in Eq.~(\ref{updownevol}) are 
$$
 {{\sqrt{N^2 \cos^2(\alpha t)+(N-2)^2\sin^2(\alpha t)}}\over{N}}
\ \ 
{\rm and}
\ \  
{{2|\sin(\alpha t)|}\over{N}},
$$
respectively. Comparing these values to Eq (\ref{subradiant}), it can be 
shown that for arbitrary  $N$, there exists a time instant $t_m$ 
when
\begin{eqnarray}
|\phi(t)\rangle={{1}\over{\sqrt{N(N-1)}}}\Big[(N-1)e^{i \varphi}
|100\ldots 0 \rangle- 
\nonumber \\ 
\sum_{k=2}^N |0\ldots 0\stackrel{k}{1}0 
\ldots 0\rangle  \Big],
\label{correctmod}
\end{eqnarray}
which differs from the subradiant state $|2\rangle$
only in the phase factor $e^{i\varphi}$ of the first term.
Combination of the previous two equations and Eq.~(\ref{subradiant})
yields the following requirement for $t_m$:
\be
{{\sqrt{N^2 \cos^2(\alpha t_m)+(N-2)^2\sin^2(\alpha t_m)}}
\over{\left| 2\sin(\alpha t_m)\right|}}=N-1.
\ee 
We can find a solution of this equation for all $N>1$:
\be
\sin \alpha t_m = \sqrt{N/(4N-4)},
\label{tmod}
\ee
and also obtain 
$\cos \varphi={{N-2}\over{2N-2}}$ in Eq.~(\ref{correctmod}).

Now it is clear that at the time instant given by Eq (\ref{tmod}), 
an appropriate 
rapid change in the phase of the state $|100\ldots 0 \rangle$ 
relatively to all other states $|0\ldots 0\stackrel{k}{1}0 
\ldots 0\rangle$  leads to the subradiant state $|2\rangle$.

On the other hand, Eq.~(\ref{correctmod}) also shows that
the required phase transformation is equivalent to the
elimination of the phase difference $\varphi$ between
the $|1\rangle_c$ excited and $|0\rangle_c$ ground state of 
the control atom.
Therefore we consider the action of a strong laser pulse on the
control atom. In order to obtain precise addressing \cite{N99}, the 
laser is to be tuned in resonance with an allowed transition
$|1\rangle_c \rightarrow |e\rangle_c$, where $|e\rangle_c$
denotes a state of the control atom with much higher energy
than $|1\rangle_c$. E. g., by the appropriate choice of the phase
of the complex Rabi frequencies of two $\pi$ pulses leads to
the phase transformation required to prepare the subradiant 
state $|2\rangle$.   
Additionally, the duration of a Rabi period 
due to the strong, resonant laser pulse
is much shorter than the characteristic time
that governs the time evolution (\ref {expevol}).
We note that the idea of introducing phase transformation 
in a multilevel system by the aid of
short laser pulses has appeared in a somewhat different context
in \cite{ASW01}. 

Now we show that our scheme is independent of the state of the 
cavity field, and write more generally the initial state as: 
\be
|\phi(0)\rangle=|100\ldots 0 \rangle \otimes|\psi(t)\rangle=
|100\ldots 0 \rangle \otimes \sum_n c_n(t) |n\rangle.
\label{nonfock}
\ee
We use the fact that
the interaction Hamiltonian $V$ does not mix states 
with different number of excitation (essentially $n+m$): 
\be
\langle j,m,\lambda| \otimes \langle n| V |n^{\prime}\rangle
\otimes |j,m\pm 1,
\lambda\rangle =0,
\label{nonmix}
\ee
unless $n^{\prime}=n \mp 1$.  
This implies  that the calculations based on second order
perturbation theory can be performed for each $N$-fold degenerate 
energy level of $H_0$ corresponding to different values of $n$. 
After replacing the state $|n-1\rangle$ 
with $|\psi (t)\rangle$ in Eqs.~(\ref {symmetric}) and (\ref {kn}), 
we obtain the following result:
\be
\delta E_i-\delta E_1
={{g^2}\over{\Delta}} N \sum_n |c_n|^2=2\alpha,
\ee
which is therefore also valid in this general case.
Thus we have proven that our scheme does not require
special preparation of the cavity field.


\section{Comparison with experiments}
\label{compsec}
Although the results above can be checked analytically, the scheme is 
based on second order perturbation theory.
Clearly, there are well defined conditions limiting the
validity of the perturbative approach and the applicability of this scheme.
Now we are going to analyze these conditions in comparison with recent
experimental results. As we shall see, present day experimental techniques
allow for the preparation of decoherence-free, subradiant states in the
proposed way.

The time instants $t_{m}$ (and
the whole time evolution) is found to be independent from the cavity field
within the framework of our perturbative approach. 
However, we must require the energy corrections given by 
Eqs.~(\ref{corrections}) to be
much less than $\hbar |\Delta |$, which is the minimal difference between
the eigenvalues of $H_{0}$. 
According to  Eqs.~(\ref{corrections}), 
this can be achieved if the condition  
\begin{equation}
{\frac{{g}}{{\Delta }}}\sqrt{N}\ll 1  \label{relation}
\end{equation}
holds, and the average number of photons in the cavity field,
$\overline{n}$, is not too large. 
By cooling the apparatus and sending a train of absorbing atoms through 
the cavity before the experiment takes place \cite{O01},
average photon numbers $\overline{n} \ll 1$ can be achieved.
Therefore until the relation (\ref{relation}) holds,
the perturbative approach is valid.

In the following we shall compare the requirements of our proposal
with the parameters of the atom-cavity
experiments of Haroche and co-workers. First of all, as discussed in the
review paper \cite{HR85}, the interaction of a number of Rydberg atoms with
a single mode cavity is truly described in the framework of the Dicke model.
In more recent experiments the time evolution of an entangled state of the
cavity field and an atom \cite{BHD96} and also the entangled state of two
atoms \cite{O01} has been found to be in agreement with theoretical
predictions. From our point of view, Ref. \cite{O01} is of special
interest, because our proposal is essentially the generalization (with an
additional phase transformation) of the two-atom experiment described in that
paper to $N$ atoms. In the experiment the atoms emitted by two single atom
sources propagate with different velocities and collide inside the cavity.
It is possible to apply classical RF pulses on the outgoing atoms
independently in order to analyze the final state of the collision process.
In our scheme such RF pulses can perform the desired single qubit phase
operation and consequently prepare a subradiant state.

\begin{figure}[htbp]
\begin{center}
\includegraphics*[bb=80 40 740 520 , width=10cm]{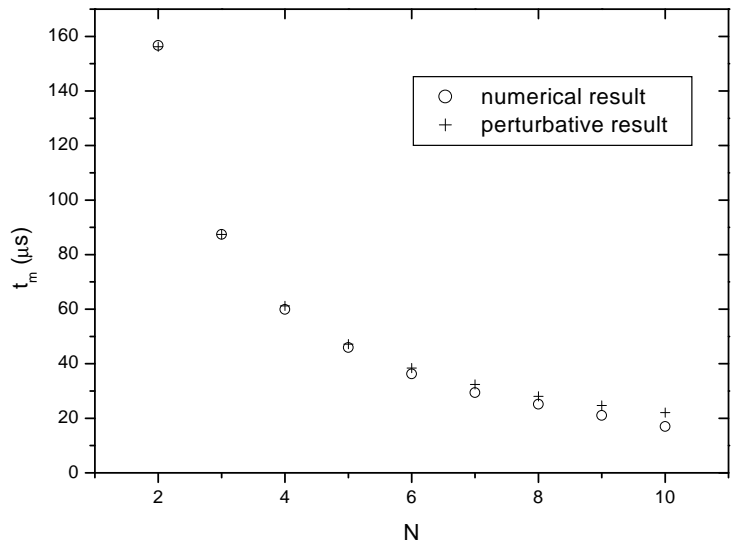}
\end{center}
\caption[The validity of the perturbative approach in the proposal for
preparing decoherence-free states in a cavity]
{The shortest $t_m$ values (introduced by Eq.~(\ref{correctmod})) 
as a function of the number of atoms. Open circles denote
the values calculated numerically while crosses  
correspond to the perturbative approximation, see Eq.  (\ref {tmod}).}
\label{tmodfig}
\end{figure}

In order to investigate the effect of the increasing number 
of atoms on the applicability of our method
we solve numerically the Schr\"{o}dinger
equation induced by the Hamiltonian (\ref{ModHam}) and compare this exact
time evolution with the perturbative one. The realistic parameters \cite
{BHD96} $g/2\pi\approx 24 kHz$ and maximal detunings $\Delta/2\pi \approx
800 kHz$ show that the relation (\ref{relation}) holds as much as for about
hundred atoms. However, it is easy to see that the smallest value of $t_m$
increases as the left hand side of Eq.~(\ref{relation}) decreases, and $t_m$
must clearly be shorter than the interaction time which has the order of
magnitude of $10\mu s$ \cite{BHD96}. To elucidate the trade-off between
this requirement and the validity of the perturbative method, we calculate
the minimal $t_m$ value for different number of atoms.
We use the above realistic value of the coupling constant 
and choose $\Delta/2\pi = 720 kHz$, thus $\Delta/g=30$.
The resulting graph is shown in Fig.~\ref{tmodfig}, where the shortest $t_m$ is
plotted versus the number of atoms, $N$. In the numerical calculation 
$t_m$ is defined as the time instant when the distance (in the sense
of the norm naturally connected to the usual inner product)
between the current state and the state given by Eq.~(\ref {correctmod})
is minimal. 
With the parameters above 
the agreement between the numerical and
the perturbative results is convincing, this minimal distance 
is small, less than $0.04$. The validity of the perturbative approach
can also be seen by comparing the values of $t_m$ obtained 
in the two different ways.

The most interesting property of Fig.~\ref{tmodfig} is that $t_m(N)$ is a 
decreasing function. This means that once our scheme is implemented 
with two atoms, the
interaction time stipulates no additional conditions on the case of more
atoms.
Concerning the numerical values appearing in the figure, the
conclusion is that the condition $\Delta/g=30$ is too strict in the sense 
that $t_m(2)$ is 
longer than the interaction time reported in \cite{O01}. 
The effects predicted by perturbation theory can also be observed 
in the case of a weaker condition.
Indeed, based on the agreement of theoretical and 
experimental results, Ref. \cite {O01} draws the limit of the perturbative 
regime at $\Delta/g>3$.


\section{Conclusions}
We have proposed a method to prepare decoherence-free, subradiant
states of a multiatomic system.  
We also compared the
theoretical requirements with the parameters of existing experimental setups
and found that the proposed scheme is feasible with presently available
experimental techniques.
\chapter*{Summary\markboth{SUMMARY}{}}
\addcontentsline{toc}{chapter}{Summary}
\emptypage
The phenomenon that is called decoherence and the
correspondence with classical mechanics 
is a fundamental question in the quantum theory since the 1930s.
Nowadays, due to the rapid development of experimental techniques,
it is possible to investigate the mechanisms of decoherence also in 
laboratories. Controlled number of ions in a trap \cite{MKT00,N99}, 
single Rydberg atoms traversing a cavity \cite{O01}
or EPR pairs of polarized photons \cite{K99} can shed new light on long-lived
theoretical questions. Besides this principal importance, state of the art
experiments in this field offer possibilities for practical applications
as well. Quantum cryptography protocols have already been implemented, 
and there is a hope also for quantum computation to 
turn into reality. All these facts underline
the importance of understanding the role of highly nonclassical quantum
states and the process of decoherence in the field of quantum optics.
   
In this work the focus was on concrete quantum systems, and the
related theoretical models were  based on the approach of the environment
induced decoherence. That is, we assumed that the reason for the emergence
of classical properties in a quantum system is the interaction with the
environment. An overview of this concept was given in Chap.~\ref{introchap},
and Chap.~\ref{methodchap} dealt with the mathematical methods that 
are used to treat open quantum systems. The most important properties
of the Wigner functions that are relevant from our point of view have also
been given, because these functions are found to visualize decoherence 
in an instructive way. In Part II of the thesis we presented 
our own results that are summarized as follows:  

\subsubsection*{Molecular wave packets in an anharmonic potential} 
\vspace {-6  pt}
The realistic vibrational potential energy of a diatomic molecule 
in a  given electronic state can be approximated by the Morse potential.
First we investigated the time evolution of wave packets in this anharmonic 
potential without the influence of the environment.
For small oscillations, the behavior of the wave packets are similar
to the harmonic case, but when anharmonicity plays an important role, 
peculiar quantum effects can be observed. The Wigner function of the
system shows that in this case there are certain stages of the time
evolution when the vibrational state of the molecule can be considered
as a highly nonclassical Schr\"odinger-cat state: 
it is the superposition of two other states that are well 
localized in the phase space.

\subsubsection*{Decoherence of molecular wave packets} 
\vspace {-6  pt}
The highly nonclassical Schr\"odinger-cat states that spontaneously
form in the Morse potential provide the motivation for the study of 
the decoherence in a diatomic molecule. 
We introduced a model where the environment is represented by a set of 
harmonic oscillators and took into account that the rate of an
environment induced transition depends on the involved Bohr frequencies,
which are different for different transitions. This model led to a master
equation, the final, steady-state solution of which represents thermal 
equilibrium with the environment. On the time scale of dissipation 
decoherence is a very fast process, and the time instant
when decoherence dominated time evolution turns into dissipation dominated
dynamics, naturally defines the characteristic time of the decoherence.
The behavior of the entropy of the molecular system reflects the separation of
the two time scales and allows us to determine the decoherence time for
different initial wave packets. We found that keeping all other parameters
fixed, decoherence becomes faster by increasing the diameter of the 
a phase space orbit corresponding to the wave packet. 
It has also been demonstrated that 
decoherence follows a general scheme in this case, it drives the molecule
into the classical mixture of states that are localized along the 
corresponding classical phase space orbit. This scheme is valid not
only for the case of wave packets as initial states, but also for energy
eigenstates. 
  
\subsubsection*{Two-level atoms and decoherence} 
\vspace {-6  pt}
A system of two-level atoms provides a model in which the 
microscopic $\rightarrow$ macroscopic (mesoscopic) transition is 
straightforward: it means the simple increase of the number of the atoms
in the ensemble. In this system, starting from superposition 
of atomic coherent states as initial states, we have shown 
that the decoherence
time can be determined in a way similar to the case of a diatomic
molecule: The behavior of the linear entropy changes its character
around the decoherence time. The uncertainty of this
operational definition decreases when we increase the number of the atoms.
We have found that certain superpositions of atomic coherent states,
the so-called symmetric Schr\"odinger-cat states, exhibit exceptionally
slow decoherence. We introduced a decoherence scheme that is able to
describe the time evolution of these symmetric states as well.  

\subsubsection*{Preparing decoherence-free states} 
\vspace {-6  pt}
The presence of a cavity around two-level atoms modifies the mode
structure of the electromagnetic field surrounding the atoms,
and the spontaneous emission rate can be different from that of 
in free space. The system consisting of the cavity field and the atoms
is coupled to the environment also in this case. 
In order to avoid the decoherence of the atomic state, we introduced a 
method that can prepare subradiant states in a cavity. 
These states have recently gained wide attention, because the interaction 
of the cavity field and the atoms in a subradiant state is negligible,
therefore cavity losses do not lead to the decoherence of the atomic state.
Therefore the subradiant states span a decoherence-free subspace,
which can be important from the viewpoint of quantum computation. 
Our method for preparing decoherence-free atomic states is based 
on the natural time evolution of the atomic system in the cavity,
and requires the individual manipulation of an initially chosen
control atom. The analysis of the experimental 
requirements of our scheme shows that it can be implemented 
with present day cavity QED setups.

\chapter*{Acknowledgements \markboth{Acknowledgements}{}}
\addcontentsline{toc}{chapter}{Acknowledgements}
\emptypage
First of all I would like to express my deepest gratitude to 
Dr.~Mihály Benedict, my supervisor. 
He, as a master, gave me an excellent introduction to
quantum physics, and our common research was always full of excitement and
pleasure. His wisdom means an indispensable help, when a young
physicist faces difficult problems in various aspects of life and work.

I have been enjoying the stimulating atmosphere of the Department of
Theoretical Physics for several years. It is a pleasure for me to
thank all the members of the Department for their help and encouragement 
throughout the years. I enjoyed and will not forget 
the lectures of Dr.~Iván Gyémánt and Dr.~László Fehér. 
I am grateful to Dr.~Iván Gyémánt, the head of the Department,
for his continuous support.
Frequent friendly discussions with Dr.~Ferenc Bartha and 
Dr.~Ferenc Bogár revealed me several important topics related to the 
physics of molecules. Dr.~Attila Czirják has always been listening to
my ideas and his kind criticism was very helpful.
I enjoyed the  discussions with my friend Balázs Molnár on physics  
as well as on other important subjects, which always helped me
to see the happier side of the world. 

I owe my parents and grandparents my warmest thanks for loving, trusting, 
supporting and encouraging me since I was born.      
I am glad to be able to thank Anikó for her love and patience.
She and my brother Gergely have outstanding sense for identifying
the really important things in life, and fortunately they never
forget to warn me when I miss one of these orientation points. 

The scientific projects I participated were financially supported by 
the Hungarian Scientific Research Fund (OTKA) under contracts
Nos.~T32920, D38267, and by the Hungarian Ministry of Education under contract
No.~FKFP 099/2001.

\newpage

\end{spacing}
\begin{spacing}{1.0}

\end{spacing}

\addcontentsline{toc}{chapter}{Bibliography}

\begin{thebibliography}{100}
\expandafter\ifx\csname url\endcsname\relax
  \def\url#1{\texttt{#1}}\fi
\expandafter\ifx\csname urlprefix\endcsname\relax\def\urlprefix{URL }\fi

\bibitem{SCH35a}
\textsc{E.~Schr\"{o}dinger}, Naturwissenschaften \textbf{23}, 807 (1935).

\bibitem{SCH35b}
\textsc{E.~Schr\"{o}dinger}, Naturwissenschaften \textbf{23}, 823 (1935).

\bibitem{H00}
\textsc{O.~Hadjar, P.~F\"{o}ldi, R.~Hoekstra et~al.}, Phys. Rev. Lett.
  \textbf{84}, 4076 (2000).

\bibitem{A99}
\textsc{M.~Arndt, O.~Nairz, J.~Voss-Andreae et~al.}, Nature \textbf{401}, 680
  (1999).

\bibitem{NC00}
\textsc{M.~A. Nielsen and I.~L. Chuang}, \emph{Quantum computation and quantum
  information} (Cambridge Univ. Press, Cambridge, 2000).

\bibitem{FBC98}
\textsc{P.~F\"{o}ldi, M.~G. Benedict and A.~Czirj\'{a}k}, Acta. Phys. Slov.
  \textbf{48}, 335 (1998).

\bibitem{FCB00}
\textsc{P.~F\"{o}ldi, A.~Czirj\'{a}k and M.~G. Benedict}, Acta. Phys. Slov.
  \textbf{50}, 285 (2000).

\bibitem{FCB01a}
\textsc{P.~F\"{o}ldi, A.~Czirj\'{a}k and M.~G. Benedict}, Phys. Rev. A
  \textbf{63}, 033807 (2001).

\bibitem{FBC01}
\textsc{P.~F\"{o}ldi, M.~G. Benedict and A.~Czirj\'{a}k}, Fortschr. Phys.
  \textbf{49}, 961 (2001).

\bibitem{FC02}
\textsc{P.~F\"{o}ldi, A.~Czirj\'{a}k, B.~Moln\'{a}r et~al.}, Opt. Express
  \textbf{10}, 376 (2002).

\bibitem{FBC02b}
\textsc{P.~F\"{o}ldi, M.~G. Benedict and A.~Czirj\'{a}k}, J. Mod. Opt.
  \textbf{49}, 1263 (2002).

\bibitem{FBC02a}
\textsc{P.~F\"{o}ldi, M.~G. Benedict and A.~Czirj\'{a}k}, Phys. Rev. A
  \textbf{65}, 021802 (2002).

\bibitem{F02}
\textsc{P.~F\"{o}ldi, M.~G. Benedict, A.~Czirj\'{a}k et~al.}, Phys. Rev. A
  \textbf{67}, Art. No. 032104 (2003), \urlprefix\url{quant-ph/0208141}.

\bibitem{ACGT72}
\textsc{F.~Arecchi, E.~Courtens, R.~Gilmore et~al.}, Phys. Rev. A \textbf{6},
  2221 (1972).

\bibitem{MS91}
\textsc{P.~Meystre and M.~Sargent}, \emph{Elements of Quantum Optics}
  (Springer-Verlag, Berlin, Heidelberg, New York, 1991), 2 edn.

\bibitem{Z70}
\textsc{H.~D. Zeh}, Found. Phys \textbf{1}, 69 (1970).

\bibitem{Z81}
\textsc{W.~H. Zurek}, Phys. Rev. D \textbf{24}, 1516 (1981).

\bibitem{Z82}
\textsc{W.~H. Zurek}, Phys. Rev. D \textbf{26}, 1862 (1982).

\bibitem{CL83}
\textsc{A.~O. Caldeira and A.~J. Leggett}, Physica A \textbf{121}, 587 (1983).

\bibitem{UZ89}
\textsc{W.~G. Unruh and W.~H. Zurek}, Phys. Rev. D \textbf{40}, 1071 (1989).

\bibitem{HPZ92}
\textsc{B.~L. Hu, J.~P. Paz and Y.~Zhang}, Phys. Rev. D \textbf{45}, 2843
  (1992).

\bibitem{D81}
\textsc{H.~Dekker}, Phys. Rep. \textbf{80}, 1 (1981).

\bibitem{SW85}
\textsc{C.~M. Savage and D.~F. Walls}, Phys. Rev. A \textbf{32}, 2316 (1985).

\bibitem{D89}
\textsc{L.~Di\'osi}, Phys. Rev. A \textbf{40}, 1165 (1989).

\bibitem{BJK99}
\textsc{S.~Bose, K.~Jacobs and P.~L. Knight}, Phys. Rev. A \textbf{59}, 3204
  (1999).

\bibitem{Z93}
\textsc{W.~H. Zurek}, Progr. Theor. Phys \textbf{89}, 281 (1993).

\bibitem{GJK96}
\textsc{D.~Giulini, E.~Joos, C.~Kiefer et~al.}, \emph{Decoherence and the
  Appearance of a Classical World in Quantum Theory} (Springer-Verlag, Berlin,
  Heidelberg, New York, 1996).

\bibitem{BHD96}
\textsc{M.~Brune, E.~Hagley, J.~Dreyer et~al.}, Phys. Rev. Lett. \textbf{77},
  4887 (1996).

\bibitem{MKT00}
\textsc{C.~J. Myatt, B.~King, Q.~A. Turchette et~al.}, Nature \textbf{403}, 269
  (2000).

\bibitem{FPC00}
\textsc{J.~R. Friedman, V.~Patel, W.~Chen et~al.}, Nature \textbf{406}, 43
  (2000).

\bibitem{K86}
\textsc{F.~K\'arolyh\'azy, A.~Frenkel and B.~Luk\'acs}, \emph{Quantum concepts
  of space and time}, pp. 109--128 (Clarendon Press, 1986).

\bibitem{P86}
\textsc{R.~Penrose}, \emph{Quantum concepts of space and time}, pp. 129--146
  (Clarendon Press, 1986).

\bibitem{EMN89}
\textsc{J.~Ellis, S.~Mohanty and D.~V. Nanopoulos}, Phys. Lett. B \textbf{221},
  113 (1989).

\bibitem{N32}
\textsc{J.~Neumann}, \emph{Mathematische Grundlagen der Quantenmechanik}
  (Springer, Berlin, 1932).

\bibitem{EPR35}
\textsc{A.~Einstein, B.~Podolsky and N.~Rosen}, Phys. Rev. \textbf{47}, 777
  (1935).

\bibitem{B87}
\textsc{J.~S. Bell}, \emph{Speakable and unspeakable in quantum mechanics}
  (Cambridge University Press, Cambridge, 1987).

\bibitem{SCH07}
\textsc{E.Schmidt}, Math. Annalen \textbf{63}, 433 (1907).

\bibitem{KZ73}
\textsc{O.~K\"{u}bler and H.~Zeh}, Ann. Phys. \textbf{76}, 405 (1973).

\bibitem{EK95}
\textsc{A.~Ekert and P.~L. Knight}, Am. J. Phys. \textbf{63}, 415 (1995).

\bibitem{CLE02}
\textsc{K.~W. Chan, C.~K. Law and J.~H. Eberly}, Phys. Rev. Lett. \textbf{88},
  100402 (2002).

\bibitem{DHR01}
\textsc{M.~J. Donald, M.~Horodecki and O.~Rudolph}, \emph{The uniqueness
  theorem for entanglement measures} (2001), \urlprefix\url{quant-ph/0105017}.

\bibitem{P98}
\textsc{J.~Preskill}, \emph{Quantum information and computation} (1998),
  \urlprefix\url{A little effort leads to the correct url \ldots}.

\bibitem{Z98}
\textsc{K.~\.{Z}yczkowski, P.~Horodecki, A.~Sanpera et~al.}, Phys. Rev. A
  \textbf{58}, 883 (1998).

\bibitem{Z99}
\textsc{K.~\.{Z}yczkowski}, Phys. Rev. A \textbf{60}, 3496 (1999).

\bibitem{WM85}
\textsc{D.~F. Walls and G.~J. Milburn}, Phys. Rev. A \textbf{31}, 2403 (1985).

\bibitem{ZHP93}
\textsc{W.~H. Zurek, S.~Habib and J.~P. Paz}, Phys. Rev. Lett. \textbf{70},
  1187 (1993).

\bibitem{D93}
\textsc{L.~Di\'osi}, Europhys. Lett. \textbf{22}, 1 (1993).

\bibitem{G90}
\textsc{C.~W. Gardiner}, \emph{Handbook of Stochastic Methods}
  (Springer-Verlag, Berlin, 1990), 2 edn.

\bibitem{WW30}
\textsc{V.~Weisskopf and E.~P. Wigner}, Z. Physik \textbf{63}, 54 (1930).

\bibitem{WW31}
\textsc{V.~Weisskopf and E.~P. Wigner}, Z. Physik \textbf{65}, 18 (1931).

\bibitem{A74}
\textsc{G.~S. Agarwal}, \emph{Quantum Statistical Theories of Spontaneous
  Emission and their Relation to Other Approaches}, vol.~70 of \emph{Springer
  tracts in modern physics}, pp. 1--129 (Springer-Verlag, Berlin, Heidelberg,
  New York, 1974).

\bibitem{N58}
\textsc{S.~Nakajima}, Prog. Theor. Phys. \textbf{20}, 948 (1958).

\bibitem{Z60a}
\textsc{R.~Zwanzig}, J. Chem. Phys. \textbf{33}, 1338 (1960).

\bibitem{Z60b}
\textsc{R.~Zwanzig}, \emph{Statistical mechanics of irreversibility}, vol.~3 of
  \emph{Boulder Lectures in Theoretical Physics}, pp. 106--141 (Interscience,
  1960).

\bibitem{WM94}
\textsc{D.~F. Walls and G.~J. Milburn}, \emph{Quantum Optics} (Springer-Verlag,
  Berlin, 1994).

\bibitem{CTDL77}
\textsc{C.~Cohen-Tannoudji, B.~Diu and F.~Lalo\"{e}}, \emph{M\'{e}chanique
  Quantique}, vol.~2 (Hermann, Paris, 1977), 2 edn.

\bibitem{SB96}
\textsc{R.~Schack and T.~A. Brun}, \emph{A C++ Library Using Quantum
  Trajectories to Solve Quantum Master Equations} (1996),
  \urlprefix\url{quant-ph/9608004}.

\bibitem{AMO98}
\textsc{O.~Arratia, M.~A. Martn and M.~A. del Olmo}, \emph{Deformation in Phase
  Space} (1998), \urlprefix\url{math-ph/9805016}.

\bibitem{CG69ab}
\textsc{K.~E. Cahill and R.~J. Glauber}, Phys. Rev. \textbf{177}, 1882 (1969),
  \emph{ibid.} 1882.

\bibitem{A81}
\textsc{G.~S. Agarwal}, Phys. Rev. A \textbf{24}, 2889 (1981).

\bibitem{S57}
\textsc{R.~L. Stratonovich}, Sov. Phys. JETP \textbf{4}, 891 (1957).

\bibitem{BL81}
\textsc{L.~C. Biederharn and J.~D. Louck}, \emph{Angular Momentum in Quantum
  Physics} (Addison-Wesley, Reading MA, 1981).

\bibitem{BC99}
\textsc{M.~G. Benedict and A.~Czirj\'{a}k}, Phys. Rev. A \textbf{60}, 4034
  (1999).

\bibitem{G63a}
\textsc{R.~J. Glauber}, Phys. Rev. \textbf{130}, 2529 (1963).

\bibitem{G63b}
\textsc{R.~J. Glauber}, Phys. Rev. \textbf{131}, 2766 (1963).

\bibitem{JDA93}
\textsc{J.~Janszky, P.~Domokos and P.~Adam}, Phys. Rev. A \textbf{48}, 2213
  (1993).

\bibitem{B90}
\textsc{M.~Brune, S.~Haroche, V.~Lefevre et~al.}, Phys. Rev. Lett. \textbf{65},
  976  (1990).

\bibitem{PS86}
\textsc{J.~Parker and C.~R. Stroud}, Phys. Rev. Lett. \textbf{56}, 716 (1986).

\bibitem{AS00}
\textsc{D.~L. Aronstein and J.~C.~R.~Stroud}, Phys. Rev. A \textbf{62},
  022102/1 (2000).

\bibitem{VE94}
\textsc{S.~I. Vetchinkin and V.~V. Eryomin}, Chem. Phys. Lett. \textbf{222},
  394 (1994).

\bibitem{BM99}
\textsc{M.~G. Benedict and B.~Moln\'{a}r}, Phys. Rev. A \textbf{60}, R1737
  (1999).

\bibitem{MF02}
\textsc{B.~Moln\'{a}r, P.~F\"{o}ldi, M.~G. Benedict et~al.}, Europh. Lett.
  \textbf{61}, 445 (2003), \urlprefix\url{quant-ph/0202069}.

\bibitem{HH79}
\textsc{K.~P. Huber and G.~Herzberg}, \emph{Molecular spectra and molecular
  structure IV. Constants of diatomic molecules} (van Nostrand Reinhold, 1979).

\bibitem{MBB01}
\textsc{B.~Moln\'{a}r, M.~G. Benedict and J.~Bertrand}, J. Phys. A: Math. Gen.
  \textbf{34}, 3139 (2001).

\bibitem{BA99}
\textsc{J.~Banerji and G.~S. Agarwal}, Opt. Exp. \textbf{5}, 220 (1999).

\bibitem{BIA01}
\textsc{J.~Bertrand and M.~Irac-Astaud}, Czech J. Phys. \textbf{51}, 1272
  (2001).

\bibitem{MBF01}
\textsc{B.~Moln\'{a}r, M.~G. Benedict and P.~F\"{o}ldi}, Fortschr. Phys.
  \textbf{49}, 1053 (2001).

\bibitem{JC63}
\textsc{E.~T. Jaynes and F.~W. Cummings}, Proc. Inst. Elect. Eng. \textbf{51},
  89 (1963).

\bibitem{ENSM80}
\textsc{J.~H. Eberly, N.~B. Narozhny and J.~J. Sanchez-Mondragon}, Phys. Rev.
  Lett. \textbf{44}, 1323 (1980).

\bibitem{AP89}
\textsc{I.~S. Averbukh and N.~F. Perelman}, Phys. Lett. \textbf{A139}, 449
  (1989).

\bibitem{LAS96}
\textsc{C.~Leichtle, I.~S. Averbukh and W.~P. Schleich}, Phys. Rev. A
  \textbf{54}, 5299 (1996).

\bibitem{DK00}
\textsc{P.~Domokos, T.~Kiss, J.~Janszky et~al.}, Chem. Phys. Lett.
  \textbf{322}, 255 (2000).

\bibitem{ER91}
\textsc{J.~Eiselt and H.~Risken}, Phys. Rev. A \textbf{43}, 346 (1991).

\bibitem{JV94}
\textsc{J.~Janszky, A.~V. Vinogradov, T.~Kobayashi et~al.}, Phys. Rev. A
  \textbf{50}, 1777 (1994).

\bibitem{BS40}
\textsc{F.~Bloch and A.~Siegert}, Phys. Rev. \textbf{57}, 522 (1940).

\bibitem{AT55}
\textsc{S.~H. Autler and C.~H. Townes}, Phys. Rev. \textbf{100}, 703 (1955).

\bibitem{YL98}
\textsc{Y.-M. Yuan and W.-K. Liu}, Phys. Rev. A \textbf{57}, 1992 (1998).

\bibitem{L66}
\textsc{M.~Lax}, Phys. Rev. \textbf{145}, 110 (1966).

\bibitem{A73}
\textsc{G.~S. Agarwal}, \emph{Master Equation methods in quantum optics},
  vol.~XI of \emph{Progress in Optics}, pp. 3--73 (North Holland, 1973).

\bibitem{R65}
\textsc{F.~Reif}, \emph{Fundamentals of Statistical and Thermal Physics}
  (McGraw-Hill, Singapore, 1965).

\bibitem{KP95}
\textsc{P.~Kasperkovitz and M.~Peev}, Phys. Rev. Lett. \textbf{75}, 990 (1995).

\bibitem{BBH00}
\textsc{D.~Braun, P.~A. Braun and F.~Haake}, Opt. Comm. \textbf{179}, 411
  (2000).

\bibitem{B01}
\textsc{C.~Brif, H.~Rabitz, S.~Wallentowitz et~al.}, Phys. Rev. A \textbf{63},
  063404 (2001).

\bibitem{KZ90}
\textsc{L.~Khundkar and A.~H. Zewail}, Annu. Rev. Phys. Chem. \textbf{41}, 15
  (1990), and see also references therein.

\bibitem{WT00}
\textsc{C.~Warmuth, A.~Tortschanoff, F.~Milota et~al.}, J. Chem. Phys.
  \textbf{112}, 5060 (2000).

\bibitem{PHBK99}
\textsc{M.~B. Plenio, S.~F. Huelga, A.~Beige et~al.}, Phys. Rev. A \textbf{59},
  2468 (1999).

\bibitem{BD00}
\textsc{C.~H. Bennett and D.~DiVincenzo}, Nature \textbf{404}, 247 (2000).

\bibitem{BSCHH71}
\textsc{R.~Bonifacio, P.~Schwendimann and F.~Haake}, Phys. Rev. A \textbf{4},
  302 (1971).

\bibitem{HR85}
\textsc{S.~Haroche and J.~M. Raimond}, vol.~20 of \emph{Advances in atomic and
  molecular physics}, pp. 347--411 (1985).

\bibitem{BCB97}
\textsc{M.~G. Benedict, A.~Czirj\'ak and C.~Benedek}, Acta. Phys. Slov.
  \textbf{47}, 259 (1997).

\bibitem{JV90}
\textsc{J.~Janszky and A.~V. Vinogradov}, Phys. Rev. Lett. \textbf{64}, 2771
  (1990).

\bibitem{APS97}
\textsc{G.~S. Agarwal, R.~R. Puri and R.~P. Singh}, Phys. Rev. A \textbf{56},
  2249 (1997).

\bibitem{GG97}
\textsc{C.~C. Gerry and R.~Grobe}, Phys. Rev. A \textbf{56}, 2390 (1997).

\bibitem{GG98}
\textsc{C.~C. Gerry and R.~Grobe}, Phys. Rev. A \textbf{57}, 2247 (1998).

\bibitem{D54}
\textsc{R.~M. Dicke}, Phys. Rev. \textbf{93}, 439 (1954).

\bibitem{DASCH94}
\textsc{J.~Dowling, G.~S. Agarwal and W.~P. Schleich}, Phys. Rev. A
  \textbf{49}, 4101 (1994).

\bibitem{CB96}
\textsc{A.~Czirj\'{a}k and M.~G. Benedict}, Quantum Semiclass. Opt. \textbf{8},
  975 (1996).

\bibitem{BM98}
\textsc{C.~Brif and A.~Mann}, J. Phys. A: Math. Gen. \textbf{31}, L9 (1998).

\bibitem{CFW99}
\textsc{S.~M. Chumakov, A.~Frank and K.~B. Wolf}, Phys. Rev. A \textbf{60},
  1817 (1999).

\bibitem{S95}
\textsc{P.~W. Shor}, Phys. Rev. A \textbf{52}, 2493 (1995).

\bibitem{LCW98}
\textsc{D.~A. Lidar, I.~L. Chuang and K.~B. Whaley}, Phys. Rev. Lett.
  \textbf{81}, 2494 (1998).

\bibitem{BBTK00}
\textsc{A.~Beige, D.~Braun, B.~Tregenna et~al.}, Phys. Rev. Lett. \textbf{85},
  1762 (2000).

\bibitem{SRAD}
\textsc{M.~G. Benedict, A.~M. Ermolaev, V.~A. Malyshev et~al.},
  \emph{Superradiance} (IOP, Bristol, 1996).

\bibitem{GH82}
\textsc{M.~Gross and S.~Haroche}, Phys. Rep. \textbf{93}, 301  (1982).

\bibitem{P85}
\textsc{D.~Pavolini, A.~Crubellier, P.~Pillet et~al.}, Phys. Rev. Lett.
  \textbf{54}, 1917 (1985).

\bibitem{DB96}
\textsc{R.~G. DeVoe and R.~G. Brewer}, Phys. Rev. Lett. \textbf{76}, 2049
  (1996).

\bibitem{WVHW99}
\textsc{M.~Weidinger, B.~T. H., Varcoe et~al.}, Phys. Rev. Lett. \textbf{82},
  3795 (1999).

\bibitem{H97}
\textsc{E.~Hagley, X.~M\^{a}itre, G.~Nogues et~al.}, Phys. Rev. Lett.
  \textbf{79}, 1 (1997).

\bibitem{R99}
\textsc{A.~Rauschenbeutel, G.~Nogues, S.~Osnaghi et~al.}, Phys. Rev. Lett.
  \textbf{83}, 5166 (1999).

\bibitem{H62}
\textsc{M.~Hamermesh}, \emph{Group Theory and its Applications to Physical
  Problems} (Dover Publications, New York, 1962).

\bibitem{CP87}
\textsc{A.~Crubellier and D.~Pavolini}, J. Phys. B \textbf{20}, 1451 (1987).

\bibitem{KSS92}
\textsc{C.~H. Keitel, M.~O. Scully and G.~S\"{u}ssmann}, Phys. Rev. A
  \textbf{45}, 3242 (1992).

\bibitem{SM99}
\textsc{A.~S{\o}rensen and K.~M{\o}lmer}, Phys. Rev. Lett. \textbf{82}, 1971
  (1999).

\bibitem{ZG00}
\textsc{S.~B. Zheng and G.~C. Guo}, Phys. Rev. Lett. \textbf{85}, 2392 (2000).

\bibitem{N99}
\textsc{H.~C. N\"{a}gerl, D.~Leibfried, H.~Rohde et~al.}, Phys. Rev. A
  \textbf{60}, 145 (1999).

\bibitem{TC68}
\textsc{M.~Tavis and F.~W. Cummings}, Phys. Rev. \textbf{170}, 379 (1968).

\bibitem{LL65}
\textsc{L.~D. Landau and E.~M. Lifshitz}, \emph{Quantum Mechanics,
  Nonrelativistic Theory} (Pergamon Press, Oxford, 1965).

\bibitem{ASW01}
\textsc{G.~S. Agarwal, M.~O. Scully and H.~Walther}, Phys. Rev. Lett.
  \textbf{86}, 4271 (2001).

\bibitem{O01}
\textsc{S.~Osnaghi, P.~Bertet, A.~Auffeves et~al.}, Phys. Rev. Lett.
  \textbf{87}, 037902 (2001).

\bibitem{K99}
\textsc{P.~G. Kwiat, E.~Waks, A.~G. White et~al.}, Phys. Rev. A \textbf{60},
  R773 (1999).

\end{thebibliography}
\end{document}